\documentclass[useAMS,usenatbib]{mn2e}

\usepackage{graphicx}
\usepackage{times}%
\usepackage{natbib}
\usepackage{graphics}
\usepackage{epsfig}
\usepackage{array}
\usepackage{stfloats}
\usepackage{fixltx2e}
\usepackage{amsmath}

\newcommand{\lesssim}{\la} 
\newcommand{\lya}{Lyman-$\alpha$~}

\newcommand{\ie}{{i.e.}}
\newcommand{\eg}{{e.g.}}
\newcommand{\gsim}{\,\lower2truept\hbox{${>\atop\hbox{\raise4truept\hbox{$\sim$}}}$}\,}

\def\eg{{\rm e.g.$\,$}}
\def\ie{{\rm i.e.$\,$}}

\newcommand{\be}{\begin{equation}}
\newcommand{\ee}{\end{equation}}
\newcommand{\bea}{\begin{eqnarray}}
\newcommand{\eea}{\end{eqnarray}}

\renewcommand{\vec}[1]{ {\bmath #1} } 

\def\ltsima{$\; \buildrel < \over \sim \;$}
\def\simlt{\lower.5ex\hbox{\ltsima}}
\def\gtsima{$\; \buildrel > \over \sim \;$}
\def\simgt{\lower.5ex\hbox{\gtsima}}

\title[Time dependent couplings in the dark sector]{Time dependent couplings in the dark sector: from background evolution to nonlinear structure formation}

\author[M. Baldi]{Marco Baldi
\\Excellence Cluster Universe, Boltzmannstr.~2, D-85748 Garching, Germany
\\University Observatory, Ludwig-Maximillians University Munich, Scheinerstr. 1, D-81679 Munich, Germany}

\begin{document}


\pagerange{\pageref{firstpage}--\pageref{lastpage}} \pubyear{2010}

\maketitle

\label{firstpage}

\begin{abstract}	

  We present a complete numerical study 
  of cosmological models with a time dependent
  coupling between the dark energy component driving the present accelerated expansion of the Universe and the Cold Dark Matter (CDM) fluid.
  Depending on the functional form of the coupling strength, these models show 
  a range of possible intermediate behaviors between the standard $\Lambda $CDM background evolution
  and the widely studied case of interacting dark energy models with a constant coupling. 
  These different background evolutions play a crucial role in the growth of cosmic structures, 
  and determine strikingly different effects of the coupling on the internal dynamics of nonlinear objects. 
  By means of a suitable modification of the cosmological N-body code {\small GADGET-2} we have performed 
  a series of high-resolution N-body simulations of structure formation in the context of 
  interacting dark energy models with variable couplings.
  Depending on the type of background evolution, the halo density profiles 
  are found to be either less or more concentrated with respect to $\Lambda $CDM, contrarily to what happens 
  for constant coupling models where concentrations can only decrease.
   However, for some specific 
  choice of the interaction function the reduction of halo concentrations can be larger than in constant coupling scenarios.
  We also find that different types of coupling evolution determine 
  specific features in the growth of large scale structures, like peculiar distortions of the matter power spectrum shape 
  or different time evolutions of the halo mass function.
  Furthermore, also for time dependent couplings 
  baryons and CDM develop a bias already on large scales, which is progressively enhanced
  for smaller and smaller scales, and the effect can be significantly larger compared to constant coupling scenarios.
  The same happens to the baryon fraction of halos, which can be more significantly reduced below its universal value 
  in variable coupling models with respect to constant coupling cosmologies.
  In general, we find that time dependent interactions between dark energy and CDM can in some cases 
  determine stronger effects on structure formation as compared to the constant coupling case, with a significantly weaker impact on the
  background evolution of the Universe, and might therefore provide a more viable possibility to 
  alleviate the tensions between observations and the $\Lambda $CDM model on small scales than the constant coupling scenario.
\end{abstract}



\section{Introduction}
\label{i}

According to the present interpretation of the vast amount of cosmological data that have been collected over the last years, the Universe in which we live has a nearly flat spatial geometry, it expands with a rate of $\sim 70 $ km s$^{-1}$ Mpc$^{-1}$, and the total amount of matter it contains accounts only for $\sim 27\%$ of the critical energy density that is required to justify its spatial flatness. Furthermore, the expansion seems to have entered an accelerated phase since about $6$ billion years, and the missing $\sim 73\%$ of the critical energy density is assumed to be in the form of some dark energy (DE) component able to drive such accelerated expansion. This detailed picture can be obtained by combining a wide variety of different and complementary cosmological datasets, ranging from Cosmic Microwave Backround (CMB) \citep{wmap5,wmap7}, to Large Scale Structure surveys \citep{Percival_etal_2001,Cole_etal_2005,Reid_etal_2010}, to observations of Type Ia Supernovae (SnIa) \citep{Riess_etal_1998,Perlmutter_etal_1999,SNLS,Kowalski_etal_2008} and Baryon Acoustic Oscillations (BAO) \citep[\eg][]{Percival_etal_2009}. 
While the matter fraction of the Universe is known to be composed only for $\sim 15\%$ by baryonic particles, with the remaining mass in the form of some collisionless, weakly interacting Cold Dark Matter (CDM) component, the nature of the DE fraction is yet completely unknown, and its understanding constitutes one of the major challenges in modern cosmology.

The simplest possibility of a cosmological constant $\Lambda $, whose energy density remains constant throughout the whole expansion history of the Universe,  is in good agreement with a very large amount of cosmological and astrophysical data, and this is the reason for the establishment of the $\Lambda $CDM scenario as the present standard cosmological model. Nevertheless, the nature of the cosmological constant raises two fundamental questions concerning the very finely tuned value of its energy density (the ``fine tuning problem") and the beginning of its domination over CDM only at relatively recent cosmological epochs (the ``coincidence problem").
For this reason, alternative possibilities of dynamically evolving DE components have been proposed, in particular models where the DE is identified with a classical scalar field as for the case of quintessence \citep{Wetterich_1988,Ratra_Peebles_1988} or k-essence \citep{ArmendarizPicon_etal_2000,kessence}.

As a further extension of these dynamical DE scenarios, different possible forms of interaction between the DE component and the matter sector of the Universe have been suggested and investigated in the literature, as \eg the generalized Chaplygin gas \citep[see \eg][]{Kamenshchik_etal_2001, Bilic_Tupper_Viollier_2002,Bento_etal_2002,Carturan_Finelli_2003,Amendola_etal_2003b}, unified dark matter models \citep{Mainini_Bonometto_2004,Bertacca_etal_2007,Bertacca_Bartolo_Matarrese_2010}, extended quintessence models \citep{Perrotta_etal_2000,Baccigalupi_Matarrese_2000,Pettorino_etal_2005}, and coupled DE \citep{Wetterich_1995,Amendola_2000,Amendola_2004,Pettorino_Baccigalupi_2008}. Any form of interaction between the DE sector and other matter species, like CDM or massive neutrinos \citep[as in \eg ][]{Amendola_Baldi_Wetterich_2008} would leave distinctive features in the background expansion history of the Universe and in the growth of cosmic structures \citep[see \eg ][]{Brax_etal_2010} which could provide new ways to tackle the problem of the nature of DE. 
It is therefore essential to understand the impact that the interaction would have on observable quantities as \eg the properties of CMB \citep{Amendola_etal_2003,Amendola_Quercellini_2003,Bean_etal_2008,LaVacca_etal_2009}, of large scale structure formation \citep{Koivisto_2005,Bean_etal_2008,Baldi_Pettorino_2010,Baldi_Viel_2010}, and of the nonlinear newtonian dynamics at small scales \citep{Perrotta_etal_2003,Mainini:2006zj,Maccio_etal_2004,Saracco_etal_2010,Baldi_etal_2010}. 

In the present work, in particular, we consider the case of a cosmological scenario where the DE scalar field interacts with the CDM fluid with a coupling strength that evolves in time, therefore generalizing the very widely studied case of constant couplings for the DE interaction. Other forms of effectively time dependent couplings, where the interaction depends on linear or nonlinear combinations of the energy densities of the interacting fluids and of their time derivatives have been studeid in \eg \citet{Barrow_Clifton_2006,CalderaCabral_2009,Chimento_2010}. Here we want to focus on general functional forms of the coupling strength in the context of coupled quintessence models where the interaction term is proportional to the energy density of the matter coupled fluid. One of the main motivations behind the introduction of a time dependence of the DE coupling to the matter sectors -- besides the fact that an evolving coupling is {\em per se} a more general and natural assumption than a constant interaction strength -- lies in the recent discovery \citep{Baldi_etal_2010} that the effects of the DE-CDM interaction on the formation and evolution of structures at small scales, in particular in the nonlinear regime, might help alleviating the tensions between the $\Lambda $CDM model and a series of astrophysical observations. These range from  \eg the abundance of satellites in CDM halos \citep{Navarro_Frenk_White_1995}, to the observed low baryon fraction in large galaxy clusters \citep{Ettori_2003,Allen_etal_2004,Vikhlinin_etal_2006,LaRoque_etal_2006,McCarthy_etal_2007}, to the so called ``cusp-core" problem for the density profiles of the CDM halos of dwarf galaxies \citep{Moore_1994,Flores_Primack_1994,Simon_etal_2003}, of spiral galaxies \citep{Navarro_steinmetz_2000,Salucci_Burkert_2000,Salucci_2000,Binney_Evans_2001}, and of galaxy clusters \citep{Sand_etal_2002,Sand_etal_2004,Newman_etal_2009}, or to the so called ``Dark Flow" problem \citep{Watkins_etal_2008}. Furthermore, some new potential challenges to the $\Lambda $CDM scenario have been recently reported based on the detection of very massive high-redshift clusters \citep[see \eg][]{Jee_etal_2009,Rosati_etal_2009} or on the observed dynamical properties of CDM halo satellites \citep{Lee_Komatsu_2010,Lee_2010}.

In their recent study \citet{Baldi_etal_2010} showed by means of a series of high-resolution N-body simulations how the DE-CDM interaction -- for the case of constant coupling -- could alleviate some of these problems; in particular, it was shown how the interaction can reduce the ``cuspyness" of massive CDM halos, thereby going in the right direction for a solution of the ``cusp-core" problem. Nevertheless, the magnitude of this effect is strongly limited by the tight observational constraints on constant coupling models \citep[as \eg][]{Bean_etal_2008,LaVacca_etal_2009} that put a firm bound to the maximum allowed value of the coupling. 
It is therefore natural to speculate about the possibility that a time dependent coupling with a large value during the late stages of structure formation and a progressively smaller value at high redshift might increase significantly the impact of the new physics introduced by the interaction on \eg the density profiles of CDM halos without perturbing the overall evolution of the Universe beyond the present observational limits.

The present work constitutes the natural extension of the analysis done by \citet{Baldi_etal_2010} to the case of time dependent couplings. In this paper we perform a complete numerical study of some quite general classes of coupling functions, starting from their background evolution up to the nonlinear regime of structure formation, and we present the first high-resolution hydrodynamical N-body simulations of structure formation in the context of interacting DE models with a time dependent coupling to date.

The manuscript is organized as follows. In Section~\ref{cde} we describe the main features of the coupled DE models under investigation with a particular stress on the differences with respect to the standard constant coupling models that have been widely studied in the literature. More specifically, in Sec.~\ref{bkg} we discuss the background equations and in Sec.~\ref{integration} we illustrate the numerical methods used to integrate such equations backwards in time. In Sec.~\ref{obs} we discuss observational constraints on the coupled DE scenario and a possible way to use the bounds derived for constant coupling models to check the viability of the variable coupling cosmologies investigated in the present work. In Sec.~\ref{prt} we study linear perturbations equations and we discuss the evolution of linear matter density fluctuations in variable coupling models.\\
In Sec.~\ref{sim} we briefly summarize the numerical methods used in the N-body simulations, and in Sec.~\ref{results} we present and discuss the results of our runs.
Finally, in Sec.~\ref{concl} we draw our conclusions.

\section{Coupled dark energy cosmologies with time dependent couplings} \label{cde}

Coupled DE cosmologies have been widely studied in the last decade for what concerns their background and linear perturbations evolution.
In particular, we refer here to the derivations presented in \citet{Amendola_2000, Amendola_2004, Pettorino_Baccigalupi_2008, Baldi_etal_2010} (BA10, hereafter) and detailed references therein. Although these models of DE interactions have been mainly investigated for the simplified case of a constant coupling strength, the most natural scenario allows for a variation of the coupling along with the dynamic evolution of the scalar field \citep[see \eg][]{Amendola_2004}, and in the present work we want to investigate in detail this more general situation. To this end, we discuss here the most relevant features of coupled DE cosmologies, highlighting the main differences that arise for the case of time dependent couplings with respect to the standard case of a constant coupling.
For this reason we will keep explicit the dependence of the coupling on the scalar field, and we will propose below some possible expressions for its functional form.\\

We consider a DE model based on the dynamical evolution of a classical scalar field $\phi $ rolling down a self-interaction potential $V(\phi )$, such that its intrinsic energy density and pressure can be expressed as:
\begin{eqnarray}
\rho _{\phi } = \frac{1}{2}g^{\mu \nu }\partial _{\mu }\phi \partial _{\nu }\phi +V(\phi ) \,,  \\
p _{\phi } = \frac{1}{2}g^{\mu \nu }\partial _{\mu }\phi \partial _{\nu }\phi -V(\phi ) \,,
\end{eqnarray}
where $g^{\mu \nu }$ is the metric tensor.
The interaction of the scalar field with other fluids can be expressed as a source term in the conservation equations for the different components of the Universe:
\begin{eqnarray}
\label{coupled_cons1}
\nabla_{\mu }T^{\mu }_{(i)\nu} &=& -Q_{(i)}(\phi )T_{(i)}\nabla _{\nu}\phi \,, \\
\label{coupled_cons2}
\nabla_{\mu }T^{\mu }_{(\phi)\nu} &=& \sum _{i}\left[ Q_{(i)}(\phi )T_{(i)}\right] \nabla _{\nu}\phi \,,
\end{eqnarray}
where $\nabla _{\mu }$ represents a covariant derivative, and $T^{\mu}_{(i)\nu}$ is the stress-energy tensor -- and $T_{(i)}$ its trace -- of the $i$-th component of the Universe, with $i = c$ for CDM, $b$ for baryons, $\gamma $ for radiation, $n$ for neutrinos.
Since the system of Eqs.~(\ref{coupled_cons1},\ref{coupled_cons2}) does not violate the conservation of the total stress-energy tensor of the Universe:
\begin{equation}
\nabla _{\mu } \sum_{i}T^{\mu }_{(i)\nu } = 0 \,,
\end{equation}
 this type of interaction is consistent with general covariance and does not modify Einstein equations.

It is also interesting to notice that radiation and relativistic neutrinos always remain uncoupled, since the stress-energy tensor of relativistic particles (subscript $r$) is traceless, $T_{(r)}=0$. For massive neutrinos, instead, if $Q_{n}(\phi ) \ne 0$ the coupling term would become effective as soon as they become nonrelativistic, and such mechanism could provide solutions to the cosmic ``coincidence problem", as proposed in \citet{Amendola_Baldi_Wetterich_2008}.

A coupling of DE to baryonic particles is tightly constrained by Solar System tests of scalar-tensor theories \citep[see \eg][]{Ellis_etal_1989,Carroll_1998,Will_2001}. However, following the idea first proposed by \citet{Damour_Gibbons_Gundlach_1990}, one could consider models with different coupling strengths to baryons and CDM, for which the observational constraints become much broader.
Therefore, since in this paper we are mainly interested in investigating the effects of a possible coupling between DE and CDM, we assume the baryonic coupling $Q_{(b)}(\phi )$ to be vanishing at all times. 

To further simplify the system, we also assume neutrinos to be massless, such that they remain effectively uncoupled at all times and can therefore be considered part of the fraction of relativistic particles of the Universe, together with photons, such that $\rho _{r}=\rho _{\gamma }+\rho _{n}$. However, \citet{LaVacca_etal_2009} showed how releasing this assumption might have a relevant impact on the observationally allowed range of coupling values, and we will discuss this possibility in Sec.~\ref{obs}.

With all these assumptions we are left with only one non-vanishing coupling function, the DE-CDM coupling $Q_{c}(\phi )$, and the system of Eqs.~(\ref{coupled_cons1},\ref{coupled_cons2}), in a flat FLRW metric described by the line element \footnote{We assume the speed of light $c$ to be unity.}:
\begin{equation}
ds^{2} = -dt^{2} + a^{2}(t) \left(\delta _{ij}dx^{i}dx^{j}\right) \,, 
\end{equation}
results in the following set of  dynamic equations:
\begin{eqnarray}
\label{dynamic_equations}
\ddot{\phi } + 3H\dot{\phi } +\frac{dV}{d\phi } &=& \sqrt{\frac{2}{3}}\beta _{c}(\phi ) \frac{\rho _{c}}{M_{Pl}} \,, \nonumber \\
\dot{\rho }_{c} + 3H\rho _{c} &=& -\sqrt{\frac{2}{3}}\beta _{c}(\phi )\frac{\rho _{c}\dot{\phi }}{M_{Pl}} \,, \nonumber \\
\dot{\rho }_{b} + 3H\rho _{b} &=& 0 \,, \\
\dot{\rho }_{r} + 4H\rho _{r} &=& 0\,, \nonumber \\
3H^{2} &=& \frac{1}{M_{Pl}^{2}}\left( \rho _{r} + \rho _{c} + \rho _{b} + \rho _{\phi} \right) \nonumber \,,
\end{eqnarray}
where an overdot represents a derivative with respect to the cosmic time $t$, $H\equiv \dot{a}/a$, $M_{Pl}\equiv 1/\sqrt{8\pi G}$ is the reduced Planck mass, and 
where we have followed the notation of \citet{Amendola_2000} by defining\footnote{Please notice that the definition of the coupling $\beta _{c}$ in the present work is $\sqrt{3/2}$ times the one adopted in BA10.}:
\begin{equation}
\beta _{c}(\phi ) \equiv \sqrt{\frac{3}{2}} M_{Pl} Q_{c}(\phi ) \,.
\end{equation}

An immediate consequence of the coupling source terms in Eqs.~(\ref{dynamic_equations}) is that the density of coupled matter species (CDM in our case) does not scale like $a^{-3}$, but follows an evolution given by the equation:
\begin{equation}
\label{density_variation}
\rho _{c}(a) = \rho _{c}(a_{0})a^{-3}e^{-\sqrt{2/3}\int \beta _{c}(\phi )d\phi /M_{Pl}} \,,
\end{equation}
where the exponential extra factor accounts for the direct exchange of energy between DE and CDM.
If one assumes that the number density of matter particles is conserved (\ie particels are neither created nor destroyed), the direct consequence of Eqn.~(\ref{density_variation}) is that the mass of coupled matter particles has to change in time according to the evolution of the scalar field:
\begin{equation}
\label{mass_variation}
M_{c}(a) = M_{c}(a_{0})e^{-\sqrt{2/3}\int \beta _{c}(\phi )d\phi /M_{Pl}} \,.
\end{equation}

In this study we will always assume an exponential form for the scalar field self-interaction potential:
\begin{equation}
V(\phi )\equiv Ae^{-\sqrt{2/3}\alpha \phi /M_{Pl}} \,,
\end{equation}
where $\alpha > 0$ is a free parameter of the model. We normalize the scalar field evolution such that $\phi (t_{0}) = 0$, which implies $A=V(t_{0})$.

Exponential potentials for the self interaction of a scalar field have been proposed in the context of inflation by \eg \citet{Lucchin_Matarrese_1985, Lucchin_Matarrese_1984} and for the case of late-time acceleration by \eg \citet{Wetterich_1988, Ferreira_Joyce_1998, Amendola_2000}. In both cases they determine the existence of attractor solutions that are almost independent from the scalar field initial conditions.

\subsection{Background evolution} \label{bkg}

Following and generalizing the approach devised by \citet{Copeland_etal_1998}, and subsequently largely applied to the study of coupled quintessence models \citep[see \eg][]{Amendola_2000, Amendola_2004}, we can rewrite the system of Eqs.~(\ref{dynamic_equations}) as an autonomous dynamical system by introducing the dimensionless variables:
\begin{equation}
\label{dimensionless_variables}
x\equiv\frac{\phi '}{M_{Pl}\sqrt{6}} \,, y\equiv\frac{\sqrt{V(\phi )/3}}{M_{Pl} H} \,, r\equiv\frac{\sqrt{\rho _{r}/3}}{M_{Pl} H} \,, v\equiv\frac{\sqrt{\rho _{b}/3}}{M_{Pl} H} \,,
\end{equation}
where now a prime represents a derivative with respect to the e-folding time $N\equiv \ln (a)$.
With these new variables the dimensionless density parameters $\Omega _{i}\equiv \rho _{i}/(3H^{2}M_{Pl}^{2})$ of the different cosmic components can be expressed as:
\begin{equation}
\Omega _{b} = v^2 \,, \Omega _{r} = r^{2} \,, \Omega _{\phi }=x^{2}+y^{2} \,.
\end{equation}
Furthermore, in this work we will always assume the Universe to be exactly flat, such that $\Omega _{c}=1 - x^{2} - y^{2} - r^{2} - v^{2}$.
We can then recast the system of Eqs.~(\ref{dynamic_equations}) in the form:
\begin{eqnarray}
\label{dimensionless_system}
x'&=&  \frac{x}{2}\left( 3x^{2} - 3y^{2} + r^{2} - 3 \right) + \alpha y^{2} \nonumber \\
& &  + \beta _{c}(\phi )\left( 1 - x^{2} -y^{2} - r^{2} -v^{2} \right) \,,\nonumber \\
y'&=&  \frac{y}{2}\left( 3x^{2} - 3y^{2} + r^{2} + 3\right) - \alpha x y \,, \nonumber \\
r'&=&   \frac{r}{2}\left( 3x^{2} -3y^{2} + r^{2} - 1\right) \,, \\
v'&=&  \frac{v}{2}\left( 3x^{2} -3y^{2} + r^{2}\right) \,, \nonumber \\
H'&=&  -\frac{H}{2}\left( 3x^{2} - 3y^{2} + r^{2} +3 \right) \,.\nonumber 
\end{eqnarray}
The dynamical evolution of the system (\ref{dimensionless_system}) has been thoroughly studied in the past, and analytic solutions for the critical points of the system and for their stability have been found for the case of a constant coupling function $\beta _{c}(\phi ) = \beta _{c}$ \citep{Amendola_2000}. One of the most prominent features of these cosmologies consists in the existence of a matter dominated scaling regime -- which has been called a ``$\phi${\em -Matter Dominated Epoch}" ($\phi$MDE, hereafter) in \citet{Amendola_2000} -- between the DE scalar field and the coupled matter component, where the two fluids keep a constant ratio of energy densities before the final accelerated attractor is reached.
Here we generalize the system (\ref{dimensionless_system}) to the case of non-constant couplings investigated in the present work.
For variable couplings, in fact, we need to rewrite the coupling function $\beta _{c}(\phi )$ in terms of the dimensionless variables (\ref{dimensionless_variables}) in order to close the system.

In this work, we will consider three possible different forms for the time evolution of the coupling function $\beta _{c}(\phi )$ which are discussed below, and for each of these forms we will solve numerically the system (\ref{dimensionless_system}) to get the background evolution of each specific model.

\subsubsection{Coupling proportional to a power of the scale factor}
\label{a_beta}

We start exploring a rather phenomenological form for the time evolution of the coupling strength, where $\beta _{c}(\phi )$ evolves as a power of the cosmological scale factor $a$:
\begin{equation}
\beta _{c}(\phi (a)) \equiv \beta _{0}a^{\beta _{1}} \,.
\end{equation}
This is a completely phenomenological parametrization of the time evolution of the coupling, since it does not depend directly on the dynamics of the scalar field $\phi $, and might therefore look quite unphysical, but is particularly easy to implement and integrate, and can already give an idea of the main features that characterize the dynamics of a variable coupling model.
For this case, in order to close the system (\ref{dimensionless_system}), we can just substitute $\beta _{c}(\phi )$ with its expression in terms of the dimensionless time variable of our system, which is the e-folding time $N$, as:
\begin{equation}
\beta _{c}(\phi ) \rightarrow \beta _{0}e^{\beta _{1} N}\,.
\end{equation}

\subsubsection{Coupling proportional to $\Omega _{\phi }$}

Another phenomenological possibility for the time variation of the coupling is to relate the coupling strength to the fractional DE density during cosmic evolution.
In this case, the evolution of the coupling depends on the dynamics of the scalar field as:
\begin{equation}
\beta _{c}(\phi ) \equiv \beta _{0}\frac{\Omega _{\phi }}{\Omega _{\phi }(t_{0})} \,,
\end{equation}
and one can implement this type of coupling in the system (\ref{dimensionless_system}) just by substituting $\beta _{c}(\phi )$ as:
\begin{equation}
\beta _{c}(\phi ) \rightarrow \beta _{0}\frac{x^{2}+y^{2}}{x_{0}^{2}+y_{0}^{2}}\,,
\end{equation}
where we have indicated with a subscript $0$ the value at the present time $N=0$.

\subsubsection{Exponential coupling}
\label{exp_beta}

The most natural form for the evolution of the coupling is given by a direct dependence of $\beta _{c}$ on the value of the scalar field $\phi $, as we stressed above. In particular, we consider here an  exponential coupling in the form:
\begin{equation}
\label{exponential_coupling}
\beta _{c}(\phi ) \equiv \beta _{0}e^{\beta _{1}\phi /M_{Pl}} \,.
\end{equation}
Exponential forms of the coupling strength between DE and CDM have been proposed in \eg \citet{Amendola_2004}.
For this choice, a direct implementation of the coupling function $\beta _{c}(\phi )$ into the system (\ref{dimensionless_system}) is not possible since none of our dimensionless variables (\ref{dimensionless_variables}) represents the scalar field itself.

One possible way to include the coupling (\ref{exponential_coupling}) into our system of dimensionless equations consists in expressing $\beta _{c}(\phi )$ in terms of the potential variable $y$, such that:
\begin{eqnarray}
\beta _{c}(\phi ) &=& \beta _{0}e^{\beta _{1}\phi /M_{Pl}} \rightarrow \beta _{0}\left(\frac{V(\phi )}{A}\right) ^{\sqrt{3/2}\beta _{1}/\alpha} \nonumber \\
&=& \beta _{0}\left( \frac{H^{2}y^{2}}{H^{2}_{0}y^{2}_{0}}\right) ^{\sqrt{3/2}\beta _{1}/\alpha} \,.
\end{eqnarray}

An alternative possibility is to add a further equation to the system (\ref{dimensionless_system}), and to treat the scalar field $\phi $ as an independent degree of freedom.
If we define
\begin{equation}
\xi \equiv \phi /M_{Pl} \,,
\end{equation}
from the definition of the variable $x$ we get that
\begin{equation}
\label{additional_equation}
\xi ' = \sqrt{6} x\,.
\end{equation}
By including Eqn.~(\ref{additional_equation}) into the system (\ref{dimensionless_system}) we can then express the coupling function $\beta _{c}(\phi)$ as:
\begin{equation}
\beta _{c}(\phi) = \beta _{0}e^{\beta _{1}\xi } \,,
\end{equation}
and close the system.

These two approaches are both possible, but we found the latter be simpler to handle numerically, and therefore we will prefer it for the numerical integration of the cosmological background evolution equations described in the next section.

\subsection{Numerical integration of the background equations}
\label{integration}

In order to study the background evolution of cosmological models with a time dependent coupling between DE and CDM, we integrate numerically the system of ordinary differential equations (\ref{dimensionless_system}) plus the additional equation (\ref{additional_equation}) for a series of possible models with the three different types of coupling evolution discussed in Sec.~\ref{bkg}.
Since we want to find viable solutions for the expansion history, \ie we want to realize a cosmological background evolution that ends up at the present time with the observed values of the cosmological parameters, we integrate the system backwards in time, assuming the fiducial values of the cosmological parameters at $z=0$ as our initial conditions.
This procedure is not straightforward, and deserves to be briefly discussed.
The main problem relies in the fact that the stability of the critical points of the autonomous system (\ref{dimensionless_system}) is inverted under time inversion: stable points become unstable and vice versa. This makes the system be attracted towards the ``wrong'' solution in the backwards integration. 

To overcome the problem, we have devised an algorithm that starting from the desired cosmological parameters at $z=0$ integrates the equations backwards in time, and at the end of the integration checks whether the system has reached the desired solution, which is given by a radiation dominated phase ($r\sim1$). If the correct solution has not been found, the initial conditions at $z=0$ are adjusted according to some specific prescription and the system is integrated again. This procedure is repeated until the correct solution is found. 
The adjustment of the initial conditions for the system at $z=0$ consists in changing the ratio of kinetic to potential energy for the scalar field $\phi $, consequently changing the present value of its equation of state parameter $w_{\phi }$, defined as:
\begin{equation}
w_{\phi }\equiv \frac{p _{\phi }}{\rho _{\phi }} = \frac{x^{2} - y^{2}}{x^{2} + y^{2}} \,.
\end{equation}
Therefore, all our final ``correct solutions'' will share the same cosmological parameters at $z=0$ except for $w_{\phi }$ which is allowed to vary in a range $[-1.0,-0.9]$.

For our integrations we assume the most updated set of maximum likelihood cosmological parameters of WMAP-7 combined with BAO measurements and the {\em Hubble Space Telescope} (HST) determination of the Hubble constant as reported in \citet{wmap7}, which are listed in Table \ref{cosmological_parameters}.

\begin{center}
\begin{table}
\begin{center}
\begin{tabular}{cc}
\hline
Parameter & Value\\
\hline
$H_{0}$ & 70.2 km s$^{-1}$ Mpc$^{-1}$\\
$\Omega _{\rm CDM} $ & 0.227 \\
$\Omega _{\rm DE} $ & 0.728 \\
$\sigma_{8}$ & 0.807\\
$ \Omega _{b} $ & 0.0455 \\
$n_{s}$ & 0.961\\
\hline
\end{tabular}
\end{center}
\caption{Cosmological parameters for our set of models, 
  consistent with the WMAP 7 year results for a $\Lambda $CDM cosmology \citep{wmap7}. }
\label{cosmological_parameters}
\end{table}
\end{center}

In order to check the convergence of our method we first apply our integration algorithm to the fiducial $\Lambda$CDM model by imposing $w_{\phi } = -1$ (\ie $x_{0} = 0 \,, y_{0}=\sqrt{\Omega _{DE}}$), and then we extend it to the known cases of an uncoupled scalar field, and a few coupled scalar field models with constant couplings. The numerical solutions of the backwards integrations for these well known models reproduce all the predicted features of uncoupled and coupled quintessence models respectively, like \eg the existence of a ``$\phi$-MDE" phase for the models with a constant coupling.
Finally, we can use our algorithm to integrate a number of variable coupling models, and obtain for each of them the detailed background evolution. 

\begin{figure}
\begin{center}
\includegraphics[scale=0.4]{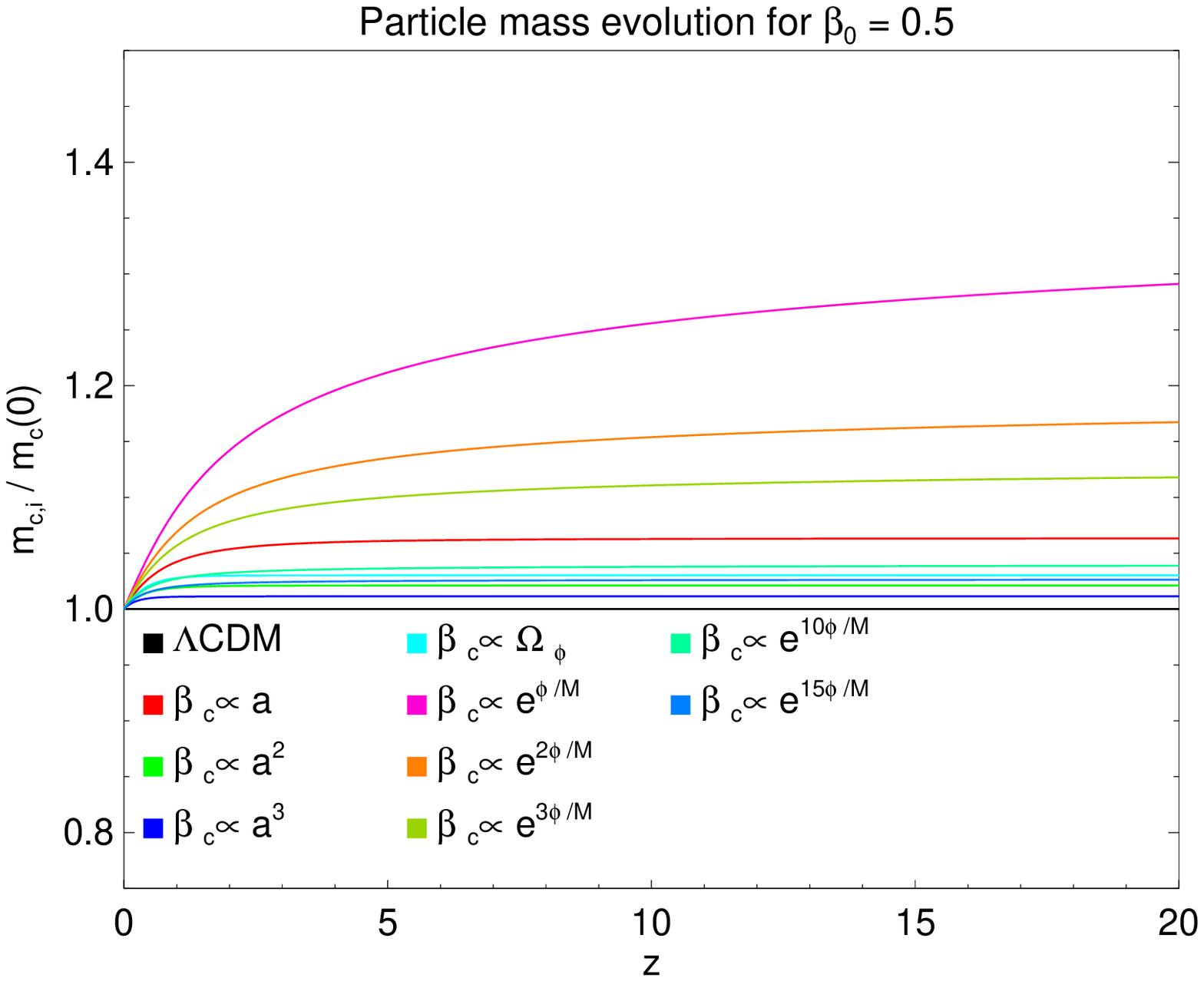}\\
\includegraphics[scale=0.4]{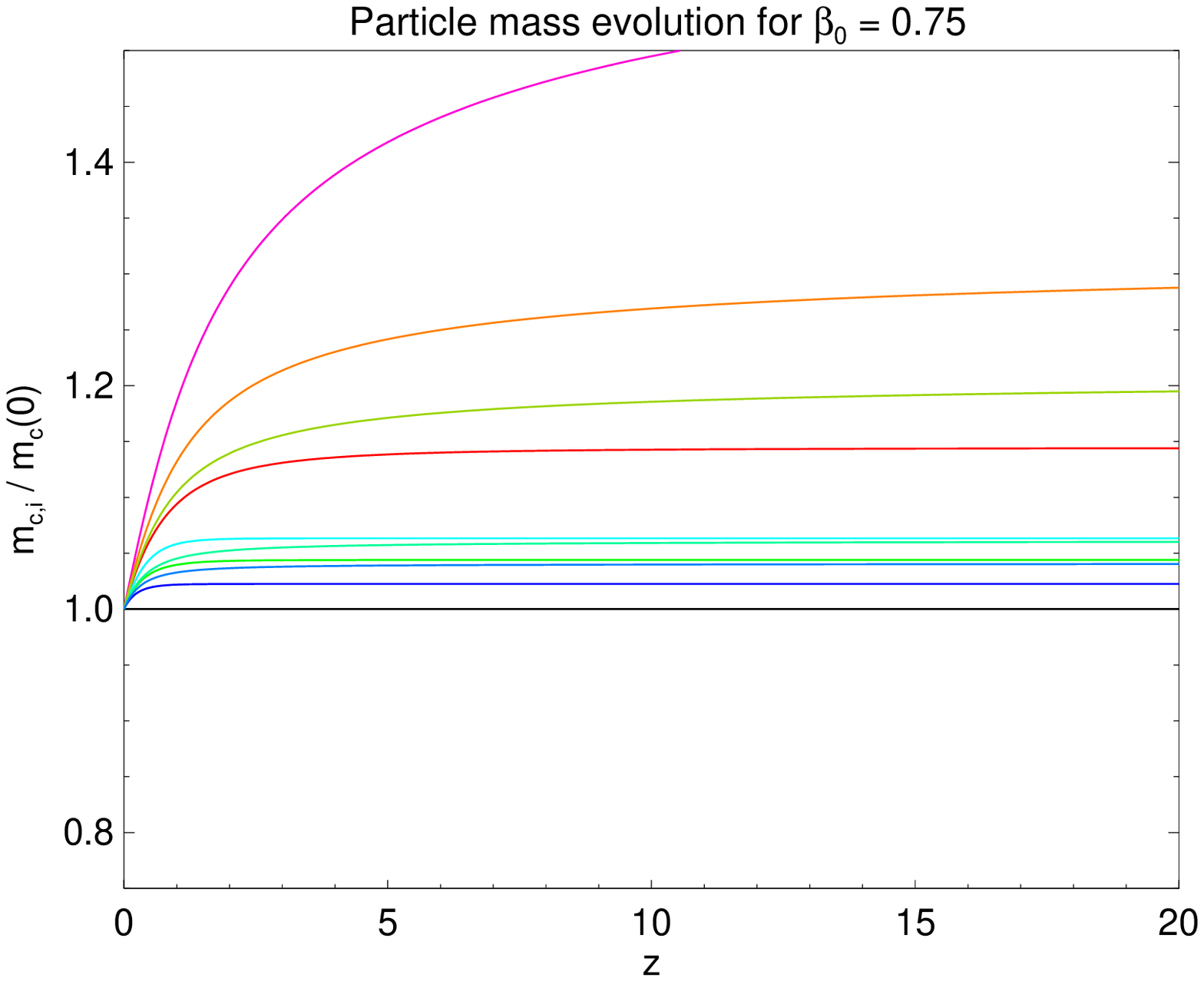}\\
\includegraphics[scale=0.4]{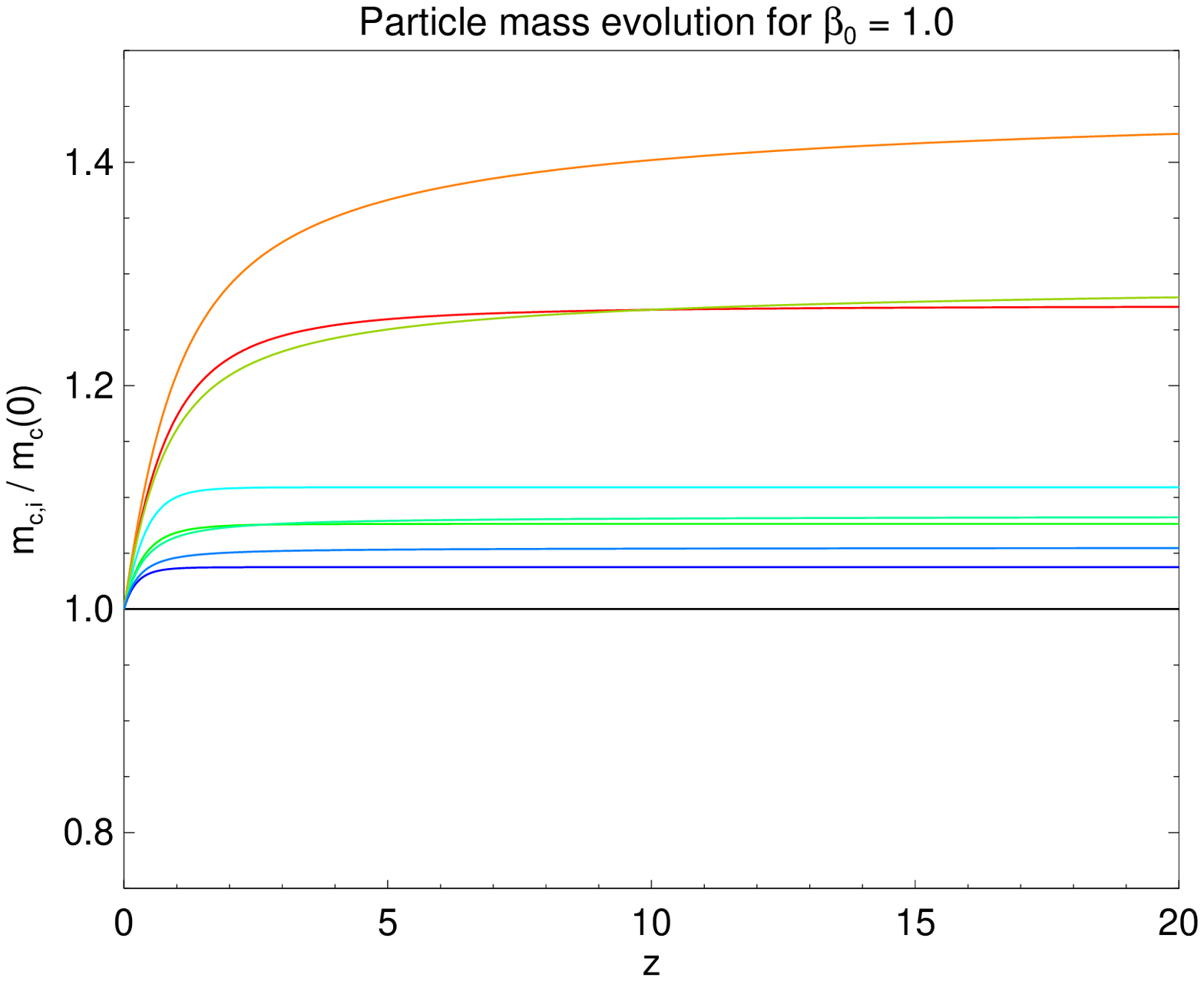}
  \caption{Mass correction of CDM particles as a function of redshift for the different interacting DE models under investigation. The three panels refer to models with three different  values of the coupling at $z=0$, respectively $\beta _{c}(\phi _{0}) = 0.5$ ({\em upper panel}), $\beta _{c}(\phi _{0}) = 0.75$ ({\em middle panel}), and $\beta _{c}(\phi _{0}) = 1.0$ ({\em lower panel}). The color legend for all the models is given in the {\em upper panel}.}
\label{fig:mass_correction}
\end{center}
\end{figure}

All the models, with the specific values of the coupling parameters $\beta _{0}$ and $\beta _{1}$, and the value of the resulting equation of state parameter $w_{\phi }$ at $z=0$ are listed in Table~\ref{cosmological_models}. The potential slope parameter has the same value $\alpha = 0.1$ in all the cosmologies.
The evolution with redshift of the mass correction factor given by Eqn.~\ref{mass_variation} is shown in Fig.~\ref{fig:mass_correction} for all the models listed in Table~\ref{cosmological_models}.

\begin{center}
\begin{table}
\begin{center}
\begin{tabular}{lcccc}
\hline
Model & Type of coupling & $\beta _{0}$ & $\beta _{1}$ & $w_{\phi }(z=0)$\\
\hline
$\Lambda$CDM & -- & -- & -- & -1.0 \\
 EXP000 &  -- & -- & -- & -0.999\\
 \hline
 EXP002 &  $\beta _{c}(\phi ) = \beta _{0}$ & 0.1& -- & -0.995\\
 EXP005 &  '' & 0.25 & -- & -0.984 \\ 
 EXP010 &  '' & 0.5 & -- & -0.945\\
 BE08 &  '' & 0.067 & -- & -0.997 \\
 LV09 &  '' & 0.17 & -- & -0.991\\
 \hline
  EXP010a &  $\beta _{c}(\phi) = \beta _{0}a^{\beta _{1}}$& 0.5 & 1.0 & -0.981 \\
  EXP010a2 &  '' & 0.5 & 2.0 & -0.990 \\
  EXP010a3 &  '' & 0.5 & 3.0 & -0.993 \\
  EXP015a&  '' & 0.75 & 1.0 &  -0.963 \\
  EXP015a2&  '' & 0.75 & 2.0 &  -0.981\\
  EXP015a3&  '' & 0.75 & 3.0 & -0.988 \\
  EXP020a &  '' & 1.0 & 1.0 & -0.938 \\
  EXP020a2 &  '' & 1.0 & 2.0 & -0.970 \\
  EXP020a3 &  '' & 1.0 & 2.0 & -0.981 \\
\hline
  EXP010DE &  $\beta _{c}(\phi )=\beta _{0}\Omega _{\phi }/\Omega _{\phi ,0}$ & 0.5 & -- & -0.986\\
  EXP015DE &  '' & 0.75 & -- & -0.975 \\  
  EXP020DE &  '' & 1.0& -- & -0.960 \\
\hline  
  EXP010e &  $\beta _{c}(\phi )=\beta _{0}e^{\beta _{1}\phi /M}$ & 0.5 & 1.0 & -0.966 \\
  EXP010e2 &  '' & 0.5 & 2.0 & -0.972 \\
  EXP010e3 &  '' & 0.5 & 3.0 & -0.976 \\  
  EXP010e10 &  '' & 0.5 & 10.0 & -0.987 \\  
  EXP010e15 & '' & 0.5 & 15.0 & -0.989 \\  
  EXP015e & '' & 0.75 & 1.0 & -0.935 \\
  EXP015e2 & '' & 0.75 & 2.0 & -0.952 \\
  EXP015e3 & '' & 0.75 & 3.0 & -0.960 \\  
  EXP015e10 & '' & 0.75 & 10.0 & -0.979 \\  
  EXP015e15 & '' & 0.75 & 15.0 & -0.984 \\  
  EXP020e2 & '' & 1.0 & 2.0 & -0.926 \\
  EXP020e3 & '' & 1.0 & 3.0 & -0.941 \\  
  EXP020e10 & '' & 1.0 & 10.0 & -0.971\\  
  EXP020e15 & '' & 1.0 & 15.0 & -0.976 \\      
\hline        
\end{tabular}
\end{center}
\caption{List of all the cosmological models considered in the present background evolution analysis. The models are divided based on the type of coupling function $\beta _{c}(\phi )$ between the DE scalar field $\phi $ and the CDM fluid. The last column indicates the value of the equation of state of the scalar field $w_{\phi }$ at $z=0$ obtained from the numerical integration of the background evolution, and represents the closest value of $w_{\phi }$ to the cosmological constant value of $-1$ that is possible to obtain for each model and for the set of cosmological parameters listed in Table~\ref{cosmological_parameters}. The potential slope $\alpha $ is equal to $0.1$ in all the cosmologies.}
\label{cosmological_models}
\end{table}
\end{center}
\normalsize

\begin{figure*}
\includegraphics[scale=0.45]{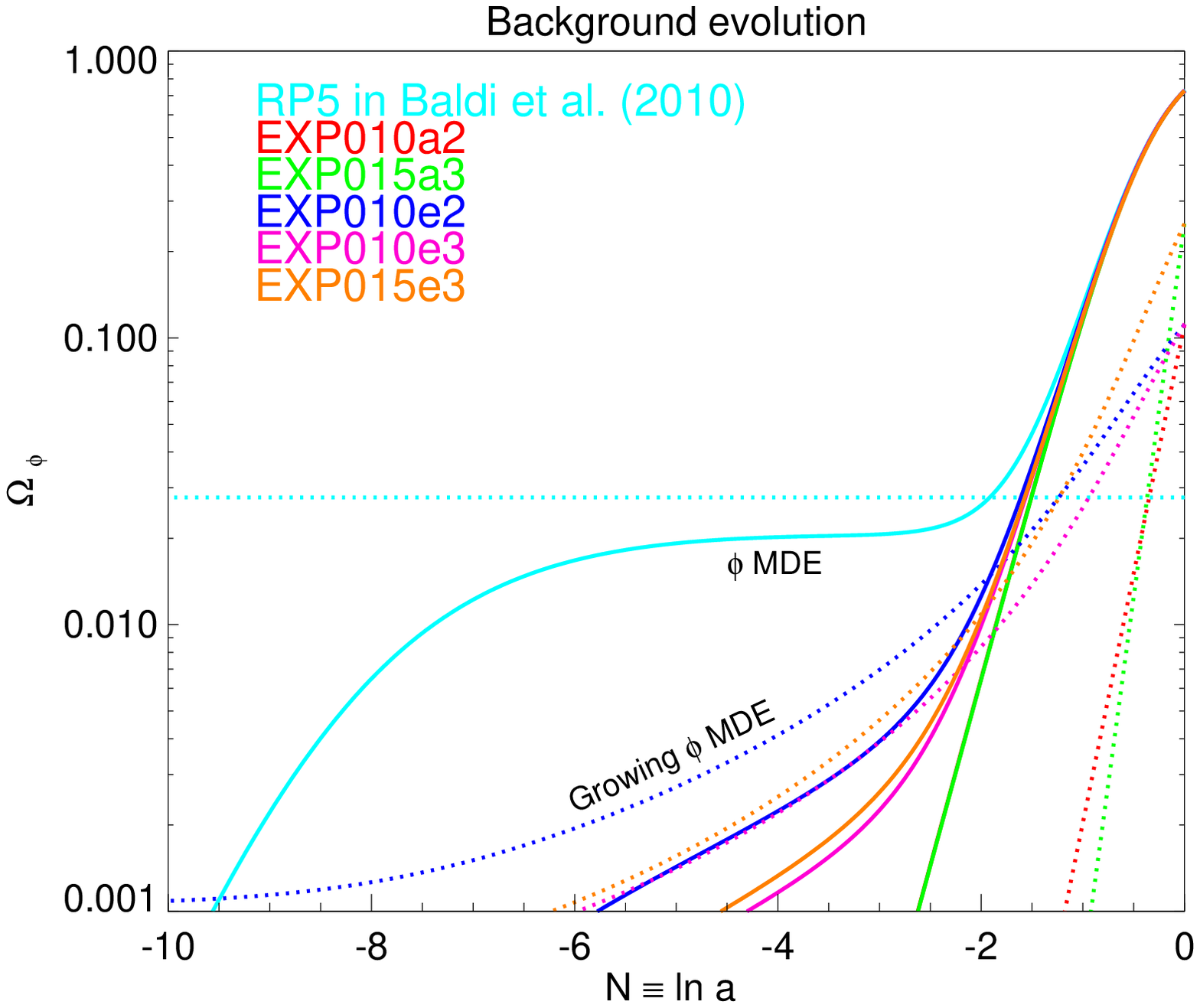}
\includegraphics[scale=0.45]{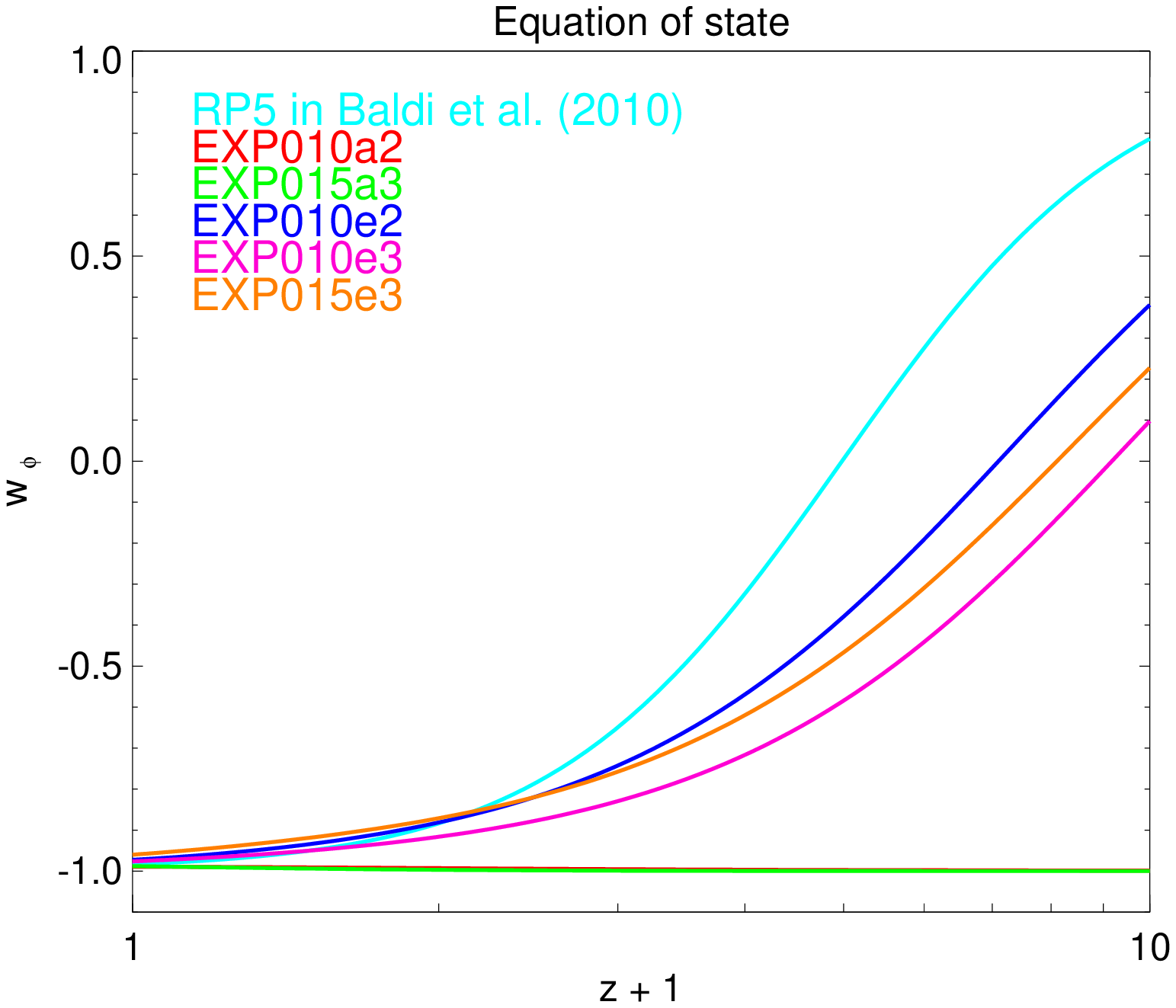}
\caption{{\em Left Panel:} The evolution of the DE fractional energy density $\Omega _{\phi }$ (solid lines) for a few models of variable coupling, and in addition for the constant coupling models RP5 studied in BA10. The ``$\phi $MDE" phase is clearly visible for the RP5 model, and is completely absent for the phenomenological coupling models EXP010a2 and EXP015a3, that show no significant difference with respect to a $\Lambda $CDM evolution. The the exponential coupling models have the intermediate behavior of a ``Growing $\phi $MDE". Dotted lines represent the value of the analytic ``$\phi $MDE" solution \citep{Amendola_2000} extended to the case of growing couplings. {\em Right Panel:} Equation of state parameter $w_{\phi }$ for the same models as in the {\em left panel}. The phenomenological coupling models have always an equation of state parameter very close to the cosmological constant case, while the exponential coupling models follow a similar behavior to the constant coupling model RP5.}
\label{fig:background}
\end{figure*}
\normalsize

Even in the absence of analytic solutions for the cosmological background evolution of variable coupling models, our numerical integrations allow to identify some relevant features of these cosmologies and to compare them with the familiar case of constant couplings. In particular, it is very interesting to notice how the time variation of the coupling affects the background evolution during matter domination. As already mentioned above, a constant coupling between DE and CDM gives rise to a metastable scaling solution during matter domination where the scalar field $\phi $ and the coupled matter fluids (CDM in our case) evolve with a constant ratio of energy densities, called ``$\phi $MDE", which represents one of the most peculiar features of the coupled DE scenario, besides providing a way to put constraints on the models due to the presence of a sizeable fraction of early DE \citep{EDE,EDE1,EDE2} at high redshifts. 

For the idealized case of a Universe containing only DE and CDM interacting with a constant coupling $\beta _{c}$, the DE fractional energy density $\Omega _{\phi }$ during the ``$\phi $MDE" phase has the constant value:
\begin{equation}
\label{phiMDE}
\Omega _{\phi }(\phi \text{MDE}) = \frac{4}{9}\beta _{c}^{2} \,.
\end{equation} 
In our variable coupling models, where the coupling is large at the present time and progressively decreases at higher redshifts, we find a quite different picture for the background evolution. In particular, there is no ``$\phi $MDE" solution in our cosmologies, and no scaling regime between the scalar field $\phi $ and CDM is found. 
Among the different models investigated, we can distinguish two quite different behaviors for the different types of couplings considered in our integrations. The situation is illustrated in Fig.~\ref{fig:background}, where the time evolution of the scalar field fractional energy density $\Omega _{\phi }$ and the redshift evolution of the scalar field equation of state parameter $w_{\phi }$ are plotted for some of the models under investigation. For an easier readability of the plots, and with no loss of generality,  we show in Fig.~\ref{fig:background} only the models that will be used for our N-body simulations since they clearly show the different impact on the background evolution between the exponential coupling models ($\beta _{c}(\phi )\propto e^{\beta _{1}\phi /M}$) and  the scale factor models ($\beta _{c}(\phi ) \propto a^{\beta _{1}}$). The remaining class of couplings ($\beta _{c}(\phi )\propto \Omega _{\phi }$) is found to have a very similar behavior to the latter one. In addition, we also plot for comparison the same quantities for one of the constant coupling scenarios studied in BA10, the RP5 model, which features a coupling of $\beta _{c}=0.25$.

In the left panel of Fig.~\ref{fig:background} the solid lines represent the evolution of the DE fractional energy density $\Omega _{\phi }$. The ``$\phi $MDE" phase is clearly visible for the constant coupling model RP5, while none of the variable coupling models follows a similar evolution. However, while the EXP010a2 and EXP015a3 models, where the coupling scales like a power of $a$, present a constant decay with redshift of the DE fractional energy density, with no significant difference from a $\Lambda $CDM evolution, the exponential coupling models EXP010e2, EXP010e3, and EXP015e3 show an intermediate behavior, which we call a ``Growing $\phi $MDE" solution, where the fraction of DE $\Omega _{\phi }$, although not constant during all matter domination, evolves significantly slower than in the $\Lambda $CDM scenario. 
In order to compare with the constant coupling case, we have also plotted (as dotted lines) the analytic solution for $\Omega _{\phi }(\phi $MDE$)$ given by Eqn.~(\ref{phiMDE}), derived for the case of constant $\beta _{c}$, also for the variable coupling models. The gap between the theoretical value and the actual evolution of $\Omega _{\phi }$ for the constant coupling RP5 model is due to the inclusion in our integrations of the uncoupled baryonic component that perturbs the solution given by Eqn.~(\ref{phiMDE}) for the idealized case of $\Omega _{b} =0$. 

It is very interesting to notice that during the ``Growing $\phi $MDE" phase, the models with exponential couplings follow the same evolution given by the extension of the solution (\ref{phiMDE}) to the case of a variable coupling $\beta _{c}(\phi )$, with a comparable gap as for the RP5 model due to the presence of uncoupled baryons. The denomination ``Growing $\phi $MDE" therefore seems appropriate since during this phase the system evolves as for the case of the standard ``$\phi $MDE" but with a growing value of the coupling.
\\

The reason for the strong qualitative difference in the background solutions found for the different classes of coupling functions can be understood by looking at the right panel of Fig.~\ref{fig:background}, where the DE equation of state parameter $w_{\phi }$ is plotted as a function of redshift. While the scale factor models (EXP010a2, EXP015a3) always have an equation of state parameter very close to $-1$, the exponential coupling models show a significant variation of $w_{\phi }$ with redshift, with a similar trend to the constant coupling model RP5. 

This strikingly different behavior of the equation of state clearly shows that the former class of models requires the same level of fine tuning of the $\Lambda $CDM concordance cosmology, and therefore cannot be expected to provide any dynamical solution to the DE ``fine tuning problem". On the contrary, the ``fine tuning problem" might still be alleviated by the latter class of models, due to their sensibly slower evolution of the scalar field energy density with respect to $\Lambda $CDM, thereby retaining one of the main motivations that are behind the whole interacting DE scenario. 

However, as we will discuss in detail in Sec.~\ref{results}, the different dynamics of these two types of variable coupling models is not relevant only for the cosmic background evolution, but will also have a strong impact on the structure formation processes taking place in the context of these different cosmologies, in particular for what concerns the highly nonlinear regime of structure formation.

\subsection{Observational constraints and model selection criteria}
\label{obs}

The observational constraints on the DE-CDM interaction have become ever tighter in the last decade. This is mainly due to the inclusion of low redshift probes and {\em Large Scale Structure} (LSS) data in the analysis rather than to an improved quality of CMB data, which have been for a long time the main, if not the only, source of constraints for interacting DE models. 
In fact, the most stringent constraints to date on the coupling strength for constant coupling models have been presented in \citet{Bean_etal_2008} (BE08, hereafter) and come from a combined analysis of high and low redshift probes of the expansion history of the Universe as CMB data, HST measurements of the Hubble constant $H_{0}$, Baryon Acoustic Oscillations data, SnIa measurements of the low redshift expansion history, and {\em Sloan Digital Sky Survey} (SDSS) measurements of the matter power spectrum at low redshifts. By combining all these datasets, BE08 came up with a constraint of $|\beta _{c}| < 0.067$ at the 95\% C.L. for a constant coupling interacting DE scenario. However, it is interesting to notice that the same authors quote a limit of $|\beta _{c}| < 0.13$ from the analysis of CMB data alone, which is no much tighter constraint than the  upper bound of $|\beta _{c}|< 0.15$, still based on CMB data only, reported by \citet{Amendola_2000} almost a decade before. This shows, as confirmed by \eg \citet{Xia_2009, Valiviita_etal_2009}, that CMB data alone are not able to constrain much further the value of the coupling, and the inclusion of low redshift probes as well as LSS data is essential to tighten the bounds down to the accuracy reported by BE08. In addition to these combined constraints a new and completely independent bound on the coupling based on the comparison of detailed hydrodynamical N-body simulations with the observed properties of \lya absorption systems has been recently presented by \citet{Baldi_Viel_2010}, with a 2-$\sigma $ limit on the coupling of $\beta \lesssim 0.15$ based on \lya observables only.

An interesting discovery has nevertheless been recently reported by \citet{LaVacca_etal_2009} (LV09, hereafter), that found a degeneracy between the coupling amplitude and the average neutrino mass $M_{\nu }$, showing how a value of $M_{\nu }\simeq 1 $ eV could broaden the allowed range for the coupling up to $|\beta _{c}| < 0.17$. Therefore, by dropping the assumption of massless neutrinos that we adopted in Sec.~\ref{cde}, it is possible to allow larger coupling values without running into conflict with present observational constraints on the background expansion history. The inclusion of massive neutrinos with an average mass up to $M_{\nu } \sim 1$ eV would not significantly change the results we have derived so far for the background evolution as long as the neutrinos remain uncoupled, \ie $\beta _{n}(\phi ) = 0$.

We will assume here the values of $\beta _{c} = 0.067$ reported by BE08 and of $\beta _{c} = 0.17$ derived by LV09 as two limiting values for the allowed coupling in a constant interaction model, for the cases of massless and massive ($M_{\nu } \simeq 1.0$ eV) neutrinos, respectively. These values correspond to the BE08 and the LV09 models in Table \ref{cosmological_models}.

Some attempts to put observational constraints on more general classes of interacting DE models, including some forms of variable couplings, have been presented in \citet{Cai_Su_2010, Guo_etal_2007,Costa_Alcaniz_2010} and further generalized by \citet{Wei_2010}. However, such constraints are not directly applicable to the scalar field models under investigation in the present work, due to the assumption of a constant DE equation of state parameter that is made in the analysis of all the cited authors. Also, the fraction of uncoupled baryonic mass should always be included in the analysis in order to derive meaningful constraints.

Therefore, a full likelihood analysis of variable coupling models against presently available data is still missing, and will be clearly necessary to fully assess the viability of these models and put constraints on the coupling function. However, such an effort goes beyond the scope of this paper, and we leave it for future work. Here we just assume a simple criterion to test and select our models, based on a direct comparison of their background evolution with the one derived for our assumed limiting models BE08 and LV09. We compute for all our models the evolution of the luminosity distance:
\begin{equation}
\label{luminosity_distance}
d_{L}(z) = \frac{1+z}{H_{0}}\int_{0}^{z}\frac{H_{0}dz'}{H(z)} 
\end{equation}
between $z=0$ and $z=3$, and the evolution of the angular-diameter distance:
\begin{equation}
d_{A}(z) = \frac{1}{(1+z)H_{0}}\int_{0}^{z}\frac{H_{0}dz'}{H(z)}
\end{equation}
up to last scattering surface ($z\simeq 1100$), and we compare their relative evolution with respect to the fiducial $\Lambda$CDM model to the limiting cases BE08 and LV09. The change in the angular-diameter distance at high redshift would affect CMB measurements, while a change in the evolution of the luminosity distance with redshift at small $z$ would be constrained by low redshift data as \eg SnIa. 

\begin{figure*}
\includegraphics[scale=0.5]{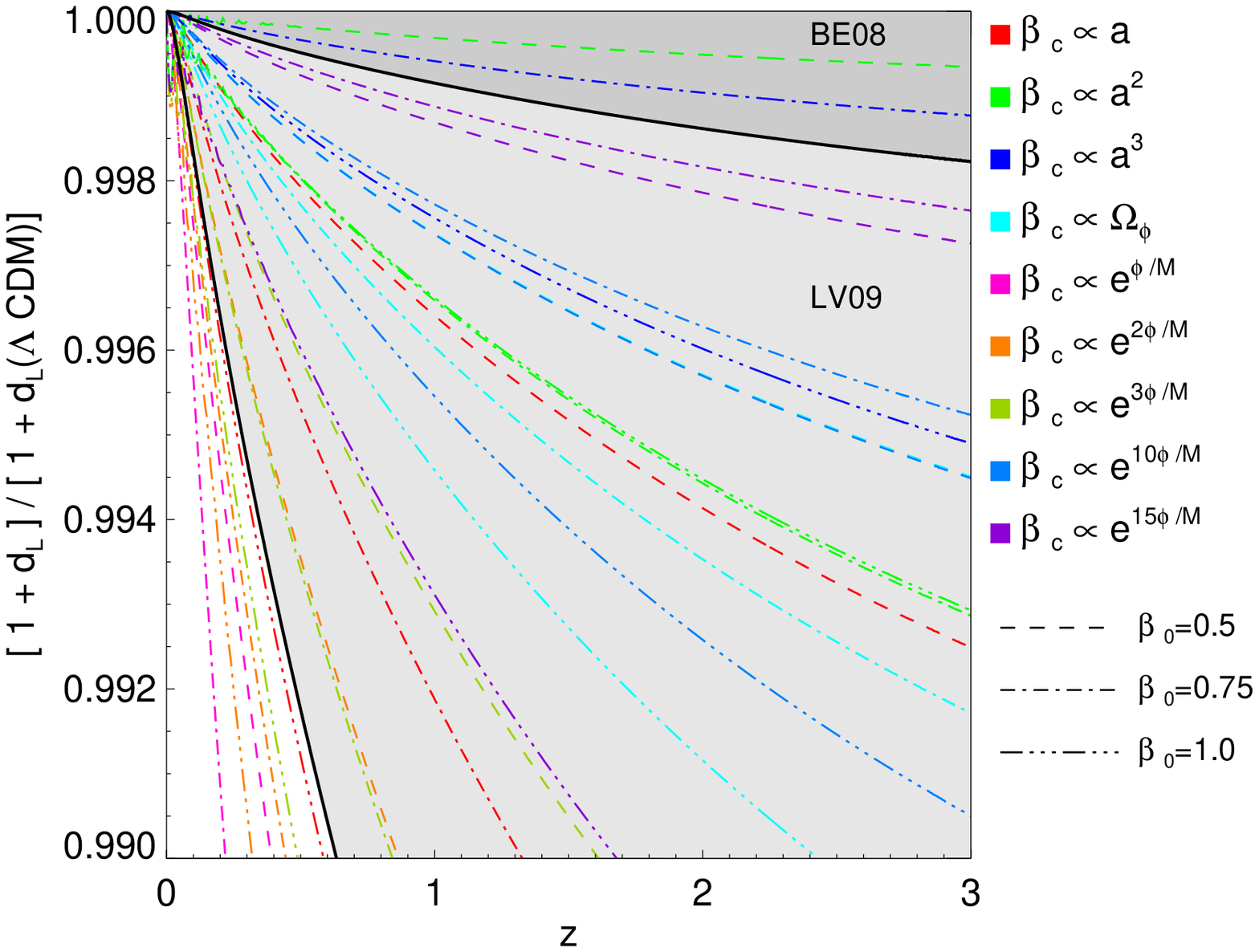}
\includegraphics[scale=0.5]{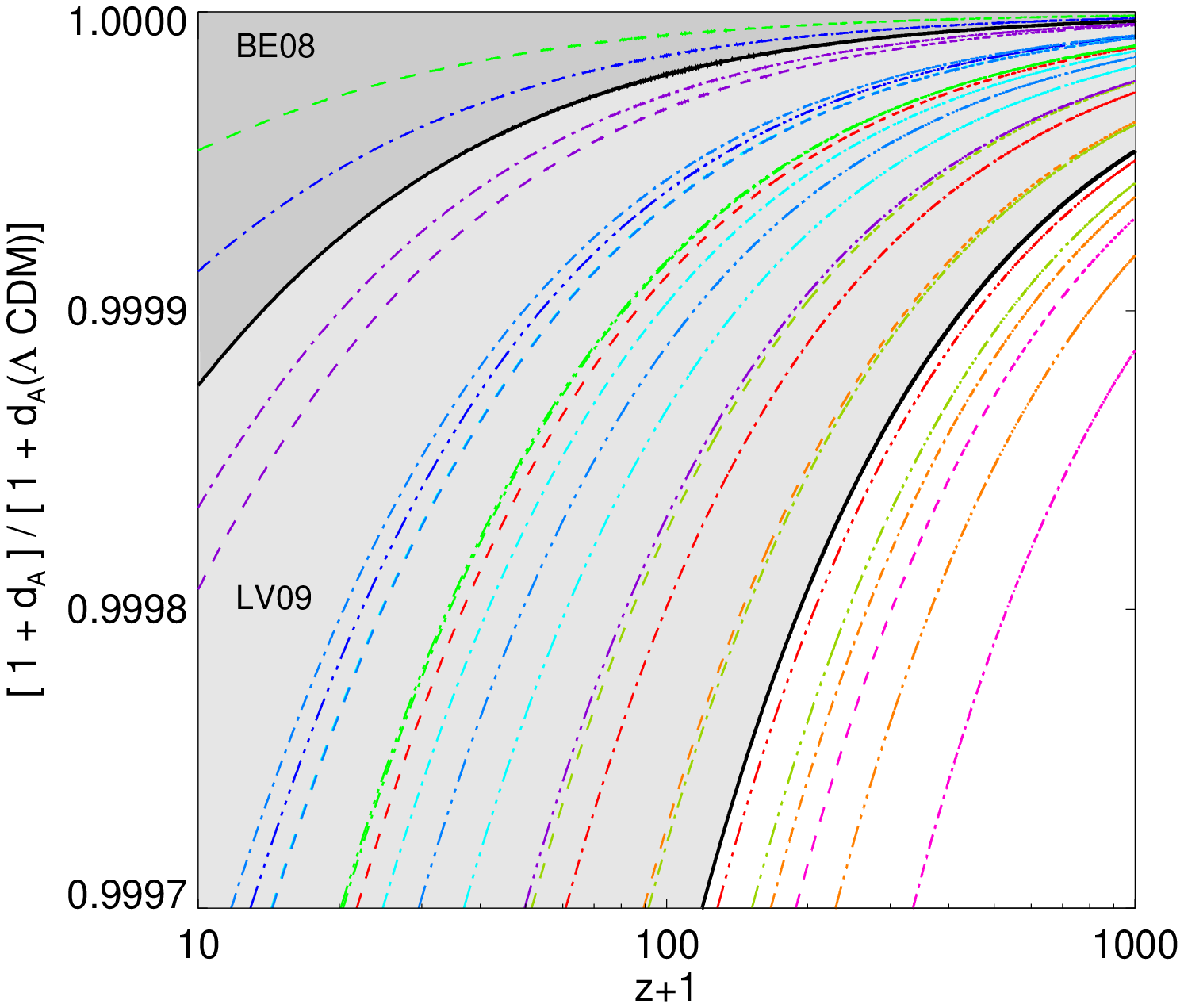}
\caption{The ratio of the luminosity distance between $z=0$ and $z=3$ ({\em left panel}) and of the angular-diameter distance between $z=10$ and $z=1000$ ({\em right panel}) to the respective $\Lambda $CDM evolution as a function of redshift. The two thick solid black lines in the two plots represent the two limiting cases considered in this work, \ie the constraints given by \citet{Bean_etal_2008} and \citet{LaVacca_etal_2009}. The dark-grey and light-grey shaded regions that lie, respectively,  between the $\Lambda $CDM value of $1$ and each of the two limiting scenarios represent the allowed regions according to the selection criterion described in the text. As it can be noticed, only a few models are found to be compatible with the constraints of \citet{Bean_etal_2008}, while a large fraction of the models considered in the present work are compatible with the bounds of \citet{LaVacca_etal_2009}.}
\label{fig:constraints}
\end{figure*}

Therefore, since most of the present  bounds on interacting DE models are based on observations of the background expansion history, we assume as a tentative selection criterion that any model that lies in between the fiducial $\Lambda $CDM cosmology and one of the two limiting scenarios BE08 and LV09 for what concerns both its luminosity distance at low redshifts and its angular-diameter distance at high redshifts should be considered compatible with observations and accepted as viable, while it should be rejected otherwise.
The situation is illustrated in the two plots shown in Fig.~\ref{fig:constraints}, where the ratio of luminosity distance (left panel) and angular-diameter distance (right panel) for all our models with respect to $\Lambda $CDM is plotted against redshift. The two limiting scenarios we are considering, BE08 and LV09, are shown as the two thick solid lines, and the allowed regions for the two limits are represented by the dark-grey and light-grey shaded areas, respectively. 

\normalsize

As Fig.~\ref{fig:constraints} shows, while only a few of our models pass the test of the most stringent constraint given by BE08, most of them are found to be compatible with the other limiting case LV09. It is nevertheless interesting to notice that none of the exponential coupling models seems to be consistent with the limit given by the analysis of BE08, for which they would probably require a much larger value of the exponential coupling slope $\beta _{1}$.\\

The selection criterion presented here is clearly not rigorous enough to be taken as a definitive check of the viability of the models under investigation, but we find it a simple and useful guideline to select on which models we should invest our computational time for the high-resolution N-body simulations described in Sec.~\ref{sim}. Based on this criterion we therefore decide to run high-resolution simulations for the two models consistent with BE08, namely the EXP010a2 and the EXP015a3 models, and for three of the more physically motivated exponential coupling models that passed at least the test of the bounds of LV09, namely the EXP010e2, EXP010e3 and EXP015e3 models. 

The choice of the latter models among all the models compatible with LV09 is not random. In fact, although all the models considered in the present work are specifically built in order to have very large values of the coupling in a redshift range relevant for the nonlinear stages of structure formation, and are therefore expected to imprint sizeable features in the properties of highly nonlinear structures, the way in which the coupling affects the newtonian dynamics of CDM particles does not depend on the coupling strength alone, but also on the evolution of the DE internal degrees of freedom like the scalar field velocity $\dot{\phi }$, as we will discuss in detail in Sec.~\ref{prt}. The models we have chosen for our N-body treatment feature substantially different behaviors of the scalar field velocity $\dot{\phi }$ (the DE kinetic degree of freedom $x$ in our dimensionless framework) which will play a significant role in determining how the DE-CDM interaction can influence the properties of collapsed structures like \eg the halo density profiles, as we will discuss in full detail in Sec.~\ref{profiles}. 

\subsection{Linear perturbations} \label{prt}

We now study the evolution of density perturbations according to linear perturbation theory for the variable coupling models introduced above. This is an interesting issue on its own since such analysis has not been done previously for the case of variable couplings. Linear perturbations equations in the context of coupled DE models have been derived in the literature (see \eg \citet{Amendola_2000, Amendola_2004,Pettorino_Baccigalupi_2008}, BA10), also for the case of a scalar field dependent coupling, but to our knowledge the solution of such equations for variable coupling models has not been presented before. 
We refer the interested reader to the cited literature, and references therein, for a detailed derivation of the perturbed equations, and we limit our discussion here to the main features of the perturbations growth that are relevant for the models considered in this work. 
To ease the readablity of lengthy equations, in this section we do not keep explicit the dependence of the coupling on the scalar field.

The direct solution of the linear density fluctuations evolution is also necessary in order to set up the initial conditions for our set of N-body simulations, since we want to be able to normalize our cosmological models to the same parameters at $z=0$, which means that we also want all the cosmologies to end up with the same $\sigma _{8}$, and therefore initial conditions have to be properly scaled with respect to each other according to the different values of their growth factor at the starting redshift of the simulations, which we assume to be $z=60$ for all of our runs.

We consider the perturbed metric in the longitudinal gauge and in the absence of anisotropic stress, given by:
\begin{equation}
ds^{2} = a^{2}(\tau ) \left[ -(1+2\Phi ) d\tau ^{2} + (1-2\Phi)\delta _{ij}dx^{i}dx^{j} \right] \,,
\end{equation}
where $\tau $ is the conformal time and $\Phi $ is the gravitational potential. The conformal Hubble function is given by ${\cal H} \equiv (da/d\tau )/a = aH$. 
Then, one can define the following perturbation variables:
\begin{eqnarray}
\delta _{c} &\equiv &\delta \rho _{c}/\rho _{c} \,, \, \delta _{b} \equiv \delta \rho _{b}/\rho _{b} \,, \\
u_{i} &\equiv &adx_{i}/({\cal H}dt)\,, \vec{\nabla } \cdot \vec{u}_{c} \equiv \theta _{c} \,, \vec{\nabla } \cdot \vec{u}_{b} \equiv \theta _{b}\\
\varphi &\equiv &\delta \phi /(M_{Pl}\sqrt{6}) \,.
\end{eqnarray}
In Fourier space we can also define the scale parameter $\lambda \equiv {\cal H}/k$.
Since we are interested here in the evolution of matter density perturbations during matter domination, we discard for simplicity the radiation density fluctuations. \\
Finally, we define the dimensionless scalar mass as in \citet{Amendola_2004}:
\begin{equation}
\hat{m}^{2}_{\phi } \equiv \frac{m^{2}_{\phi }}{H} = \frac{1}{H}\frac{d^{2}V(\phi )}{d\phi ^{2}} = 2\alpha ^{2}y^{2}\,.
\end{equation}
With all these definitions, the perturbed evolution equations for each component in Fourier space can be written as \citep[see][for a complete derivation]{Amendola_2004,Pettorino_Baccigalupi_2008}:
\begin{eqnarray}
\delta '_{c} &=& -\theta _{c} + 3\Phi ' - 2\beta _{c}\varphi ' - 2\beta '_{c}\varphi \,, \\
\theta '_{c} &=& -\left( 1 + \frac{{\cal H}'}{{\cal H}} - 2\beta _{c} x \right) \theta _{c} + \lambda ^{-2} (\Phi -2\beta _{c}\varphi ) \,, \\
\delta '_{b} &=& -\theta _{b} + 3\Phi ' \,, \\
\theta '_{b} &=& -\left( 1 + \frac{{\cal H}'}{{\cal H}}\right) \theta _{b} + \lambda ^{-2}\Phi \,, \\
\label{scalar_pert_full}
\varphi '' &+& \left( 2 + \frac{{\cal H}'}{{\cal H}} \right) \varphi ' + \left( \lambda ^{-2} +\hat{m}^{2}_{\phi } - \frac{\Omega _{c}\beta '_{c}}{x}\right)\varphi  \nonumber \\
& &- 4\Phi ' x -2y^{2}\alpha \Phi = \beta _{c}\Omega _{c}(\delta _{c} + 2\Phi) \,.
\end{eqnarray}

As it has been noticed in \citet{Amendola_2004}, it is very interesting to point out that the variation of the coupling contributes to the equation for the evolution of the scalar field fluctuations (\ref{scalar_pert_full}) as an effective mass term. 
However, we want to focus our attention here on a point that has not been sufficiently stressed in previous literature, namely the fact that this effective ``coupling mass'' 
\begin{equation}
\label{coupling_mass}
\hat{m}^{2}_{\beta _{c}} \equiv \frac{\Omega _{c}\beta '_{c}(\phi )}{x}
\end{equation}
can take both positive and negative signs depending on the signs of $x$ and $\beta '_{c}(\phi )$. As a particular case, if one assumes a direct dependence of the coupling on the scalar field, as it is the case for the exponential coupling introduced in Sec.~\ref{exp_beta}, the sign of $\hat{m}^{2}_{\beta _{c}}$ is fully determined by the derivative of the coupling function with respect to the scalar field itself:
\begin{equation}
\hat{m}^{2}_{\beta _{c}} = \sqrt{6} \Omega _{c}\frac{d\beta _{c}(\phi )}{d\phi } \,.
\end{equation}
It is therefore evident that a negative effective mass $\hat{m}^{2}\equiv \hat{m}^{2}_{\phi } - \hat{m}^{2}_{\beta _{c}}$ could arise for couplings that grow in time, as it is the case for the models considered in the present work. If this happens, large-scale instabilities might arise in the growth of density perturbations.

For the moment we ignore this potential instability issue by assuming that in the newtonian limit (small scales, $\lambda \ll 1$) that is relevant for our N-body simulations, the total mass term in Eqn.~(\ref{scalar_pert_full}) can be discarded as compared to the $\lambda ^{-2}$ term. This is true as long as $|\hat{m}^{2}| \sim {\cal O}(1)$, which will be tested later on for all the specific models under investigation.

In any case, the issue of potential large-scale instabilities due to the coupling growth should be carefully investigated in order to ensure the viability of the models, and we leave such analysis for future work \citep{Amendola_Baldi}.

By applying the newtonian limit $\lambda \ll 1$, the perturbations equations for baryons and CDM become:
\begin{eqnarray}
\label{deltac}
\delta '_{c} &=& -\theta _{c} - 2\beta '_{c}\varphi \,, \\
\theta '_{c} &=& -\left[ 1+\frac{{\cal H}}{{\cal H}} - 2\beta _{c}x\right] \theta _{c} -\frac{3}{2}\left[ \Omega _{b}\delta _{b} + \Omega _{c}\delta _{c} \Gamma _{c} \right] \,, \label{thetac} \\
\delta '_{b} &=& -\theta _{b} \,, \\
\theta '_{b} &=& -\left( 1 + \frac{{\cal H}'}{{\cal H}}\right) \theta _{b} -\frac{3}{2}\left[ \Omega _{b}\delta _{b} + \Omega _{c}\delta _{c} \right] \,,
\end{eqnarray}
where we have defined:
\begin{equation}
\label{Gamma_c}
\Gamma _{c} \equiv  1+ \frac{4}{3}\frac{\beta _{c}^{2}(\phi )}{1+\lambda^{2}\hat{m}^{2}} \,.
\end{equation}

Since from the newtonian limit of Eqn.~(\ref{scalar_pert_full}) one gets that the scalar field fluctuation $\varphi \sim \lambda ^{2}$ (see again \citet{Amendola_2004, Pettorino_Baccigalupi_2008} for a complete derivation), the term in $\beta '_{c}$ in Eqn.~(\ref{deltac}) can be dropped as long as $\beta '_{c}(\phi ) \sim {\cal O}(1) $. Similarly, the term $\lambda ^{2}\hat{m}^{2}$ at the denominator in the second term of Eqn.~(\ref{Gamma_c}) can be discarded as long as $|\hat{m}^{2}| \sim {\cal O}(1)$, in which case one gets:
\begin{equation}
\label{Gamma_c_massless}
\Gamma _{c} = 1 + \frac{4}{3} \beta ^{2}_{c}(\phi )\,.
\end{equation}
As we will show in Sec.~\ref{instability}, the former condition is always fulfilled for all the models under consideration, and we therefore assume it to hold. The latter condition, instead, due to the factor $1/x$ in the definition of the coupling mass $\hat{m}^{2}_{\beta _{c}}$ (Eqn.~\ref{coupling_mass}), might not be verified at all subhorizon scales for all our models, and has to be carefully evaluated case by case. In particular, as we will show later on, some of the models under investigation feature a slow-roll regime of the scalar field during matter domination, which makes the factor $1/x$ significantly large, with a corresponing increase of $|\hat{m}^{2}|$.  

Assuming the validity of the condition $\beta '_{c}(\phi ) \sim {\cal O}(1)$, the perturbations equations take the well known form \citep{Amendola_2000,Amendola_2004,Pettorino_Baccigalupi_2008}:
\begin{eqnarray}
\label{deltac_massless}
\delta '_{c} &=& -\theta _{c} \,, \\
\theta '_{c} &=& -\left[ 1+\frac{{\cal H}}{{\cal H}} - 2\beta _{c}x\right] \theta _{c} -\frac{3}{2}\left[ \Omega _{b}\delta _{b} + \Omega _{c}\delta _{c} \Gamma _{c} \right] \,, \\
\label{deltab_massless}
\delta '_{b} &=& -\theta _{b} \,, \\
\theta '_{b} &=& -\left( 1 + \frac{{\cal H}'}{{\cal H}}\right) \theta _{b} -\frac{3}{2}\left[ \Omega _{b}\delta _{b} + \Omega _{c}\delta _{c} \right] \,,
\end{eqnarray}
which leads, by derivation of Eqs.~(\ref{deltac_massless},\ref{deltab_massless}), to the dynamic equations for density fluctuations:
\begin{eqnarray}
\label{gf_c}
&\delta ''_{c}& + \left( 1 + \frac{{\cal H}'}{{\cal H}} -2\beta _{c} x \right) \delta '_{c} -\frac{3}{2}\left[ \Omega _{b}\delta_{b} + \Omega _{c}\delta _{c} \Gamma _{c} \right] = 0\,, \\
\label{gf_b}
&\delta ''_{b}& + \left( 1 + \frac{{\cal H}'}{{\cal H}} \right) \delta '_{b} -\frac{3}{2}\left[ \Omega _{b}\delta_{b} + \Omega _{c}\delta _{c} \right] = 0\,,
\end{eqnarray}
and to the vectorial acceleration equations in real space for baryon and CDM particles, that was first derived in its full generality in BA10:
\begin{eqnarray}
\label{accel_c}
\dot{\vec{v}}_{c} &=& -\tilde{H}\vec{v}_{c} - \vec{\nabla }\left[ \sum _{i=c} \frac{\Gamma _{c} G M_{i}(\phi)}{r_{i}} + \sum _{j=b}\frac{GM_{j}}{r_{j}}\right] \,, \\
\label{accel_b}
\dot{\vec{v}}_{b} &=& -H \vec{v}_{c} - \vec{\nabla }\left[ \sum _{i=c} \frac{GM_{i}(\phi )}{r_{i}} + \sum _{j=b}\frac{GM_{j}}{r_{j}}\right] \,.
\end{eqnarray}
In Eqs.~(\ref{accel_c},\ref{accel_b}) $\tilde{H} \equiv H[1+2\beta _{c}(\phi )x]$ and $\vec{v}_{i}$ is the peculiar velocity of a baryonic (subscript b) or a CDM (subscript c) test particle, $\vec{v}_{i} \equiv a \dot{\vec{x}}_{i}$. 
To obtain Eqs.~(\ref{accel_c},\ref{accel_b}) we have discretized the mass distribution in the Universe by assuming:
\begin{eqnarray}
\Omega _{c}\delta _{c} &=& \sum _{i=c}\frac{8\pi G M_{c}(\phi )\delta(\vec{r_{i}})}{3{\cal H}^{2}a}\,, \\
\Omega _{b}\delta _{b} &=& \sum _{j=b}\frac{8\pi G M_{b}\delta(\vec{r_{j}})}{3{\cal H}^{2}a}
\end{eqnarray}
where $\delta (\vec{r}_{i})$ stands for the Dirac distribution, $\vec{r}_{i}$ is the position of each matter particle with respect to our test particle, and the sum has to be intended as running over all the other matter particles in the Universe.

Eqs.~(\ref{gf_c},\ref{gf_b}) are the equations that we have to solve in order to compute the specific growth factors that we need for setting up the initial conditions of our N-body simulations, while Eqs.~(\ref{accel_c},\ref{accel_b}) are the modified newtonian dynamic equations that have to be implemented in the N-body algorithm in order to follow correctly the dynamics of particles in a coupled DE cosmology.

It is important to notice at this stage, as it was shown in \citet{Maccio_etal_2004} and BA10, that the acceleration equation for a CDM particle is modified in three ways, due to the presence of a DE-CDM coupling. 

\begin{figure*}
\begin{center}
\includegraphics[scale=0.33]{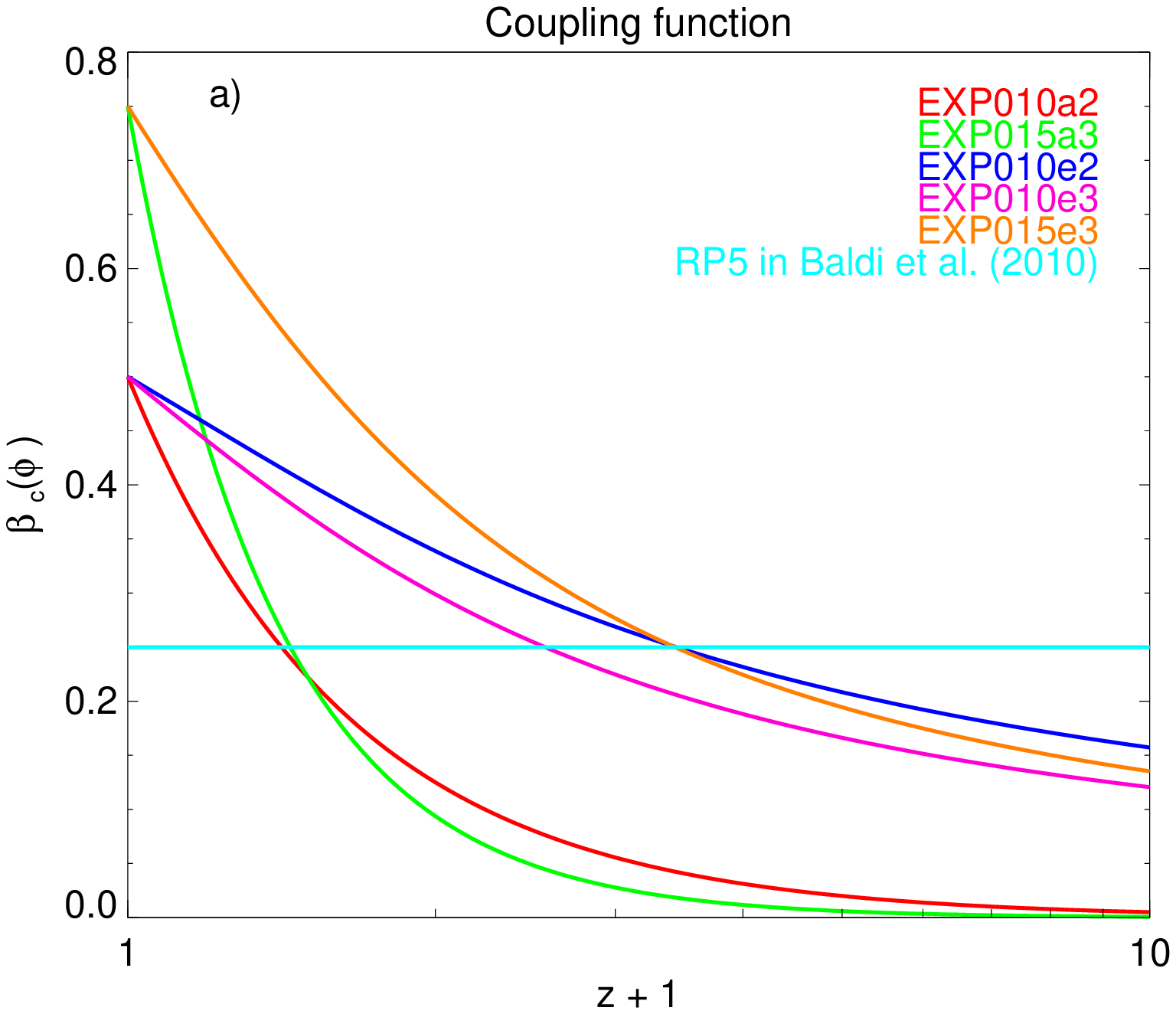}\includegraphics[scale=0.33]{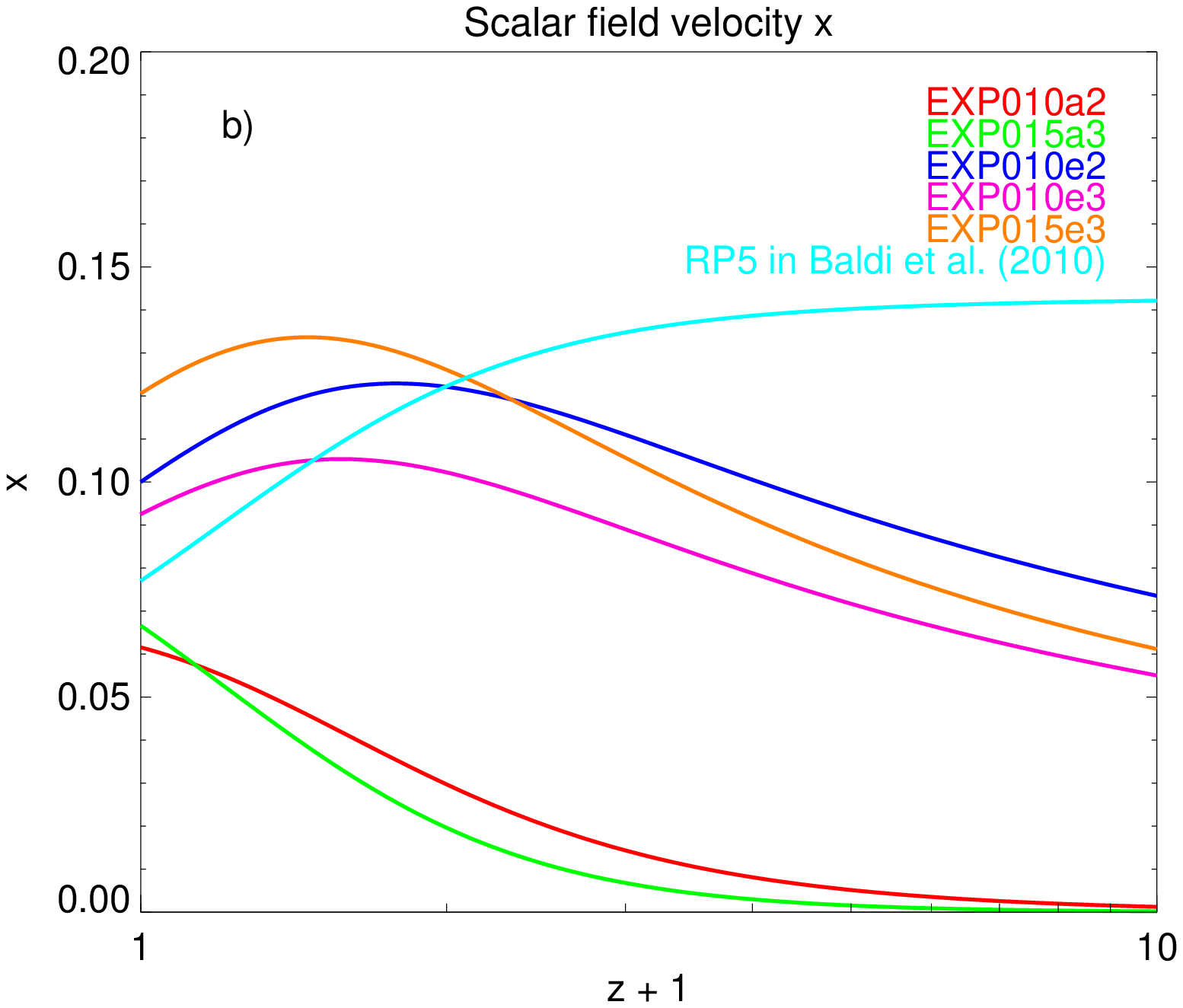}\includegraphics[scale=0.33]{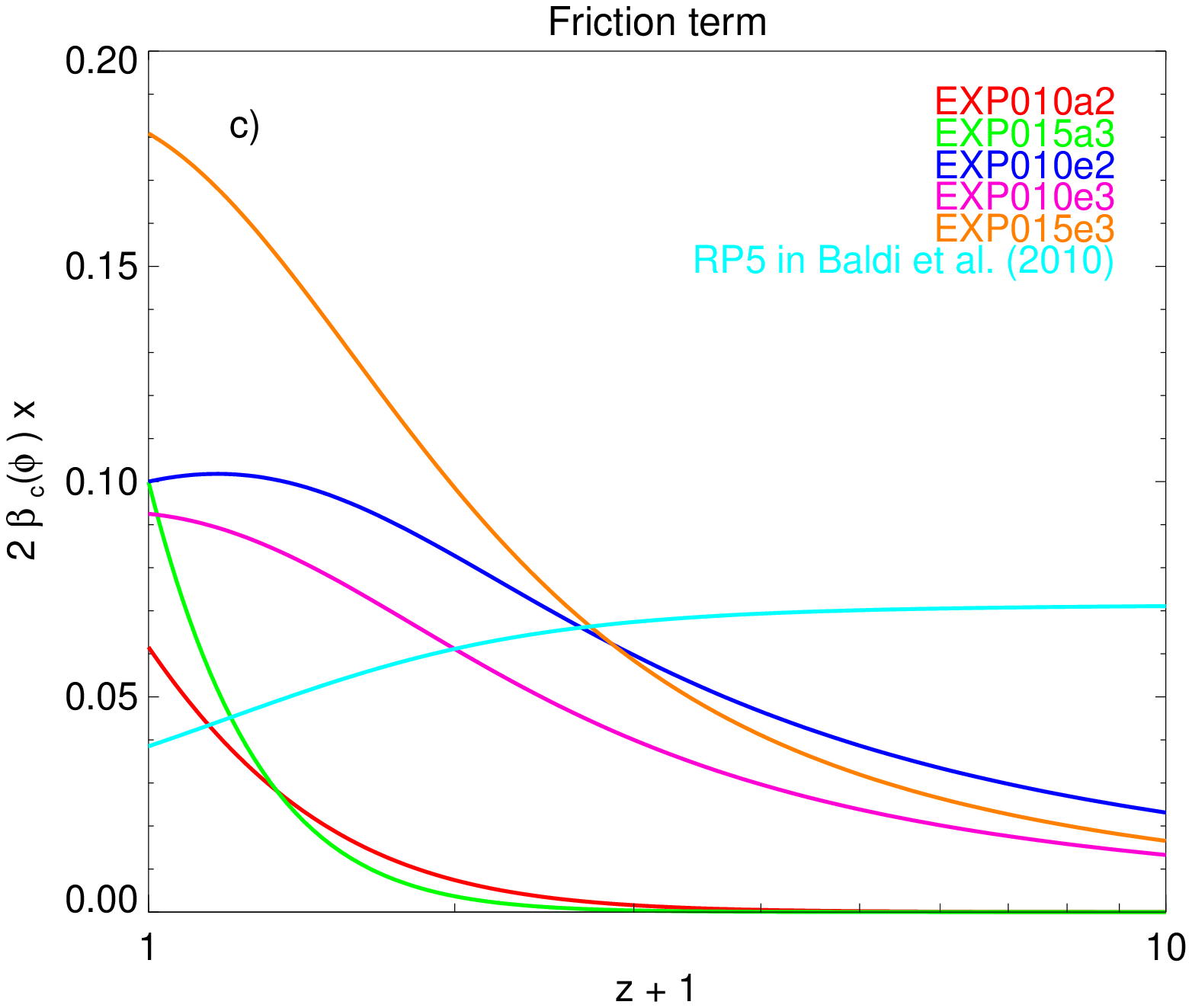}
\caption{This figure shows the evolution as a function of redshift of the coupling strength $\beta _{c}(\phi )$ ({\em panel a}), of the scalar field velocity $x$ as defined in Eqn.~\ref{dimensionless_variables} ({\em panel b}), and of the friction term coefficient given by $2 x \beta _{c}(\phi )$ ({\em panel c}), for the five coupled DE models studied with our high-resolution N-body simulations and in addition for the constant coupling model RP5 studied in BA10 which is overplotted for comparison. It is worth noticing how the dynamic evolution of the scalar field modulates the effect of the coupling in the friction term: despite a large value of the coupling $\beta _{c}(\phi )$ at low redshifts the friction term is strongly suppressed in the models which feature a too fast decrease of the coupling with redshift due to the absence of a proper ``$\phi$MDE" or of a ``Growing $\phi$MDE" phase.}
\label{fig:friction_term}
\end{center}
\end{figure*}

First, the friction term $\tilde{H}\vec{v}_{c}$ is modified by the presence of an extra contribution $2\beta _{c}(\phi )x\vec{v}_{c}$. It is interesting to notice, as anticipated in Sec.~\ref{obs}, that this extra friction does not depend on the coupling strength alone, but is modulated by the scalar field kinetic energy $x$. Therefore, the impact of the friction term on the  dynamics of CDM particles will depend substantially on the background evolution of the scalar field, which will have a strong influence on how the coupling affects the properties of nonlinear structures. In Fig.~\ref{fig:friction_term} the coupling function $\beta _{c}(\phi )$, the kinetic energy $x$, and the friction term coefficient $2\beta _{c}(\phi )x$ are plotted as a function of redshift (panels $a$, $b$, and $c$, respectively) for all the models we have selected for simulations and additionally for the RP5 model studied in BA10, which has a constant coupling of $\beta _{c}=0.25$. It is very interesting to notice how, despite the large values of $\beta _{c}(\phi )$ for all the variable coupling models at low redshift, the magnitude of the friction term can be strongly suppressed in some of the models by the faster decay or the lower initial value of the DE kinetic energy $x$. This means that the friction term becomes more efficient the more the DE equation of state $w_{\phi }$ departs from the cosmological constant value of $-1$.

\normalsize

Second, the mass of CDM particles that generate the gravitational potential in which every particle is moving changes in time, due to the evolution of the scalar filed, according to Eqn.~(\ref{mass_variation}). This variation of CDM particles mass affects also the dynamics of the uncoupled baryonic particles, as it can be seen in Eqn.~(\ref{accel_b}), which will feel a decaying gravitational potential due to the CDM mass loss.

Finally, the gravitational acceleration of CDM particles includes an extra factor $\Gamma _{c}$ accounting for the fifth-force mediated by the scalar field.
So far, we have left this extra factor in its full generality, but for the set of N-body simulations carried out in this work we will limit the analysis to the case where the effective scalar mass $\hat{m}^{2}$ is negligible, such that $\Gamma _{c}$ is given by Eqn.~(\ref{Gamma_c_massless}). This corresponds to a long-range scalar fifth-force, and we will discuss in Sec.~\ref{instability} under which circumstances this approximation is valid. It is nevertheless interesting to notice that, as discussed in \eg \citet{Amendola_2004}, the more general case where the scalar mass is not negligible, given by Eqn.~(\ref{Gamma_c}), determines a scale dependence of the fifth-force interaction between CDM particles corresponding to a Yukawa potential:
\begin{equation}
\label{Yukawa}
\tilde{\Phi}(r) = -\frac{GM}{r}\left(1 + \frac{4}{3}\beta ^{2}_{c}(\phi )e^{-\hat{m}Hr}\right) \,.
\end{equation}

Cosmological models that feature only this type of modified gravitational potential in the dark sector have been studied in the literature assuming different values of the characteristic mass scale $\hat{m}$ \citep{Gubser2004,Farrar2004,Farrar2007}, and tested with numerical N-body simulations both at galactic \citep{Kesden_Kamionkowski_2006,Keselman_Nusser_Peebles_2009} and cosmological scales \citep{Nusser_Gubser_Peebles_2005,Hellwing_etal_2010}.
Simulations of massive scalar fields coupled to CDM, which feature not only the Yukawa-like fifth-force of Eq.~(\ref{Yukawa}), but also the other modifications of newtonian dynamics arising from the DE-CDM interaction which are described above, will be carried out in an upcoming publication.

\subsection{Scalar masses and stability conditions}
\label{instability}

In the previous section we have encountered a series of conditions on the time variation of the scalar field coupling $\beta _{c}(\phi )$ that we assumed to be fulfilled in order to derive the final forms of the perturbations evolution equations (\ref{gf_c},\ref{gf_b}) and of the particle acceleration equations (\ref{accel_c},\ref{accel_b}) within our coupled DE models. In this section we want to discuss in more detail such conditions and check their validity for the models under investigation.

\begin{center}
\begin{figure}
\begin{center}
\includegraphics[scale=0.5]{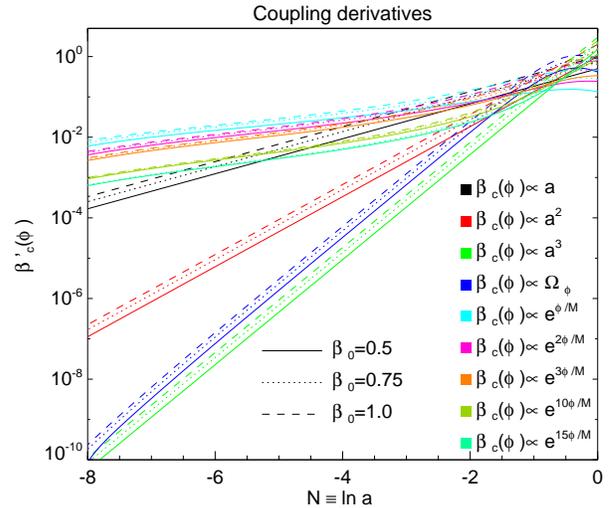}
\caption{The evolution of the coupling derivative $\beta '_{c}(\phi )$ as a function of the e-folding time $N \equiv \ln a$. The different colors represent the different types of time evolution of the coupling function $\beta _{c}(\phi )$ while the different linestyles correspond to different values of the coupling at the present time. It is important to notice that none of the models presents a value of the coupling derivative larger than ${\cal O}(1)$ at any stage of cosmic history.}
\label{fig:betaprime}
\end{center}
\end{figure}
\end{center}

First of all, in deriving Eqn.~(\ref{deltac_massless}) we have assumed that $\beta '_{c}(\phi ) \sim {\cal O}(1)$. We will show here that this condition holds for all the models considered in this work.
In particular, for the cases of an exponential coupling and of a coupling proportional to a power of the scale factor described in Sec.~\ref{exp_beta} and \ref{a_beta} respectively, it is possible to estimate analytically the magnitude of $\beta '_{c}(\phi )$:
\begin{eqnarray}
\beta _{c}(\phi ) &=& \beta _{0}a^{\beta _{1}} \rightarrow \beta '_{c}(\phi ) = \beta _{0}\beta _{1}a^{\beta _{1}} \le \beta _{0}\beta _{1} \,, \\
\beta _{c}(\phi ) &=& \beta _{0}e^{\beta _{1}\phi/M} \rightarrow \beta '_{c}(\phi ) = \sqrt{6}x\beta _{0}\beta _{1}e^{\beta _{1}\phi/M}\,.
\end{eqnarray}
For the former case, one can easily compute that the largest possible value for the models under investigation is given by $\beta '_{c}(\phi ) \sim 3$. For the latter, one can use the observation that in all the cosmologies $x \lesssim 0.15$ to put an upper limit on the coupling derivative which is $\beta '_{c}(\phi ) \sim 0.9$ for the largest assumed values of $\beta _{0}$ and $\beta _{1}$.
For the remaining case of a coupling proportional to the DE fractional density $\Omega _{\phi }$ such analytic estimation is not possible, and we compute the derivative numerically.
In any case, as shown in Fig.~\ref{fig:betaprime}, the coupling derivative $\beta '_{c}(\phi )$ is never larger than ${\cal O}(1)$. Therefore, the simplification adopted in Sec.~\ref{prt} is fully justified.

\normalsize

\begin{figure*}
\includegraphics[scale=0.32]{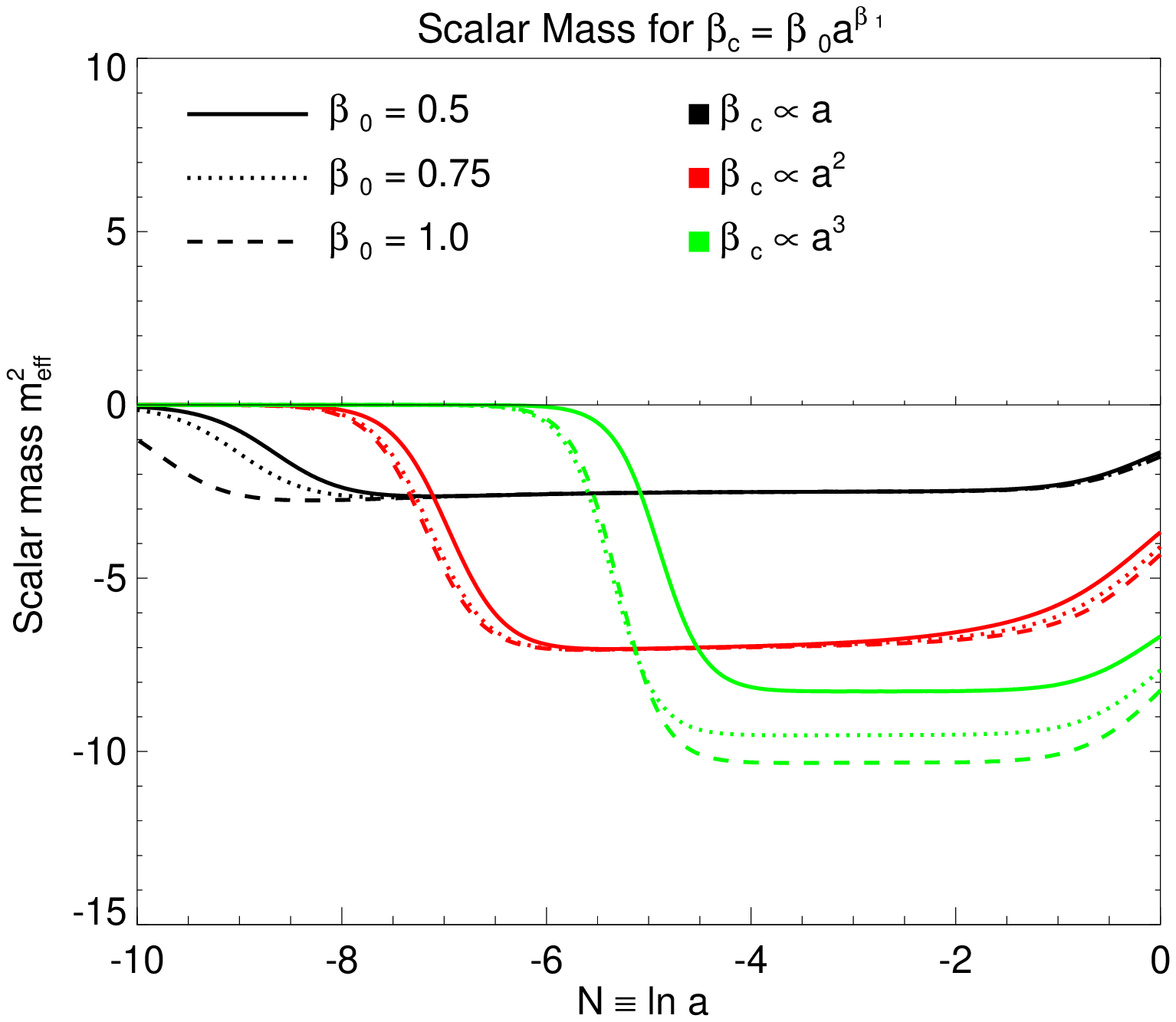}
\includegraphics[scale=0.32]{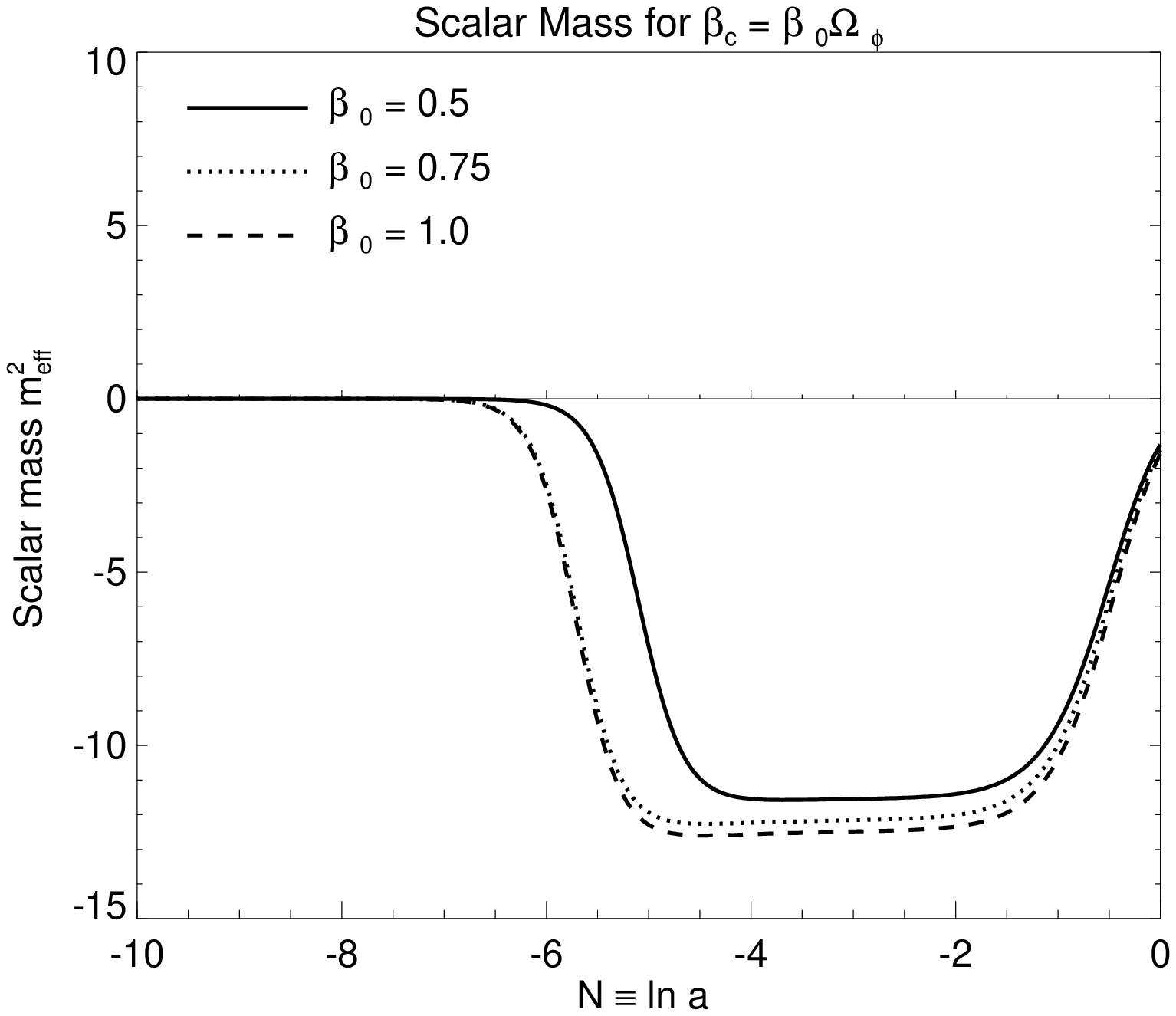}
\includegraphics[scale=0.32]{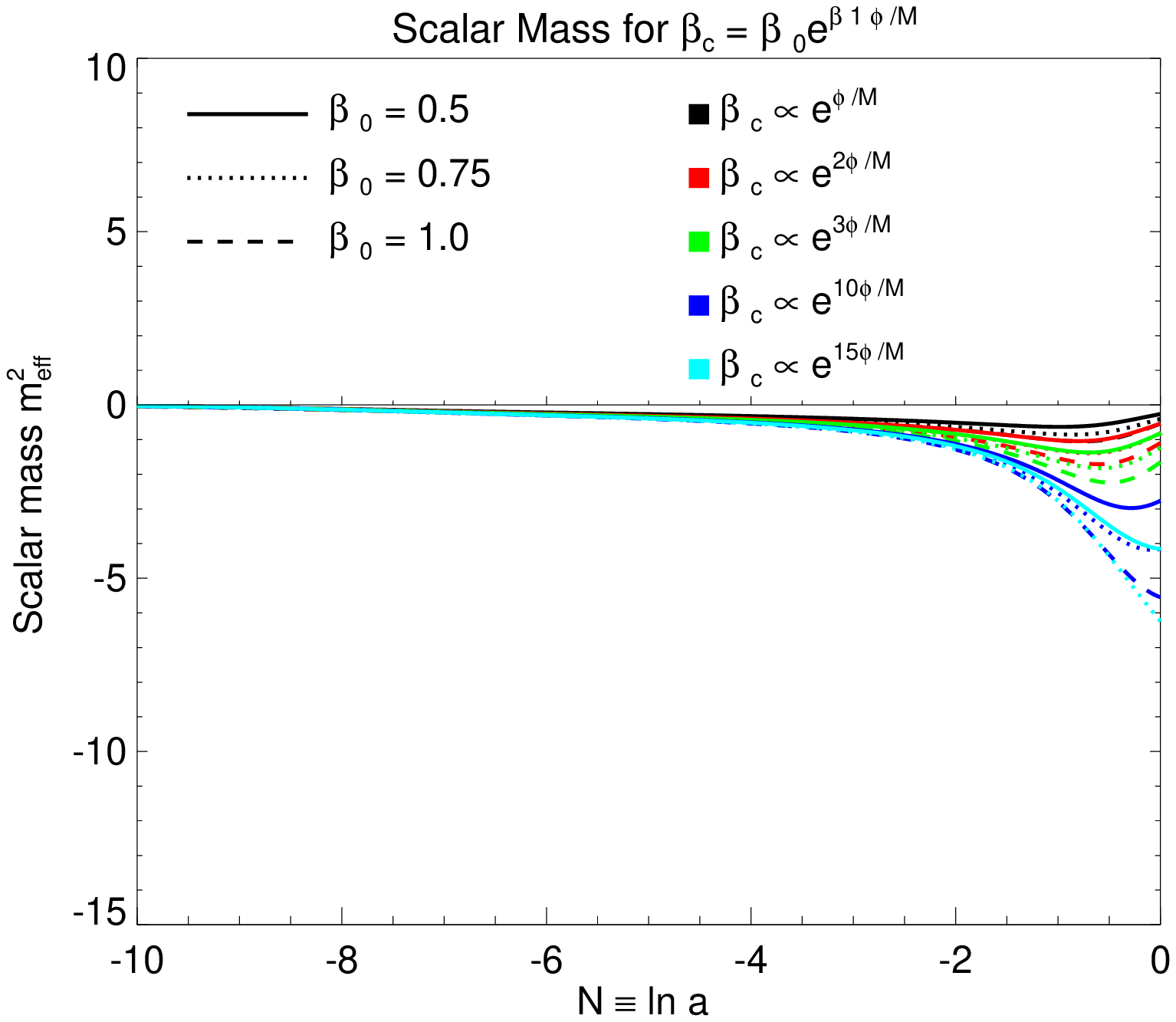}
  \caption{The evolution of the scalar mass $\hat{m}^{2} \equiv \hat{m}^{2}_{\phi } - \hat{m}^{2}_{\beta _{c}}$ as a function of the e-folding time $N$ for all the variable coupling models under investigation. The three panels refer to the total scalar mass of the models with $\beta _{c}(\phi ) \propto a^{\beta _{1}}$, $\beta _{c}(\phi )\propto \Omega _{\phi }$, and $\beta _{c}(\phi )\propto e^{\beta _{1}\phi /M}$, respectively. It is interesting to notice that the total effective mass $\hat{m}^{2}$ acquires negative values during matter domination in all the three different types of models. However, if the values of order $\sim -10$ found in the former two cases might lead to significant instabilities of the models at large scales, the exponential coupling models show very small absolute values of the total scalar mass at all redshifts, with $\hat{m}^{2}$ never exceeding ${\cal O}(1)$ for all the models considered in our high-resolution N-body simulations.}
\label{fig:scalarmass}
\end{figure*}

Secondly, as we have seen in Eqn.~(\ref{scalar_pert_full}), the variation of the coupling appears in the scalar field perturbations equation in the form of a ``coupling mass'' term $\hat{m}_{\beta _{c}^{2}}$, given by Eqn.~(\ref{coupling_mass}), such that the total effective mass takes the form:
\begin{equation}
\hat{m}^{2} = \hat{m}^{2}_{\phi } - \hat{m}^{2}_{\beta _{c}} \,.
\end{equation}
This might lead to large-scale instabilities in the scalar field perturbations if the total effective mass is negative, which is the case for all the models under investigation in this work, as we will show below. Nevertheless, if the absolute value of the total effective mass $|\hat{m}^{2}|$ is not too large, such instabilities might be confined to very large scales and the consequent growth of the large-scale gravitational potential might still be consistent with present observations. 
The total effective scalar mass for the three different types of coupling discussed in Sec.~\ref{bkg} as a function of the e-folding time $N$ is shown in the three panels of Fig.~\ref{fig:scalarmass}. As the plots show, all our models present a negative effective mass during all matter domination. Even though the absolute magnitude of the effective mass $|\hat{m}^{2}|$ is not very large, a value of $\hat{m}^{2}\sim -10$, that is realized  for the more phenomenological models of $\beta _{c}\propto a^{\beta _{1}}$ and $\beta _{\phi }\propto \Omega _{\phi }$, might become relevant at scales of a few hundred Mpc. On the other hand, the more physically motivated exponential coupling models, as one can see in the right panel of Fig.~\ref{fig:scalarmass}, feature a much smaller amplitude of the total effective scalar mass, whose absolute value is always $|\hat{m}^{2}|\sim{\cal O}(1)$, and should therefore not show instabilities at any subhorizon scale.  

The issue of potential large-scale instabilities related to the behavior of the effective scalar mass in variable coupling models will be discussed in detail in a dedicated work \citep{Amendola_Baldi}, but for the phenomenological study of variable coupling models addressed in this paper we assume the small scales of interest for N-body simulations ($\sim $ few tens of Mpc) not to be influenced by such potential large scale instabilities. Also, we will discard the Yukawa-like correction to the force law discussed in Sec.\ref{prt} due to the small size of our simulation box, and we will always consider the scalar fifth-force to have infinite range.

\normalsize

\section{The simulations} \label{sim}

\subsection{The simulations set}

The main focus of the present work is on the effects that coupled DE models with a time dependent coupling have on structure formation, and in particular on the properties of highly nonlinear structures as cluster-sized or galaxy-sized CDM halos. This is done by means of a series of high-resolution hydrodynamical N-body simulations carried out with a suitably modified version of the numerical code {\small GADGET-2} \citep{gadget-2}. 

N-body simulations of coupled DE models have been performed before for the case of a constant coupling by \citet{Maccio_etal_2004} and BA10, and this work constitutes the natural extension to a time dependent coupling of the latter publication.
In particular, we use here the same modified version of {\small GADGET-2} that was used in BA10, since the implementation presented and used in that work includes also variable coupling models, even though it had been used so far only for the simplified case of a constant coupling.

To our knowledge, this work presents the first N-body simulations of coupled DE models with time-dependent couplings carried out to date.
Previous attempts to use N-body simulations for other types of modified newtonian gravity (\ie different from the coupled DE scenario) have been discussed \eg in \citet{Nusser_Gubser_Peebles_2005,Stabenau_Jain_2006,Springel2007,Laszlo_Bean_2008,
  Sutter_Ricker_2008, Oyaizu_2008,Li_Zhao_2009, Hellwing_etal_2010,DeBoni_etal_2010}.
  
The implementation of coupled DE models in {\small GADGET-2} has already been described in full detail in BA10, and since no major modifications with respect to that setup are required to carry out the time dependent coupling simulations presented in this work, we refer the interested reader to the more detailed description presented in BA10 and we only summarize very briefly here the main features of our modified algorithm:

\begin{itemize}
\item A table of values of the Hubble function $H(a)$ that is computed for each model by integrating the system (\ref{dimensionless_system}) is read in by the code and the value of $H$ at each timestep is linearly interpolated from the table;
\item The mass of CDM particles in the simulation box is corrected at every timestep according to the mass evolution given by Eqn.~(\ref{mass_variation}), and shown for each model in Fig.~\ref{fig:mass_correction};
\item The small scale particle-particle (computed with the gravitational oct-tree algorithm implemented in {\small GADGET-2}) and the large scale particle-mesh gravitational accelerations for CDM particles are computed taking into account the additional contribution arising from the scalar-field mediated fifth-force, as described by Eqn.~(\ref{accel_c});
\item The acceleration of CDM particles computed according to the prescription given in the previous point receives at  each timestep an additional contribution from the extra friction term that appears in the expression of $\tilde{H}$ in Eqn.~(\ref{accel_c}).
\end{itemize}

As discussed in Sec.~\ref{obs}, we want to investigate the nonlinear evolution of structure formation in the context of a few selected cosmological models among all the cosmologies described in Table~\ref{cosmological_models}. To this end, we use the modified version of {\small GADGET-2} briefly summarized above to run high resolution hydrodynamical N-body simulations for the models $\Lambda$CDM, EXP010a2, EXP015a3, EXP010e2, EXP010e3 and EXP015e3. The parameters of the simulations are summarized in Table~\ref{simulations_table}. Hydrodynamical forces computed with the SPH ({\em Smoothed Particle Hydrodynamics}) algorithm implemented in {\small GADGET-2} are acting on baryonic particles in all the simulations, while non-adiabatic processes like  \eg radiative cooling, star formation, and feedback mechanisms from Supernovae or Active Galactic Nuclei are not included in any of our runs.

\begin{center}
\begin{table}
\begin{center}
\begin{tabular}{l c}
\hline
Box Size $L$&  80 $h^{-1}$ Mpc  \\
Number of baryonic particles $N_{b}$ & $512^{3}$ \\
Number of CDM particles $N_{c}$ & $512^{3}$ \\ 
Baryon Mass $M_{b}$  &  4.82 $\times $ 10$^{7} h^{-1}$ M$_{\odot }$\\
CDM mass $M_{c}(z=0)$ & 2.41 $\times $ 10$^{8} h^{-1}$ M$_{\odot }$ \\
Gravitational Softening $\epsilon_{s}$ & 3.5 $h^{-1}$ kpc \\
\hline
\end{tabular}
\end{center}
\caption{The parameters of the high-resolution hydrodynamical  N-body simulations discussed in Sec.~\ref{sim} ~and~\ref{results}.}
\label{simulations_table}
\end{table}
\end{center}

\subsection{Initial conditions}
\label{ICs}

The initial conditions for all the simulations are generated by setting up a random-phase realization of the Eisenstein \& Hu power spectrum \citep{Eisenstein_Hu_1997} according to the Zel'dovich approximation \citep{Zeldovich_1970}, where the normalization amplitude of the power spectrum is adjusted to the desired value of $\sigma _{8}$ . In doing so, we are implicitly assuming that the coupling does not affect the shape of the initial matter power spectrum. We are therefore discarding from our treatment any possible early effect of the coupling on the statistical properties of the density field. This is in general a reasonable assumption since in all our models the coupling is quite small at high redshifts. However, in particular for the exponential coupling models, such small coupling in the early Universe could still produce some slight tilt in the matter transfer functions \citep{Mainini_Bonometto_2007}. Nevertheless, at the resolution level achieved in the present work such tilt was already found to have very little impact on the subsequent evolution of structure growth as compared to the other effects related to the modified perturbations evolution (BA10). Higher resolution studies, however, might be significantly affected even by such weak distortions of the small-scale power spectrum, and should therefore include a full treatment of the early effects of the DE interactions on the matter transfer function.

Initial conditions are therefore generated by perturbing the positions of particles from a cartesian grid until the desired density field is realized, and  then by rescaling the particle displacements from the grid points with the linear growth factor $D_{+}$ of each cosmological model between $z=0$ and $z=60$, which is the starting redshift of all our runs. Although this convention on the normalization of the power spectrum amplitude is probably the most commonly used, other choices are equally valid and have been recently adopted in some related studies \citep{Baldi_Pettorino_2010,DeBoni_etal_2010}.

\begin{figure*}
\includegraphics[scale=0.32]{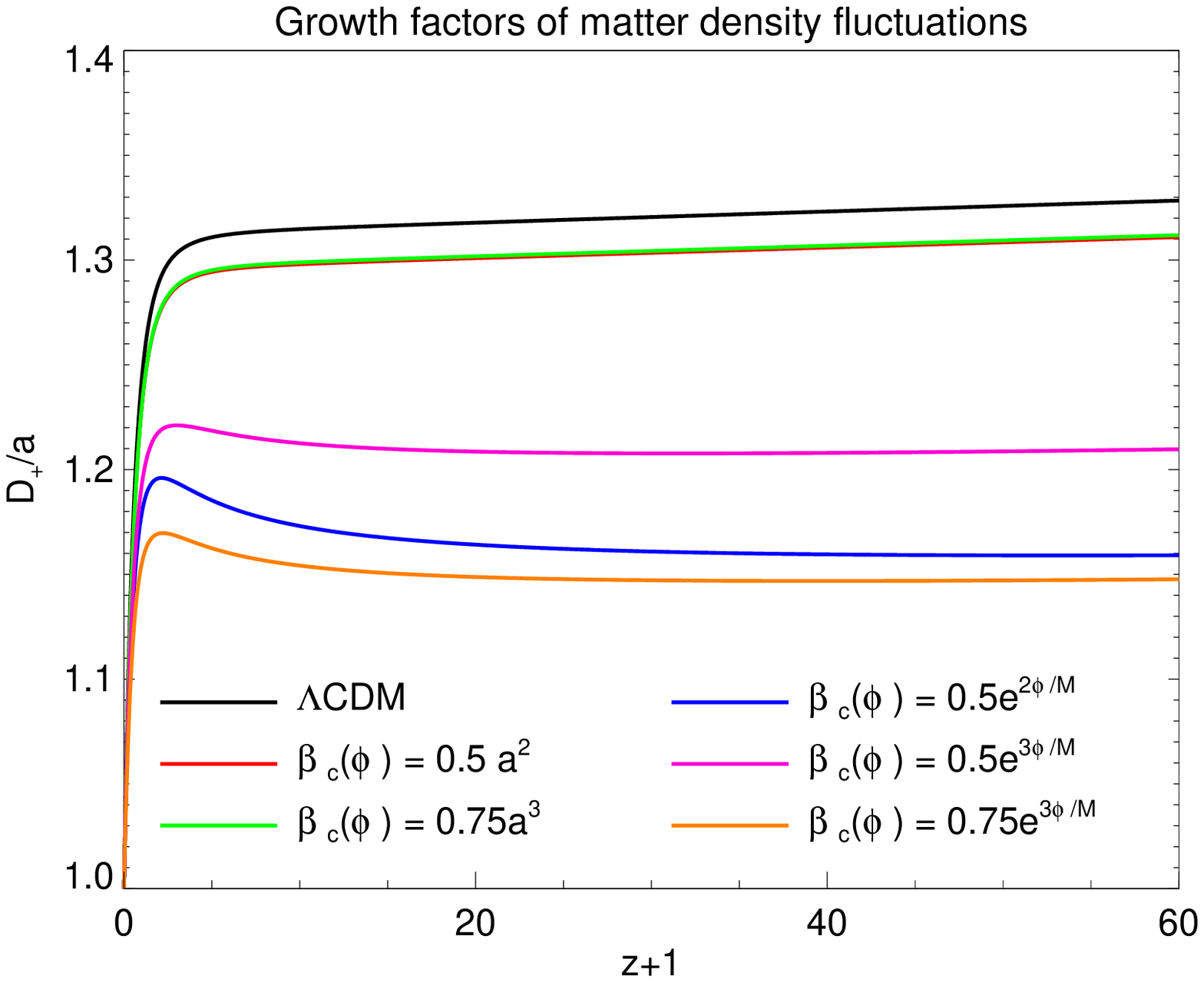}
\includegraphics[scale=0.32]{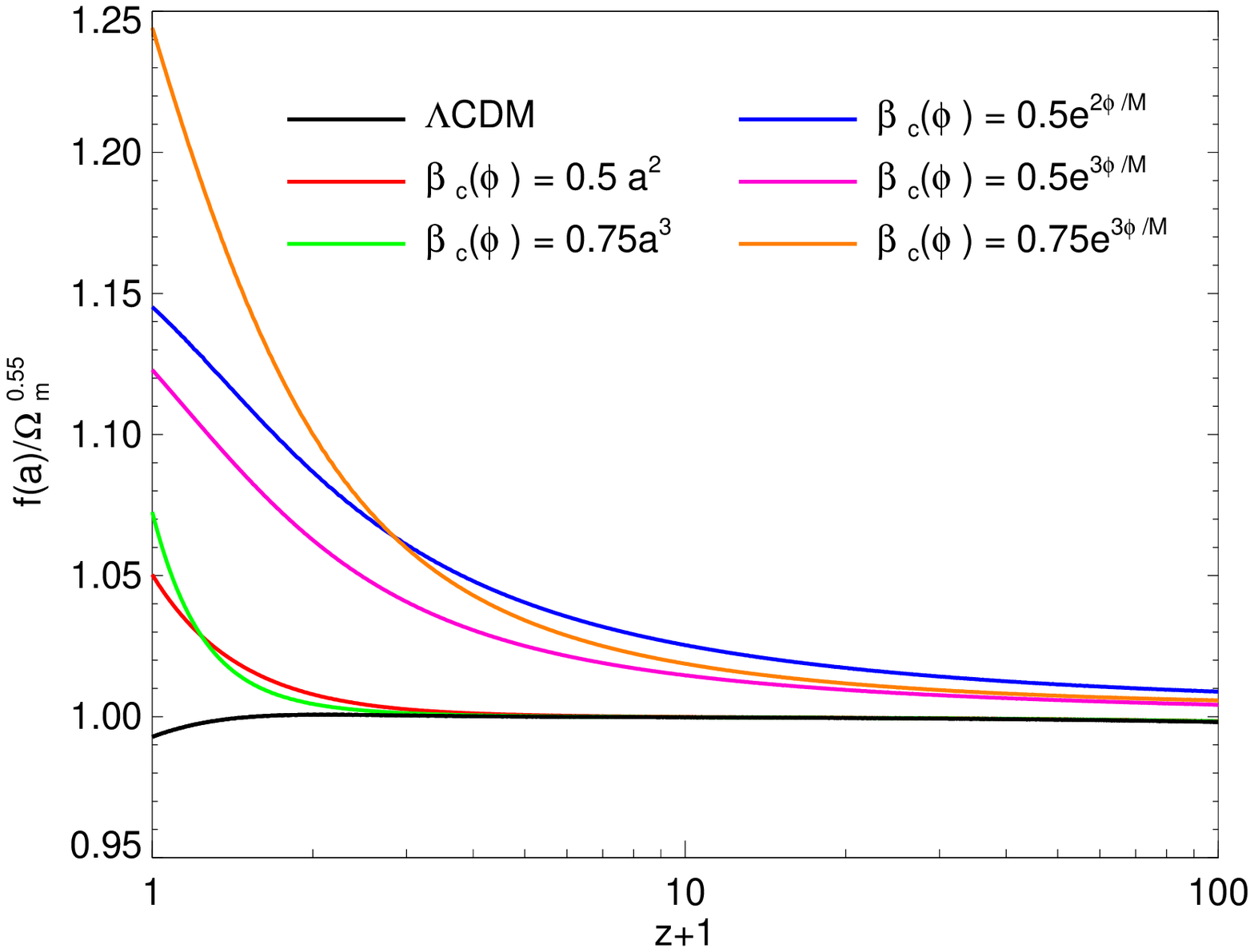}
\includegraphics[scale=0.32]{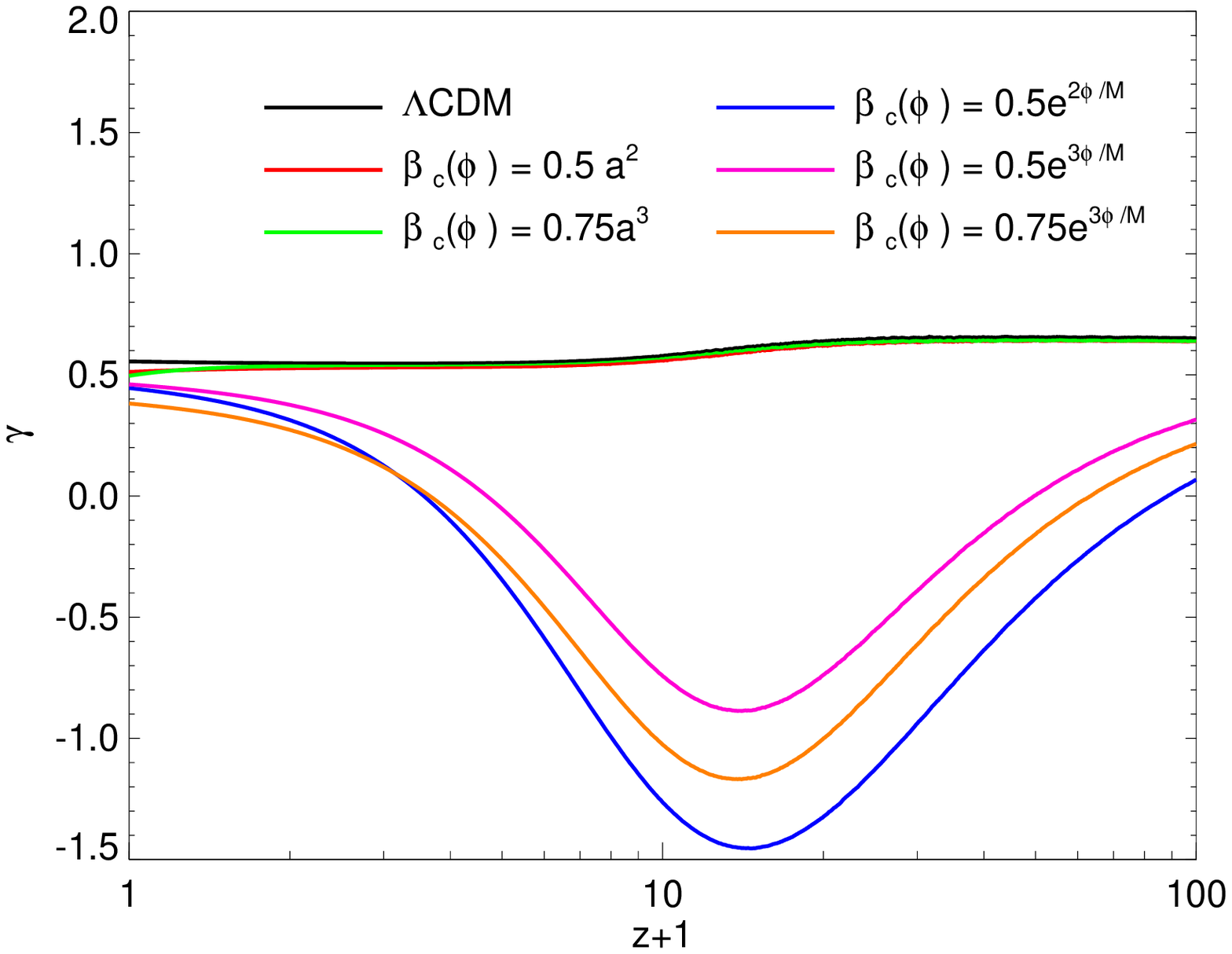}
\caption{{\em Left Panel}: The Growth Factor $D_{+}/a$ as a function of redshift for the six models selected for N-body simulations. The different colors correspond to different models, according to the legend. The red curve, corresponding to the model EXP010a2, is almost completely hidden by the green curve, corresponding to the model EXP015a3. This shows how these two models are practically indistinguishable from each other. {\em Middle Panel}: The evolution of the ratio of the growth rate function $f(a)\equiv d\ln D_{+}/d\ln a$ to the $\Lambda $CDM fitting formula $\Omega _{M}^{0.55}$ for the same set of models.  The black curve, representing $\Lambda $CDM, is consistent with a ratio of 1.0, as expected. The scale factor dependent models EXP010a2 and EXP015a3 (red and green lines, respectively), are indistinguishable from $\Lambda $CDM for most of the expansion history, but show a progressively faster growth at $z \lesssim 1-2$. The exponential coupling models, instead, have already a sensibly faster growth of perturbations at $z$ as high as $100$. {\em Right Panel}: The derived $\gamma $ index for the growth of density perturbations for the same set of models as in the previous two panels. While the scale factor dependent models show a value of $\gamma $ consistent with the $\Lambda $CDM value of $0.55$ during matter domination, with only small deviations at low redshift (the slight increase at high $z$ is due to the presence of radiation, and is an expected effect), the exponential coupling models show a completely different behavior: the $\gamma $ index decreases significantly during most of matter domination, even becoming negative at some stage, and reaches a minimum in the redshift range $z \sim 10-20$ before growing again towards values comparable, but still significantly lower, to the $\Lambda $CDM case. Similar behaviors have been recently reported for some realizations of $F(R)$ modified gravity theories \citep{Motohashi_etal_2010,Narikawa_Yamamoto_2010,Appleby_Weller_2010}.}
\label{fig:growth}
\end{figure*}

All the simulations presented in this work have the same random phases for the realization of the power spectrum in the initial conditions, and structures are therefore expected to form in the same locations in all the runs. 
Finally, the velocities of particles are set according to linear perturbation theory, by a simple relation with the computed initial overdensities, which in Fourier space reads:
\begin{equation}
\vec{v}(\vec{k},a) = if(a)aH\delta(\vec{k},a)\frac{\vec{k}}{k^{2}}\,,
\end{equation}
where the growth rate $f(a)$ is defined as
\begin{equation}
f(a) \equiv \frac{d \ln D_{+}}{d \ln a}\,.
\end{equation}
For $\Lambda $CDM cosmologies, the total growth rate is always well approximated by a power of the total matter density $(\Omega _{c}+\Omega _{b})^{\gamma }$ where $\gamma=0.55$ \citep{Peebles_1980}. This approximation is no longer valid for coupled DE models in general, as discussed in \citet{DiPorto_Amendola_2008} and BA10 where alternative phenomenological fitting formulae for constant coupling models have been proposed. Also in the case of time dependent couplings the growth rate does not follow the $\Lambda $CDM behavior, and observations of the growth of structures at different redshifts \citep[see \eg ][for a recent analysis of growth data]{Bean_Tangmatitham_2010} might have the power to detect deviations from the standard value of the growth index $\gamma $ and put constraints on variable coupling models. Such a detection would be a clear indication of some cosmological modification of standard gravity. 
The growth factor $D_{+}$ and the ratio of the growth rate $f(a)$ over the $\Lambda $CDM fitting formula $\Omega _{M}^{0.55}$, as computed from the numerical integration of Eqs.~(\ref{gf_c},\ref{gf_b}), are shown in the left and in the middle panels of Fig.~\ref{fig:growth}, respectively, for the models investigated with our high-resolution N-body simulations.
\normalsize

Additionally, in the right panel of Fig.~\ref{fig:growth} we plot for the same set of models the derived evolution of the growth index $\gamma $ computed as:
\begin{equation}
\gamma = \frac{\ln f}{\ln (\Omega _{b} + \Omega _{c})} \,.
\end{equation}
It is very interesting to notice the strikingly different behavior of $\gamma $ for the exponential coupling models EXP010e2, EXP010e3, EXP015e3, with respect to $\Lambda $CDM and the phenomenological models EXP010a2 and EXP015a3. For the latter, a quasi-constant value of gamma very close to the $\Lambda $CDM value of $0.55$ is found during matter domination, and the curves are practically indistinguishable from the fiducial $\Lambda $CDM model. On the contrary, the former models show a completely different behavior of $\gamma $, that even becomes negative for a large period of time and reaches its minimum value in the redshift range $z \sim 10-20$. 
Interestingly, a very similar evolution of the growth index parameter $\gamma $ to the one shown in the right panel of Fig.~\ref{fig:growth} has been recently reported for some $F(R)$ models of modified gravity \citep{Motohashi_etal_2010,Narikawa_Yamamoto_2010,Appleby_Weller_2010}.  Given the existence of a well known conformal correspondence between coupled DE models and modified gravity theories it would be particularly interesting to investigate whether such similar behaviors of the gamma index represent another element of connection between the two scenarios.

\section{Results of the N-body simulations}\label{results}

We present and discuss in this section the results of our analysis of the high-resolution simulations described above. In order to study the properties of collapsed structures we first need to identify groups and gravitationally bound substructures in our simulations. To this end, we apply the Friends-of-Friends (FoF) and {\small SUBFIND} algorithms \citep{Springel2001} to the particles distributions in our simulations outputs. For the FoF computation we use a linking length of $\lambda = 0.2 \times \bar{d}$ where $\bar{d}$ is the mean particle spacing. As we stressed in Sec.~\ref{ICs}, all the simulations are started from the same random realization of the Eisenstein \& Hu power spectrum, and structures are therefore expected to form in the same positions in all the runs. This will allow us to identify the same objects in all the simulations and to directly compare their properties. In BA10, due to the slight tilt in the transfer functions used to set up the initial conditions for the simulations of the different coupled DE models, it was necessary to apply a selection criterion in order to ensure that two structures identified in two different simulations could be safely considered as being the same object. This selection criterion consisted in identifying halos found in different simulations as the same object only if the most bound particle of each of them lied within the virial radius of the corresponding structure in the $\Lambda$CDM simulation. Although in the present work there is no tilt of the initial power spectrum shape, we decide nevertheless to apply the same selection criterion to our group catalogs, since the different timestepping  induced by the different evolution of the effective gravitational forces in each run, especially at low redshifts, might induce some offsets in the final positions of collapsed structures. We restrict this selection procedure to the 300 most massive halos in our catalogs, which cover a  range of CDM virial masses between $M_{200} = 3.46 \times 10^{12}~h^{-1}$ M$_{\odot }$ and $M_{200} =  3.63 \times 10^{14}~h^{-1}$ M$_{\odot }$ in the $\Lambda$CDM cosmology.
\begin{figure*}
\includegraphics[scale=0.45]{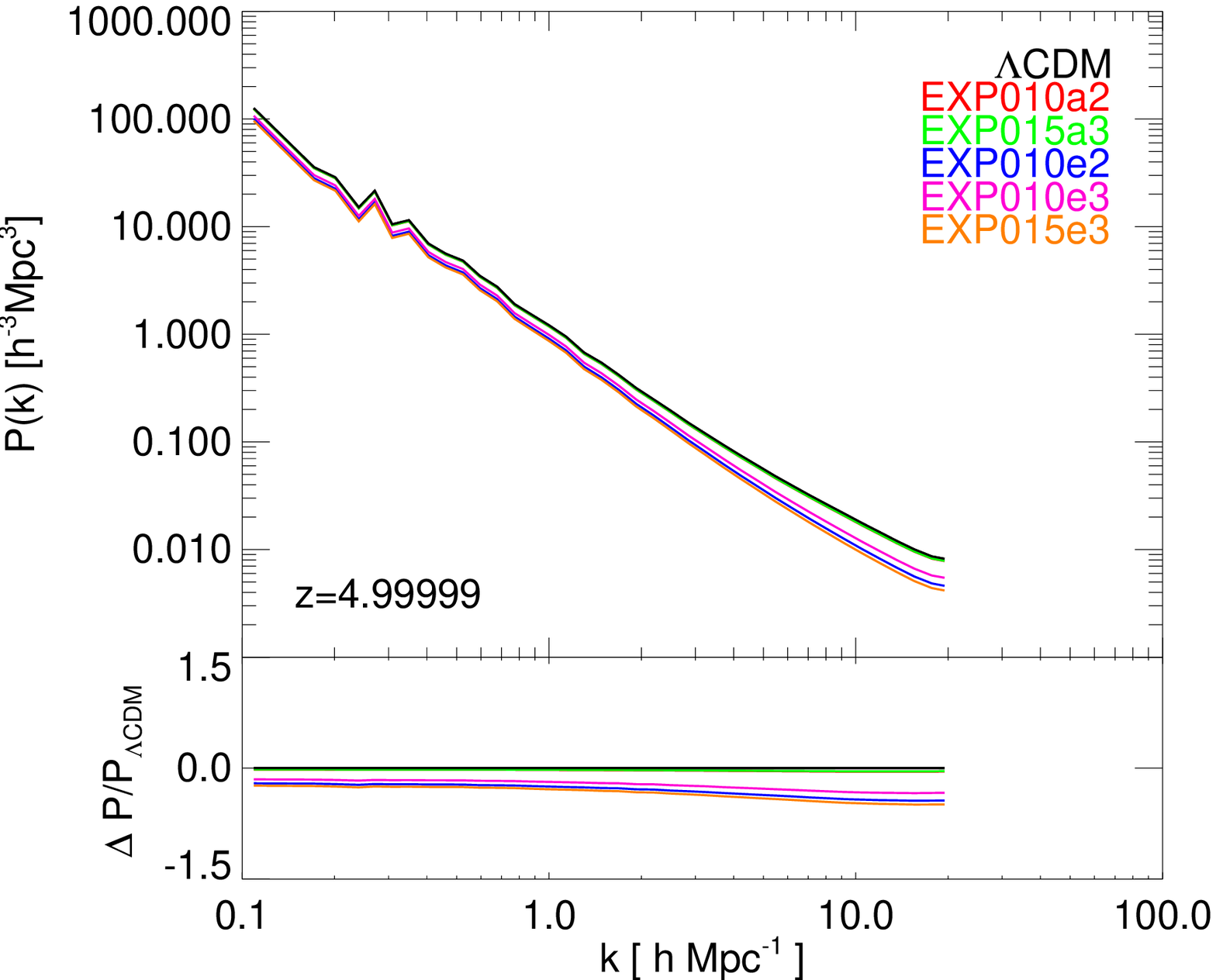}
\includegraphics[scale=0.45]{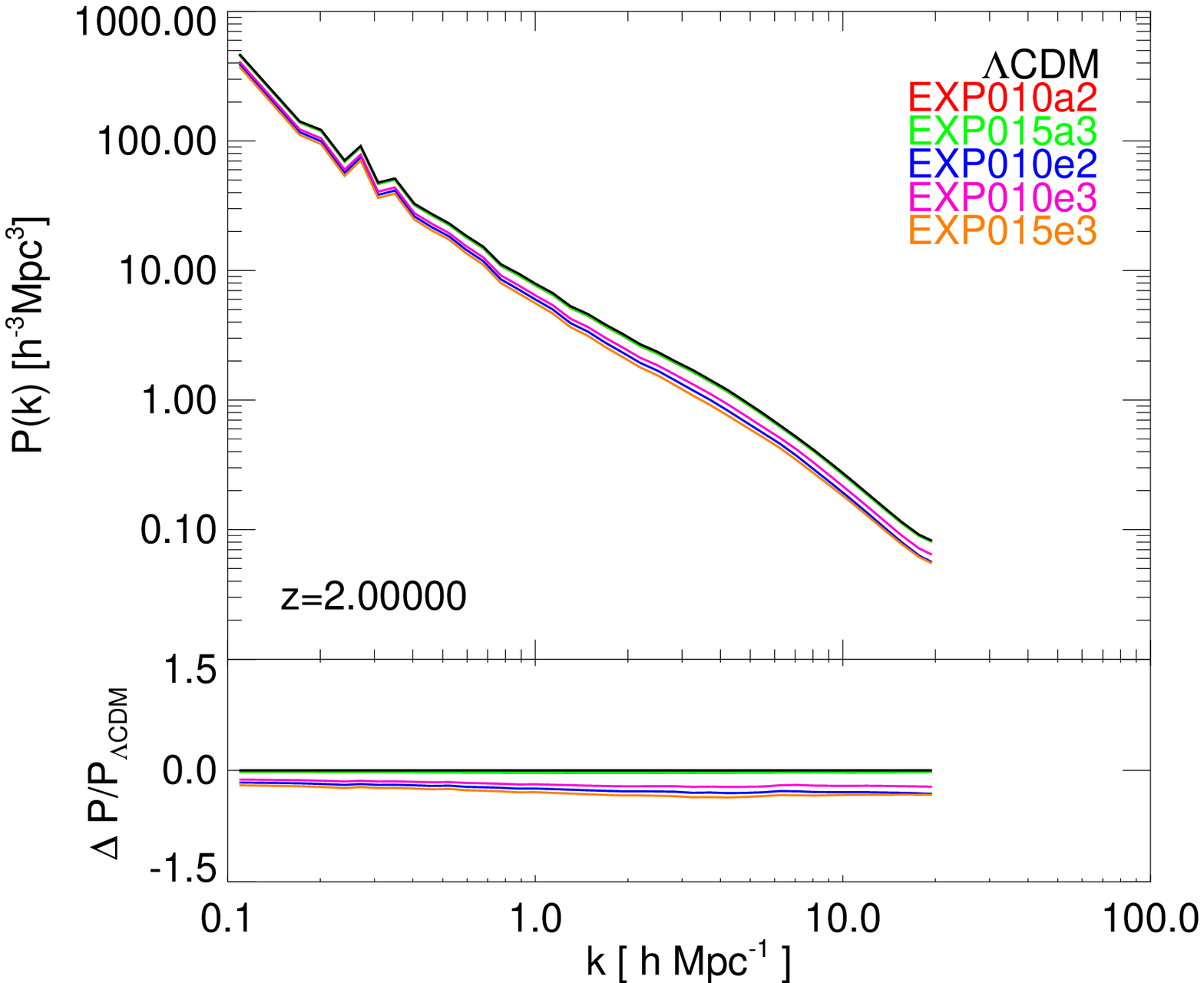}\\
\includegraphics[scale=0.45]{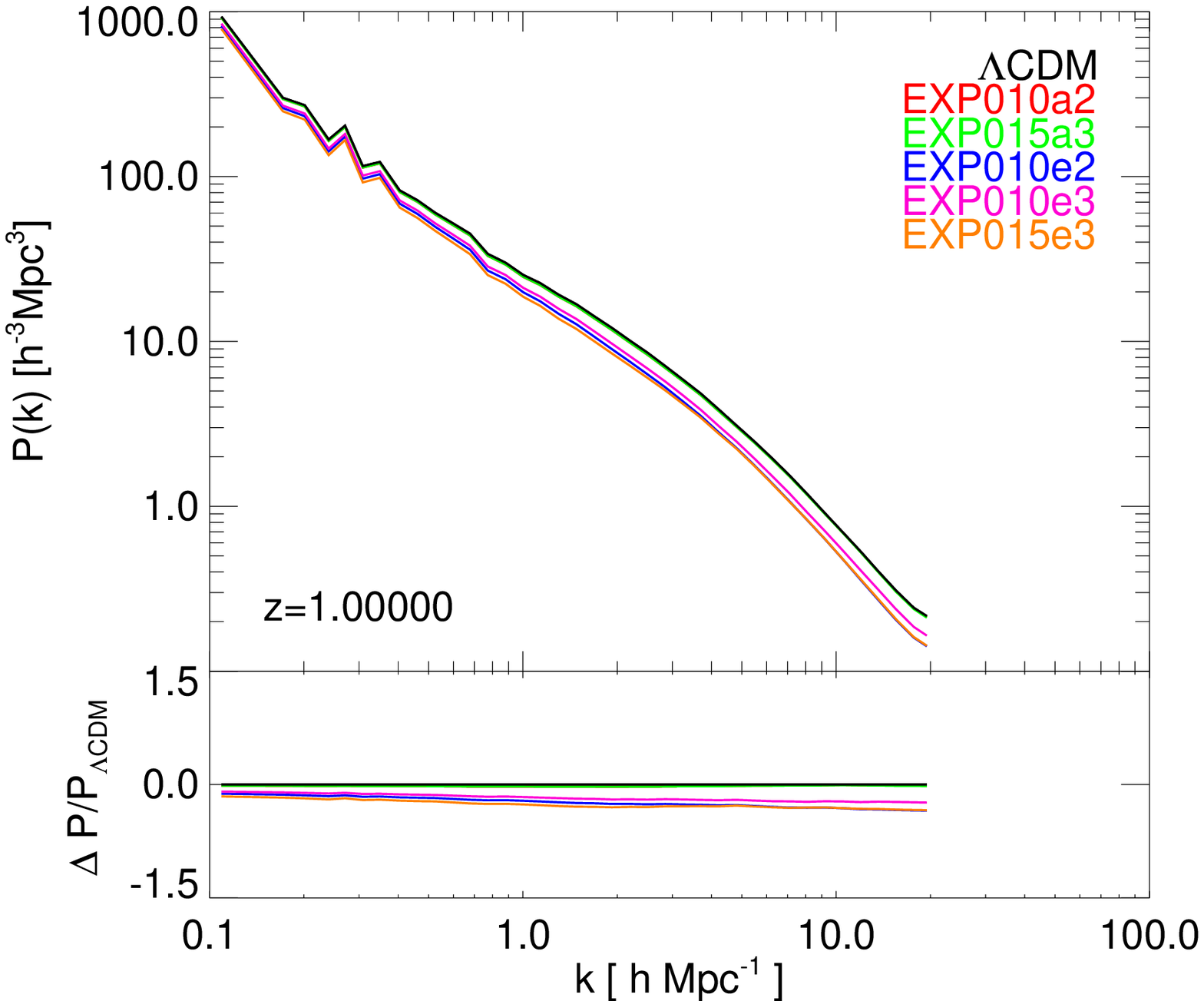}
\includegraphics[scale=0.45]{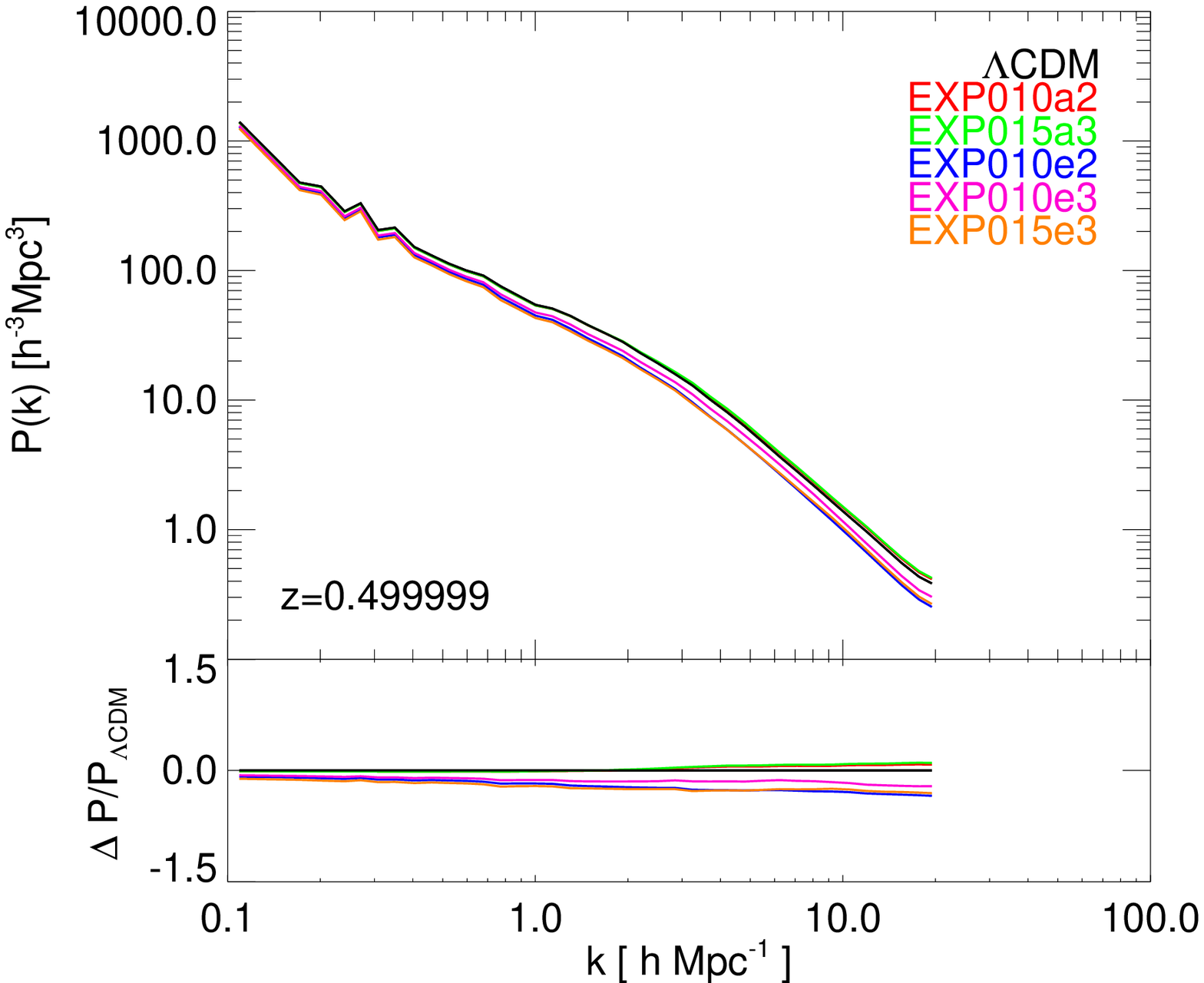}\\
\includegraphics[scale=0.45]{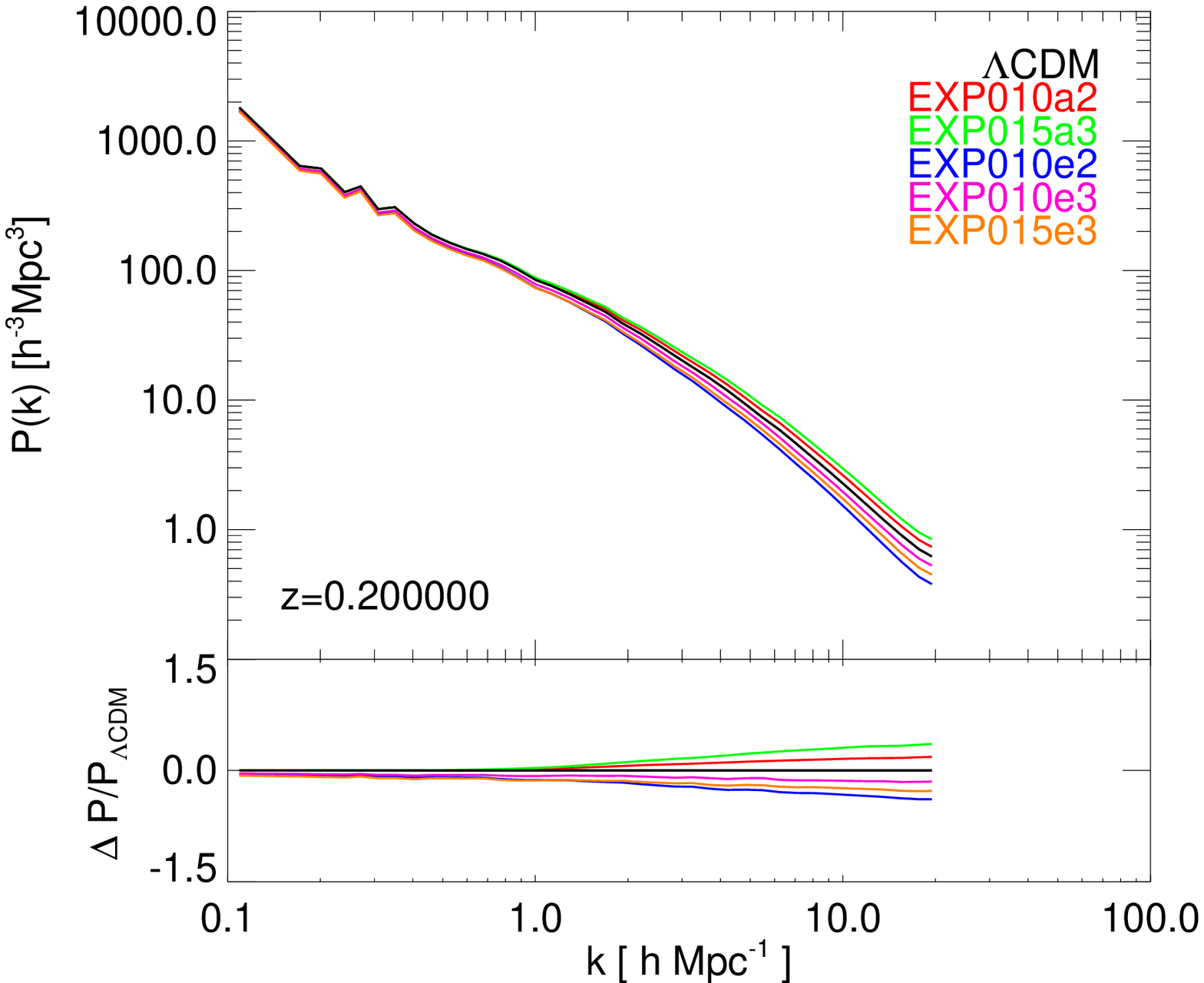}
\includegraphics[scale=0.45]{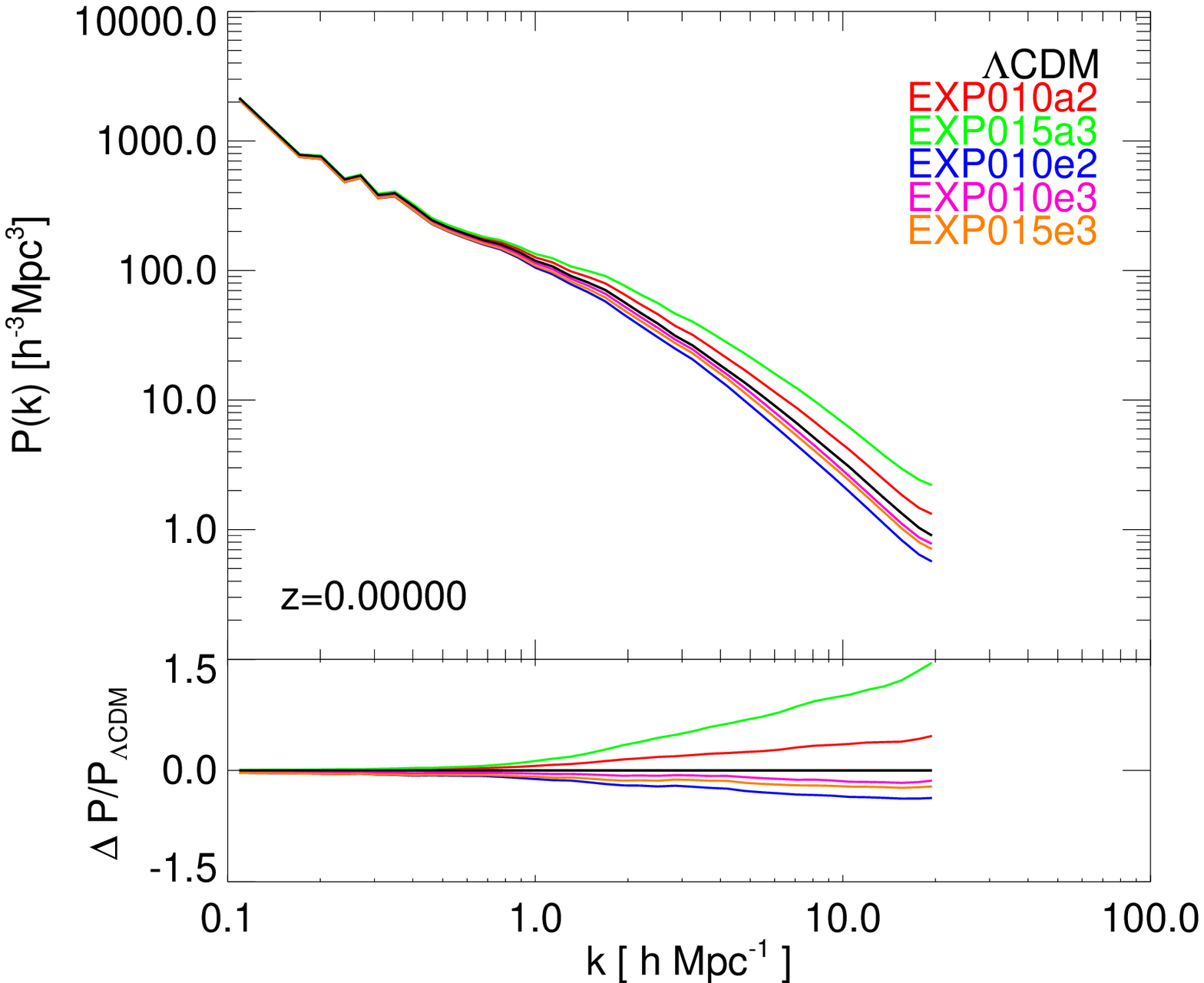}
\caption{The matter power spectrum at different redshifts for the six models studied in our set of N-body simulations. In the bottom part of each panel we plot the residuals with respect to $\Lambda $CDM for a more clear understanding of the different evolutions of the power spectrum amplitude at different length scales. The different behavior of the two types of coupling evolution, with the exponential coupling models already showing a different amplitude at high $z$, while the scale factor dependent models show a strong increase of power at small scales only at $z \lesssim 0.5$ (as illustrated in detail in the text) is clearly visible in the plots.}
\label{fig:powerspec}
\end{figure*}
\normalsize

\subsection{Matter power spectrum}
\label{power}

As a first step in the analysis of our N-body runs we compute the power spectrum of matter density fluctuations $P(k,z)$ at several different redshifts, as well as the separate density power spectra for the baryonic and the CDM components, $P_{b}(k,z)$ and $P_{c}(k,z)$. The evolution of $P(k,z)$ as a function of scale at different redshifts for all our high-resolution simulations is shown in Fig.~\ref{fig:powerspec}. 

We should remind here that all our simulations are normalized assuming the same amplitude of the large-scale linear power spectrum at the present time. This is done by normalizing with the same value of $\sigma _{8}(0)=0.807$ the present amplitude of the Eisenstein \& Hu linear power spectrum, and then scaling the values of density fluctuations to the starting redshift of the simulations with the appropriate growth factor for each cosmological model. 
As already stressed above, we have discarded early effects of the coupling on the initial power spectrum shape since these have already been shown to have a minor impact on the final results at the present level of numerical resolution (BA10) even for larger values of the early coupling than those considered here. For higher resolution works, however, such effects might become important and should be taken into account in the initial conditions of the N-body runs.

As it can be clearly seen in the first and second panels of Fig.~\ref{fig:growth}, the models where the coupling strength evolves as a power of the scale factor (EXP010a2, EXP015a3), due to the very fast decrease of the coupling towards higher redshifts, present a total growth factor which is almost indistinguishable from the $\Lambda $CDM one for most of the expansion history of the Universe, and the offset with respect to $\Lambda $CDM is due to the different growth at low redshifts ($z \lesssim 1$). On the other hand, the more physically motivated models with a coupling strength related to the evolution of the scalar field (EXP010e2, EXP010e3, EXP015e3) show a sensibly different evolution of the growth factor with respect to the $\Lambda $CDM case already at redshifts of the order of $z \sim 20$. It therefore comes as no surprise, as it can be seen in the first three panels of Fig.~\ref{fig:powerspec}, that at high redshifts the power spectra of the EXP010a2 and of the EXP015a3 models are practically indistinguishable from the $\Lambda $CDM case, while the exponential coupling models EXP010e2, EXP010e3, and EXP015e3, show a slightly lower amplitude of the power spectrum at all scales, with a weak enhancement of the effect for progressively smaller scales.

The situation is inverted at lower redshifts, as it can be seen in the last three panels of Fig.~\ref{fig:powerspec}. In fact, due to the constantly faster growth of CDM density fluctuations, the exponential coupling models show a faster evolution of the power spectrum with respect to $\Lambda $CDM, such that between $z=1.0$ and $z=0$ the gap in the power spectrum amplitude with respect to $\Lambda $CDM at large scales is progressively reduced and these models catch up the $\Lambda $CDM amplitude at the present time, while at smaller scales ($k \gsim 1.0~h$ Mpc$^{-1}$) a residual lack of power is still present at $z=0$.
On the other hand, the phenomenological models EXP010a2 and EXP015a3, due to the sudden increase of the mutual attraction of CDM particles, show a tremendous growth of density fluctuations at scales smaller than $k  \sim 1.0~h$ Mpc$^{-1}$ between $z=0.5$ and $z=0$, such that at these scales the power spectrum amplitude at low redshifts is substantially increased as compared to the $\Lambda $CDM case, with the effect becoming progressively larger for smaller scales. The most extreme case, given by the model EXP015a3, shows an amplitude of the power spectrum at $k \sim 10~h$ Mpc$^{-1}$ that is roughly twice as large as for the $\Lambda $CDM cosmology. 

This analysis therefore presents us two quite different behaviors of the two classes of models investigated in our numerical runs. On one side, the exponential coupling models have a faster growth of the power spectrum amplitude over a rather long period of time, and starting with a substantially lower power spectrum amplitude at all scales at high redshifts, end up with an amplitude of the large scale power spectrum comparable to $\Lambda $CDM at $z=0$, nevertheless showing a residual lack of power at small scales. 
On the contrary, the phenomenological parametrizations of the coupling evolution embedded in models EXP010a2 and EXP015a3 present an almost identical growth of linear density perturbations to $\Lambda $CDM during most of the expansion history of the Universe, followed by a sudden enhancement of the growth for scales below $k  \sim 1.0~h$ Mpc$^{-1}$ between $z=0.5$ and $z=0$ as the coupling strength rapidly increases towards its present value. This late faster growth turns into a strong increase of the power spectrum amplitude at small scales. 
The situation at $z=0$, shown in the last panel of Fig.~\ref{fig:powerspec}, gives a quite interesting picture, with all the models being practically indistinguishable from $\Lambda $CDM at the largest scales, as a consequence of the common normalization of the linear power spectra at $z=0$, but with a broad range of power spectrum amplitudes at small scales.

This strikingly different behavior of the evolution of linear density perturbations in these two classes of coupled DE models could provide ways to observationally distinguish among the two scenarios with present and future datasets, and to constrain the functional form of the coupling evolution.
\begin{figure*}
\includegraphics[scale=0.32]{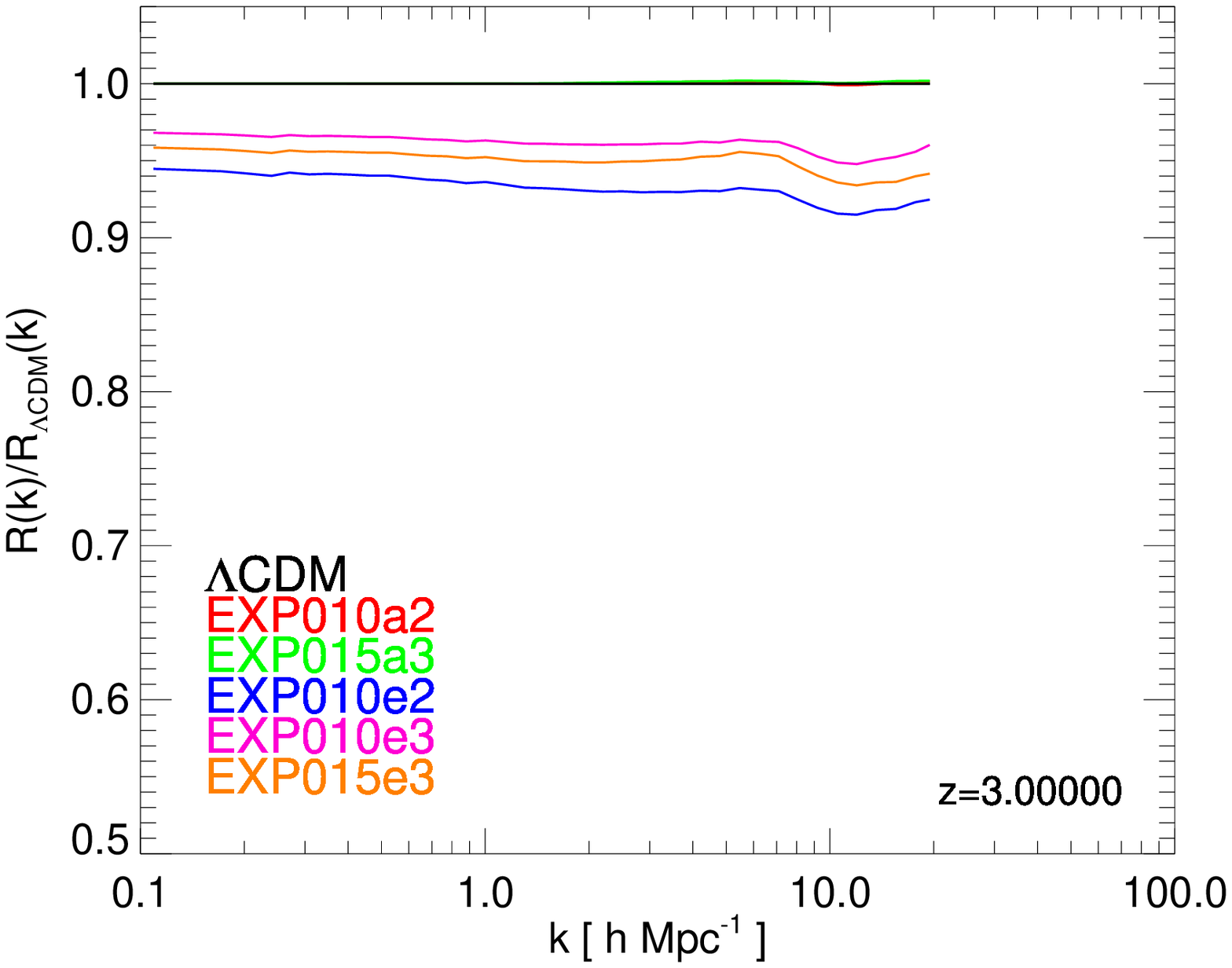}
\includegraphics[scale=0.32]{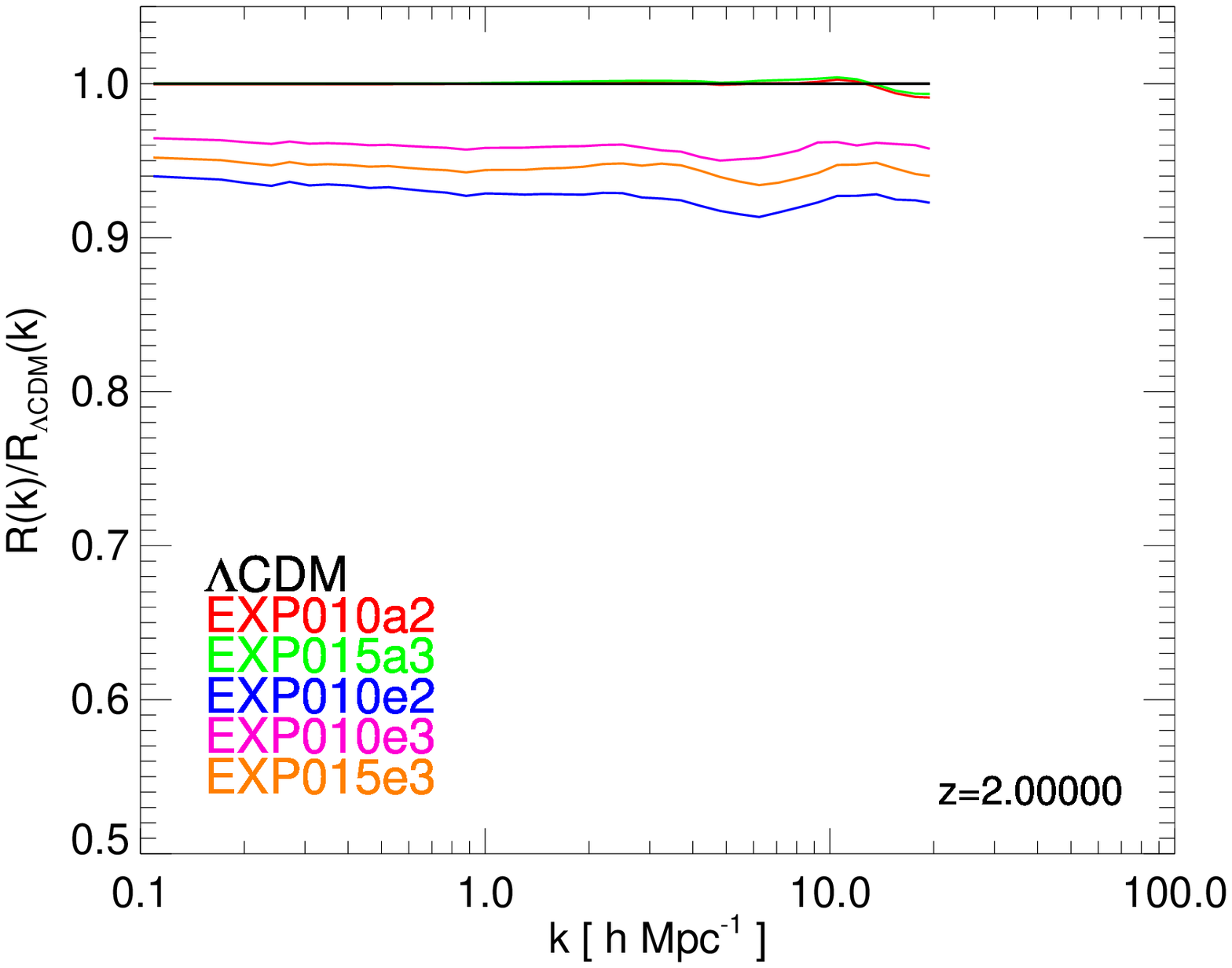}
\includegraphics[scale=0.32]{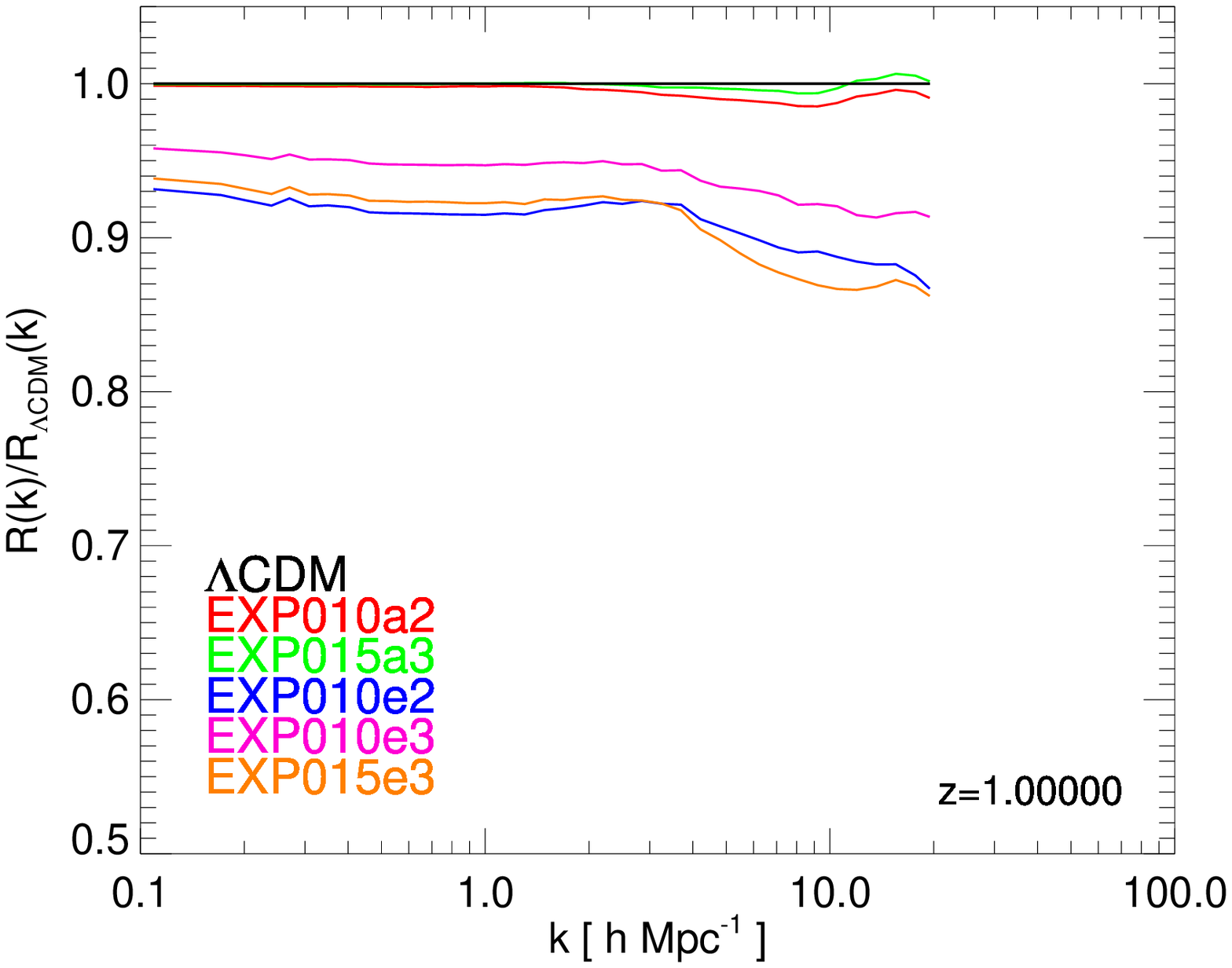}\\
\includegraphics[scale=0.32]{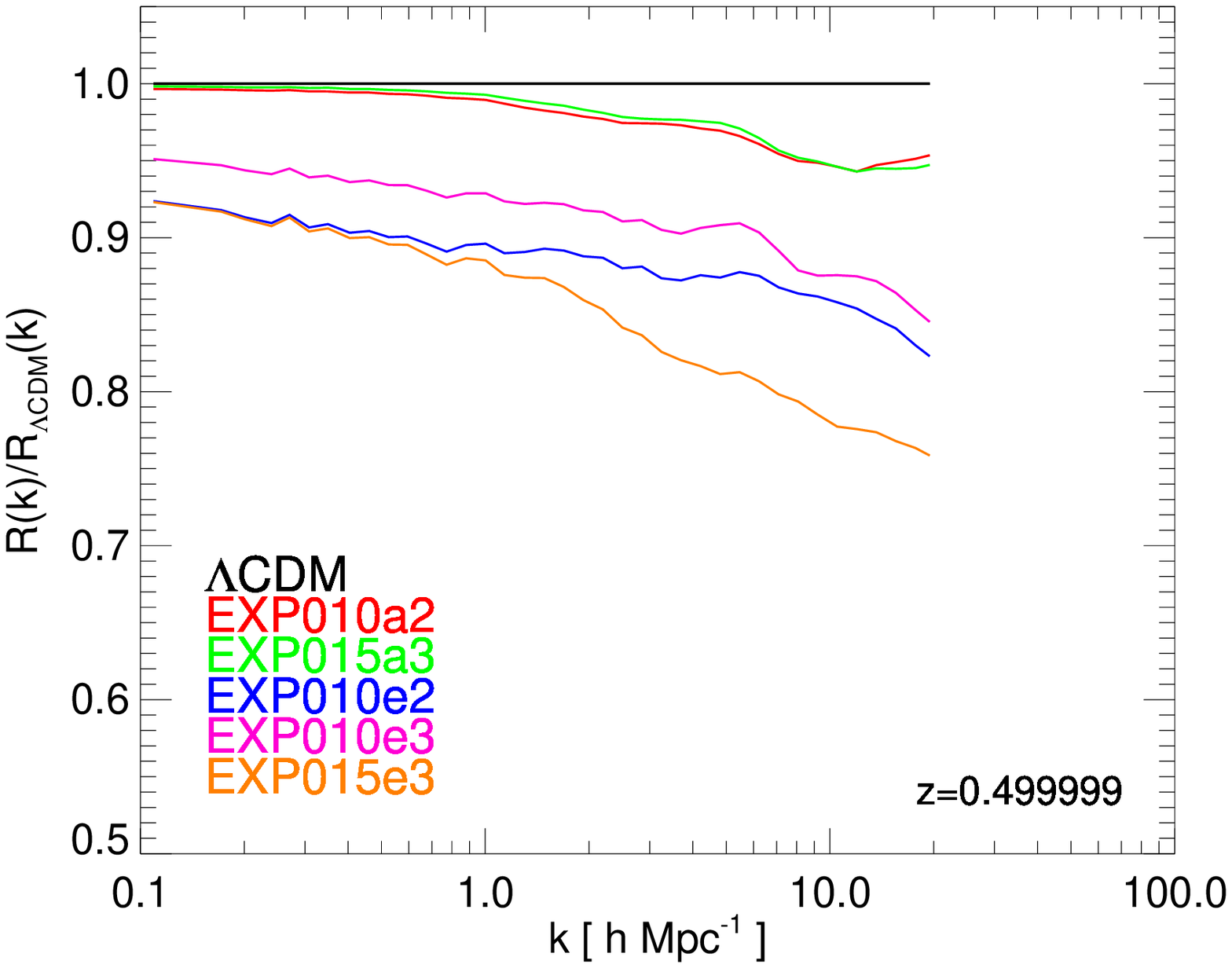}
\includegraphics[scale=0.32]{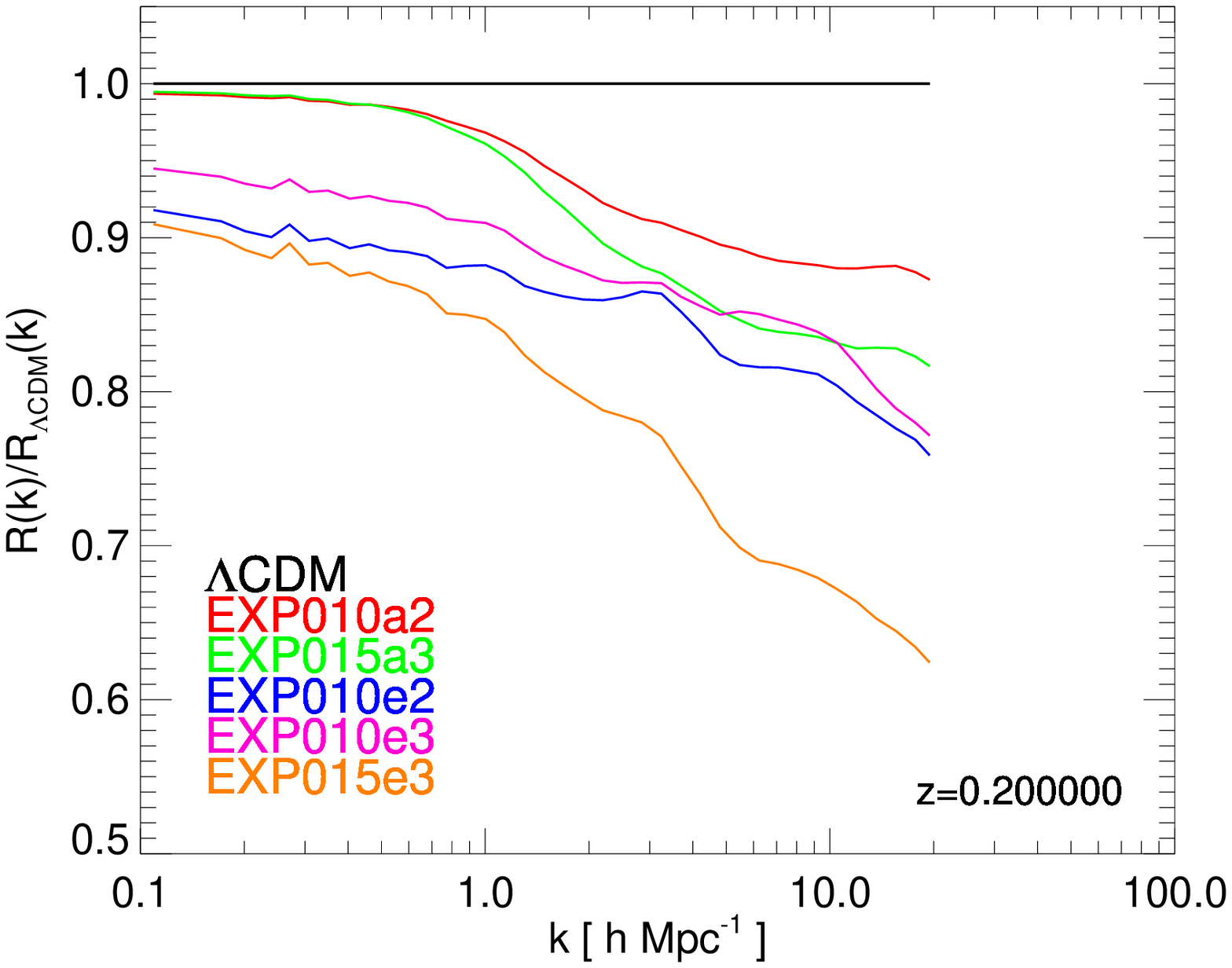}
\includegraphics[scale=0.32]{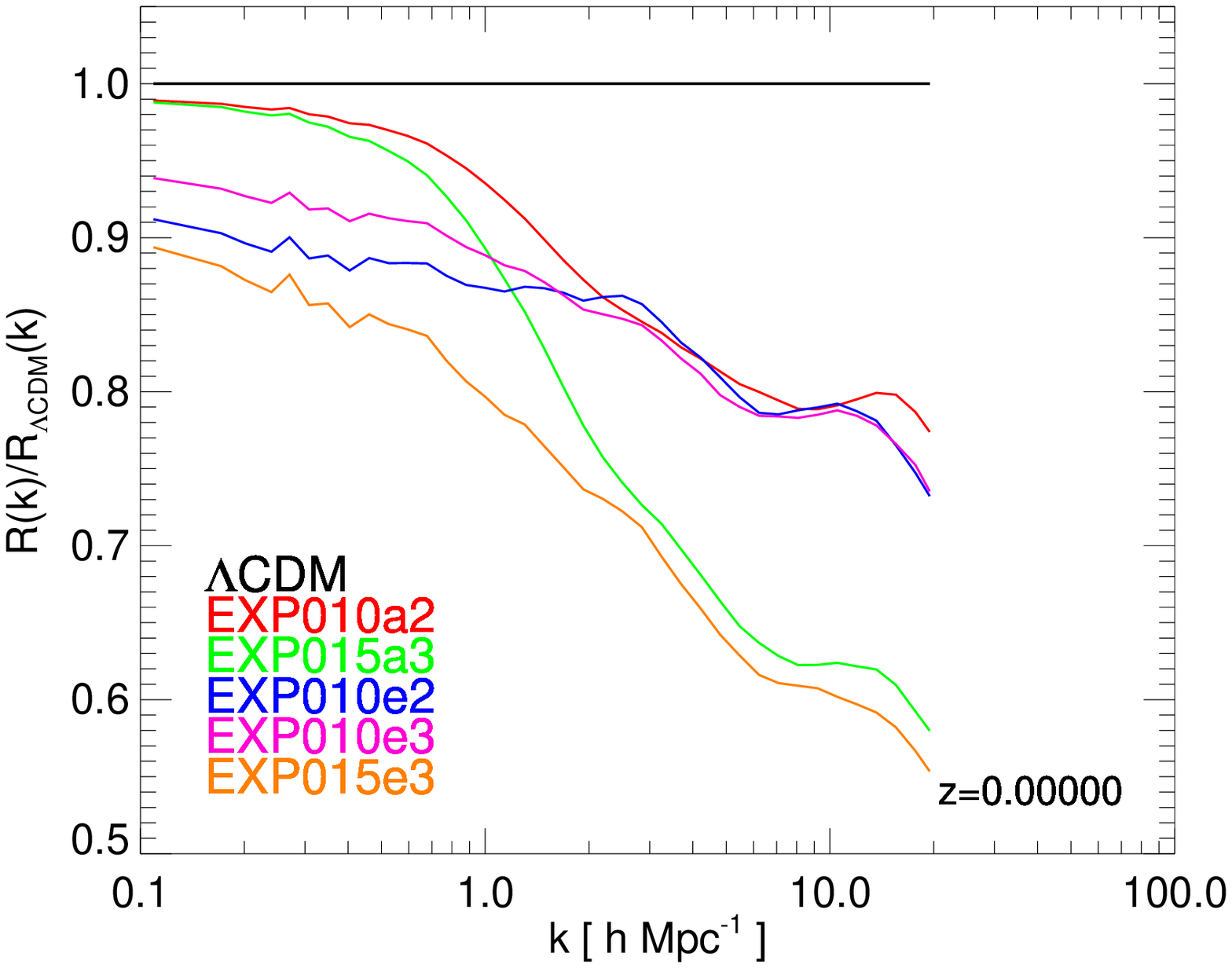}
\caption{The ratio of the bias $R(k,z)$ to the $\Lambda $CDM bias $R_{\Lambda \text{CDM}}(k,z)$ as a function of wavenumber $k$ at different redshifts. As extensively discussed in the text, at high redshifts ($z \gsim 1.0$, upper three panels) the EXP010a2 and EXP015a3 models are practically indistinguishable from $\Lambda $CDM, while the exponential coupling models already exhibit an almost constant bias at all scales, with a clear hierarchy corresponding to the different values of the coupling at high $z$. The situation changes dramatically at $z\lesssim 1$, where the scale dependence of the bias starts to appear clearly for all the models under investigation, and the hierarchy of the exponential coupling models is modified due to the more substantial growth of $\beta _{c}(\phi )$ in the EXP015e3 model. At $z=0$, all the models with the same present value of the coupling $\beta _{0}$ interestingly seem to end up with comparable values of the bias at the smallest scales available to our analysis.}
\label{fig:power_ratio}
\end{figure*}
\normalsize

\subsection{Baryon-CDM linear and mildly nonlinear bias}
\label{linear_bias}

In the context of interacting DE models with a constant coupling strength it is now a well established result \citep{Amendola_2000,Amendola_2004} that the long-range fifth-force acting between CDM particles induced by the interaction of the DE scalar field with the CDM fluid determines a different growth rate of linear density perturbations of CDM with respect to the uncoupled baryonic component. This can be clearly understood just by having a look at Eqs.~(\ref{gf_c}) and (\ref{gf_b}), where the same spatial distribution of density fluctuations in the dominant CDM component sources the CDM perturbation evolution with a strength $\Gamma _{c}$ times larger than for the baryons. In case of a constant coupling $\beta _{c}$ (\ie a constant factor $\Gamma _{c}$), this different growth rate is integrated over the whole expansion history of the Universe and induces a sizeable linear bias at all scales between the amplitude of density perturbations in the two components. This characteristic feature makes the linear bias arising from the coupling clearly distinguishable, at least in principle, from the hydrodynamical bias arising only at small scales as the Universe becomes progressively more structured.
The study of such effect within constant coupling models has been extended to the nonlinear regime by numerically following the collapse of a spherical overdensity \cite[\eg by ][]{Mainini:2006zj} or by means of N-body simulations (\citet{Maccio_etal_2004}, BA10), finding that nonlinearities enhance the bias between the two components. 
We want to extend here the analysis to the case of time dependent couplings for the linear and mildly nonlinear regimes, while the strongly nonlinear regime will be studied in Sec.~\ref{nonlinear_bias}.

The coupling-induced bias between CDM and baryon density fluctuations can be studied by comparing the ratio of the density power spectrum amplitudes of the two components in the different simulations, defined as:
\begin{equation}
R(k,z) \equiv \frac{P_{b}(k,z)}{P_{c}(k,z)} \,.
\end{equation}

In Fig.~\ref{fig:power_ratio} we plot the evolution of the ratio of $R$ to the $\Lambda $CDM case $R_{\Lambda \text{CDM}}$ for all our simulations as a function of scale, at several different redshifts. Also in this case, there is a clear difference between the exponential coupling models and the models where the coupling depends on the scale factor. For the former, as it can be seen in the upper three panels of Fig.~\ref{fig:power_ratio}, the presence of a non-negligible coupling already at high redshifts determines a moderate linear bias with a weak dependence on scale already at $z=3$ and down to $z=1$, with a clear hierarchy corresponding to the values of the couplings at high redshifts in the different models, while the latter class is practically indistinguishable from $\Lambda $CDM in this redshift range. 

At later times (lower three panels of Fig.~\ref{fig:power_ratio}) the situation changes very quickly, and the evolution of $R(k,z)$ becomes much more entangled than it has been shown to be for the case of constant coupling cosmologies. As the coupling in all the models grows towards its present value, the scale dependence of $R(k,z)$ starts to appear, showing how the bias progressively grows when moving from the linear to the mildly nonlinear regime of density fluctuations. Also in this case, as we showed in the previous section for the total power spectrum, the evolution turns out to be much faster and with a much stronger scale dependence for the EXP010a2 and EXP015a3 models as compared to the exponential coupling models. Nevertheless, the exponential coupling model EXP015e3 also shows a very quick reduction of the bias ratio between $z=1$ and $z=0$, due to the strong increase of its coupling.

It is very interesting to notice how all the models that share the same final value of the coupling $\beta _{0}$ seem to converge, at $z=0$, to very similar values of the bias at small scales, irregardless of the type of coupling evolution. The distinction between the two classes of models is nevertheless still present at the largest scales of the simulations, where the hierarchy of the exponential coupling models significantly changes with time, while the scale factor dependent models roughly retain the initial value of the large scale bias ratio $R\sim 1$.

\begin{figure*}
\includegraphics[scale=0.45]{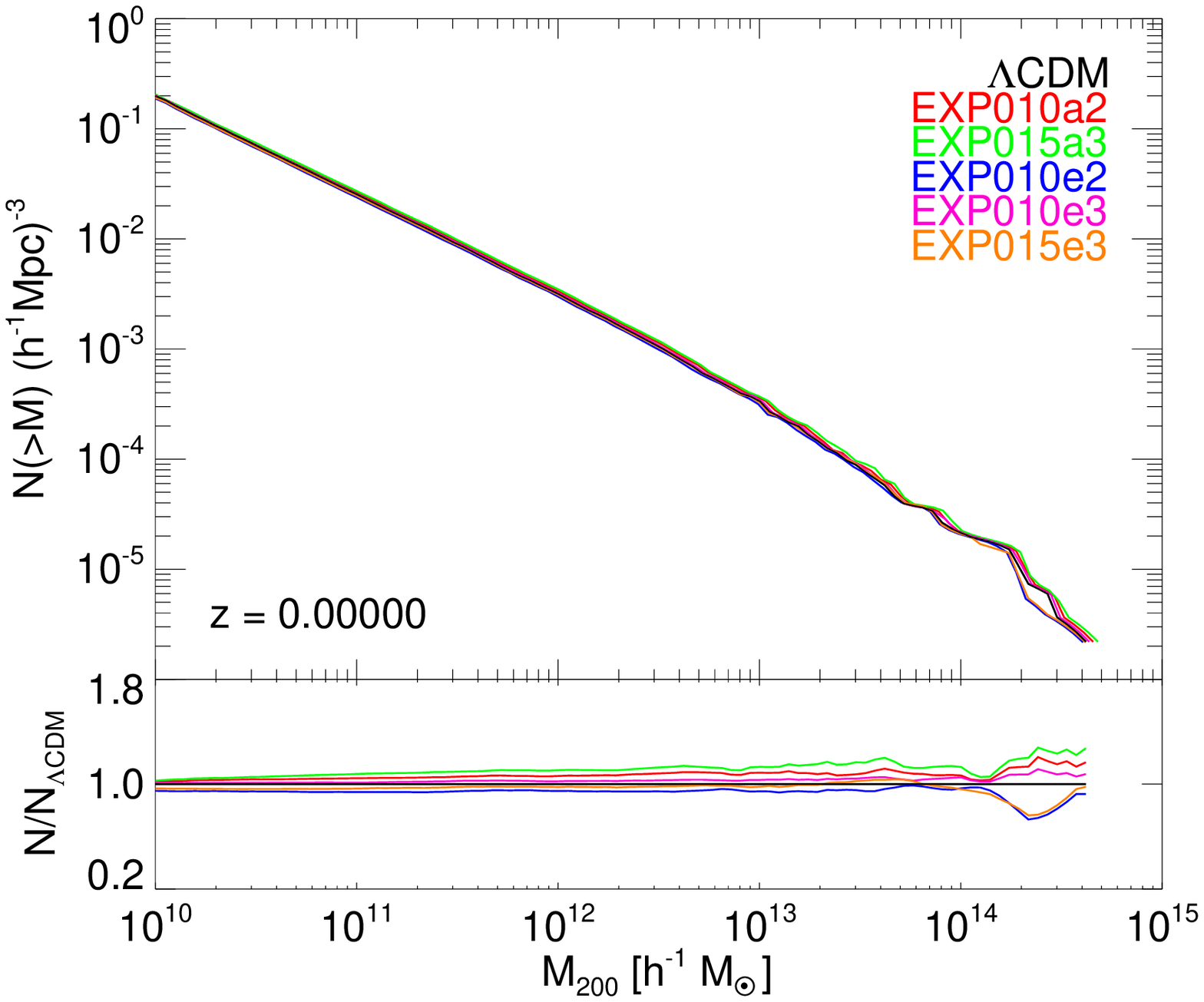}
\includegraphics[scale=0.45]{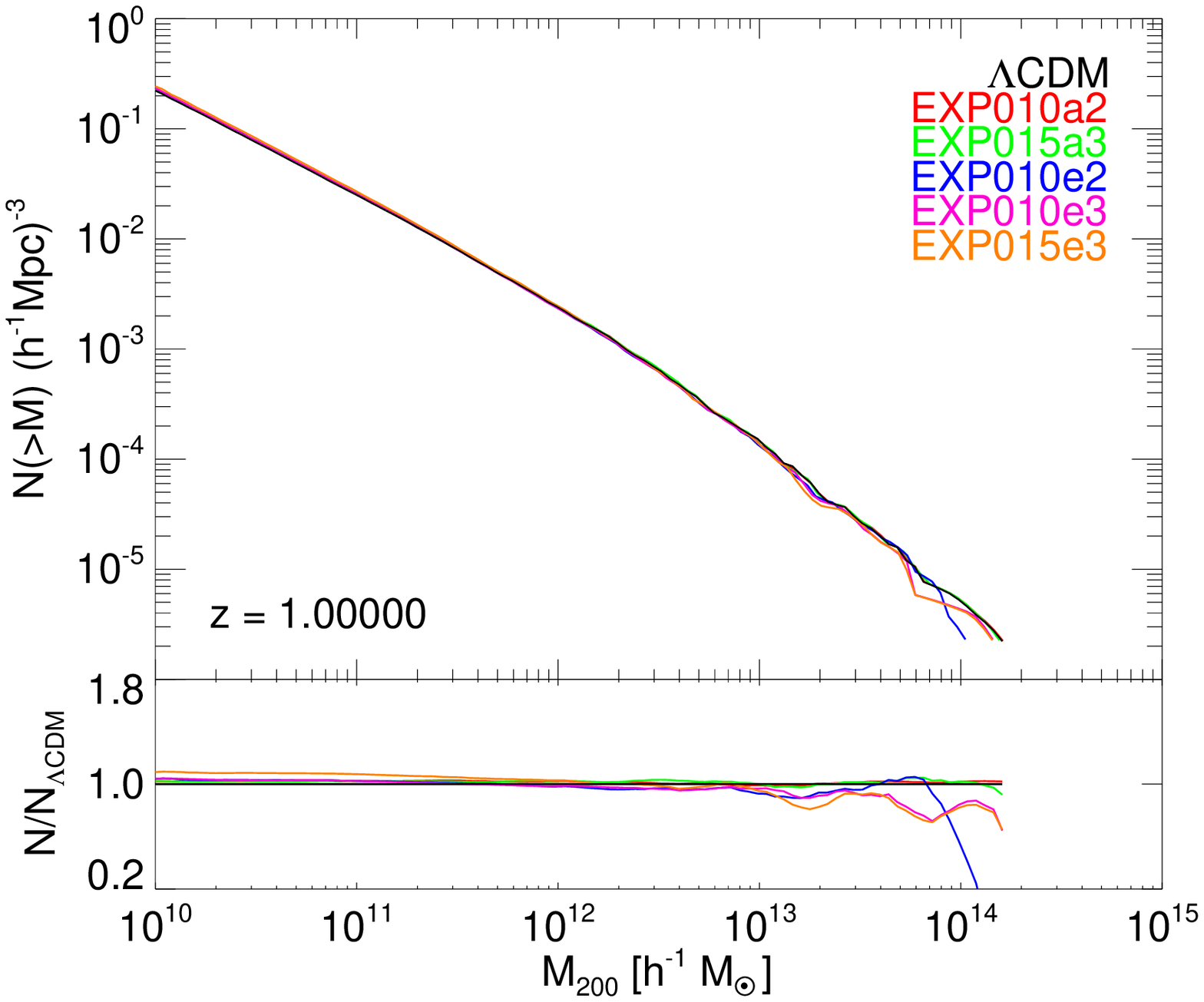}\\
\includegraphics[scale=0.45]{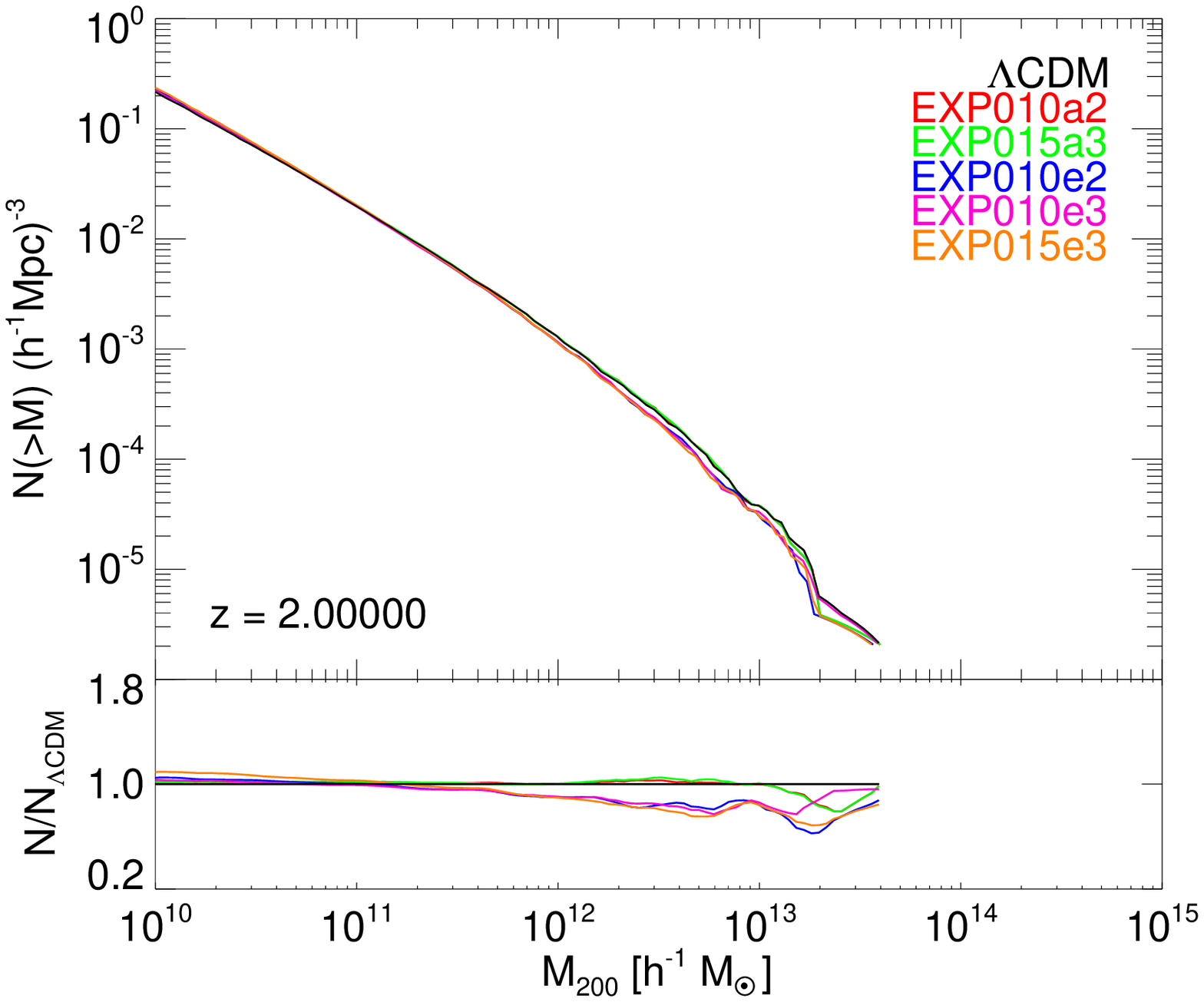}
\includegraphics[scale=0.45]{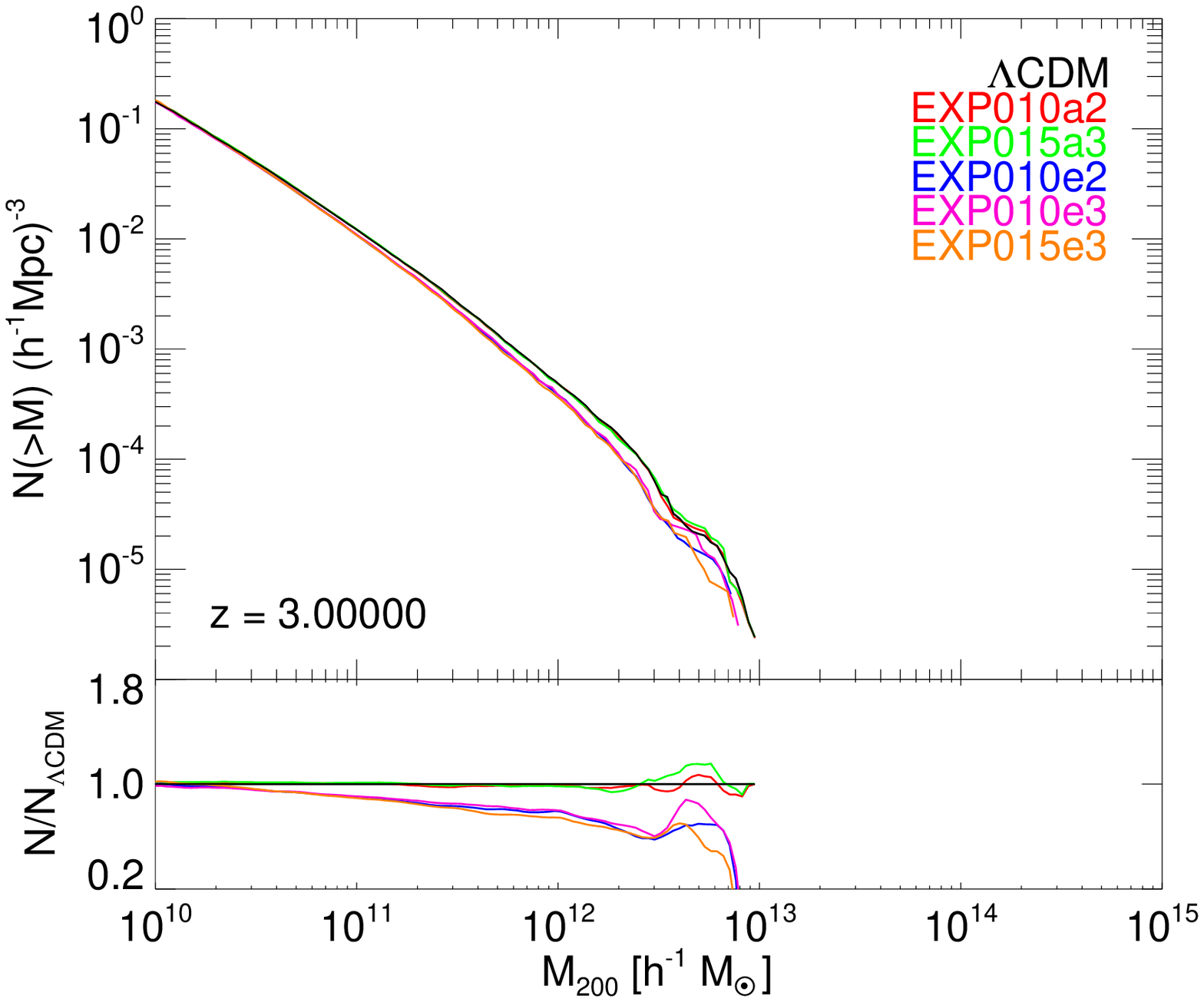}
\caption{The mass functions of all the six models studied with our set of N-body simulations, plotted at different redshifts. At the present time all the models show similar mass functions, with differences of the order of  $\sim 10 \%$ with respect to the $\Lambda $CDM case, slightly increasing at the high-mass end. At higher redshift the differences among the models become more pronounced, and the exponential coupling models EXP010e2, EXP010e3, and EXP015e3 show a slight excess of low mass objects and a considerable lack of high mass objects at redhifts in the range $z\sim 1-2$, while the remaining models do not show very significant differences with respect to $\Lambda $CDM. This picture is consistent, as explained in the text, with the fact that in the exponential coupling models structure formation starts later due to the lower initial amplitude of the power spectrum. The gap is then reduced at redshifts between 1 and 0 as a consequence of the strong increase of the growth factor in these models.}
\label{fig:massfunc}
\end{figure*}
\normalsize
\subsection{Halo mass function} 

For all of our high-resolution N-body simulations we have computed the halo mass function based on the groups identified by our FoF algorithm. The cumulative mass function for all our runs is plotted in Fig.~\ref{fig:massfunc}, where each panel refers to a different redshift. At $z=0$ all the mass functions have a similar shape and amplitude over the whole mass range covered by our catalog, with a discrepancy from model to model of the order of $\sim 10 \%$, which slightly increases at the high-mass end. In particular, it is worth noticing here how all the models except the EXP010e2 and the EXP015e3 show a slightly larger number of halos over the whole mass range with respect to $\Lambda $CDM at the present time.

At higher redshifts, instead, the models under investigation show very different mass function evolutions. On one side, the EXP010a2 and EXP015a3 models show very little differences from $\Lambda $CDM. On the other side, the exponential coupling models have always a significantly lower number of halos with respect to $\Lambda $CDM at intermediate and high masses, while a slight excess of low mass halos is clearly visible for these models at $z\sim 1-2$.
This behavior is due to the fact that in the exponential coupling models structure formation is starting later than in $\Lambda $CDM as a consequence of having normalized all the cosmologies to the same $\sigma _{8}$ at the present time. 
This shows how in these cosmologies halos of any given mass tend to form later than in $\Lambda $CDM, and how at $z\sim 1-2$ most of the small halos did not have the time yet to merge and form larger and more massive structures.
However, as it can be clearly seen by the evolution of the mass function with redshift, the gap in the number of large halos between the exponential coupling models and $\Lambda $CDM is progressively reduced as time goes by due to the higher growth rates of the coupled cosmologies, as the increase of the CDM mutual attraction speeds up the aggregation of small objects into larger structures.

\begin{figure*}
\includegraphics[scale=0.32]{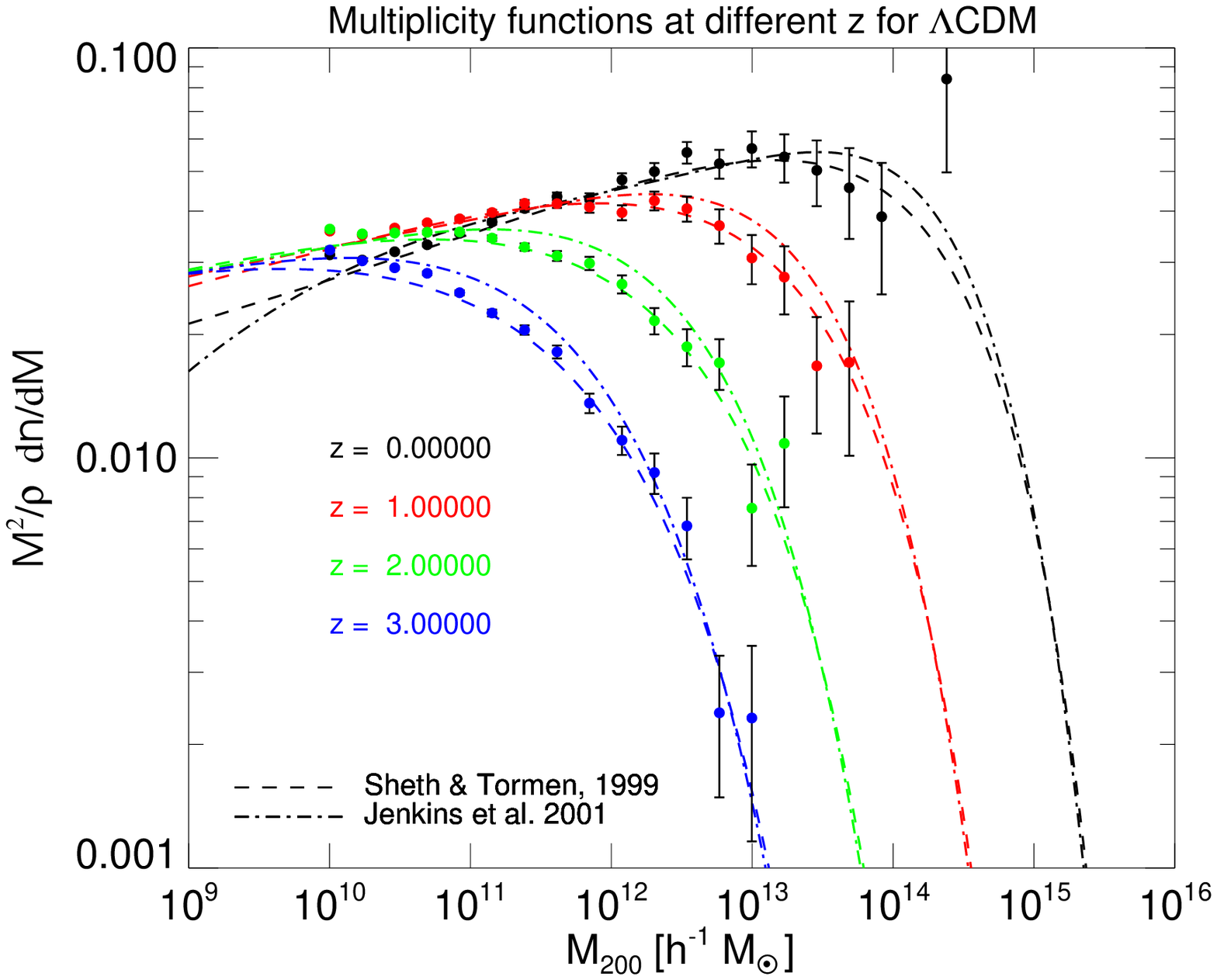}
\includegraphics[scale=0.32]{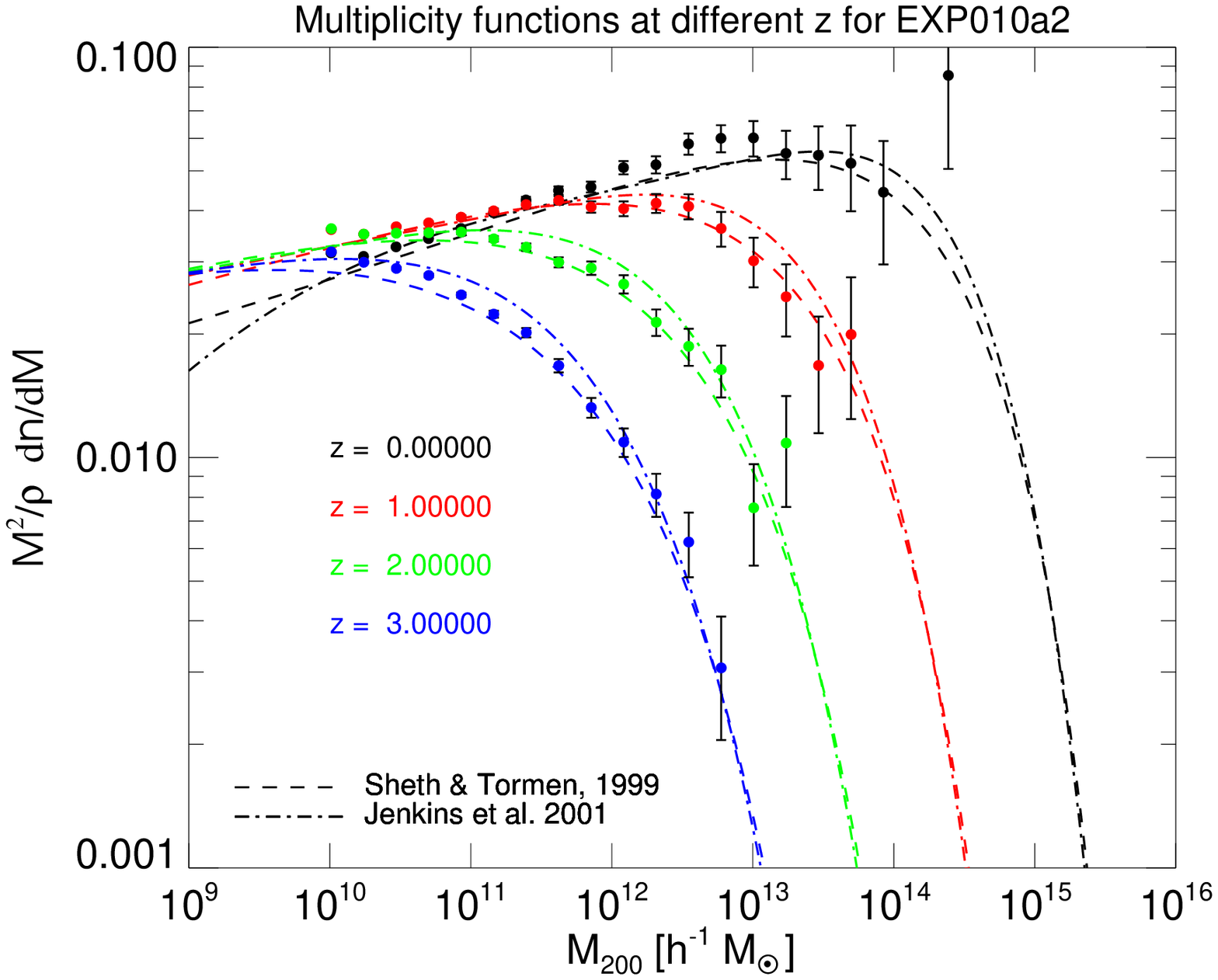}
\includegraphics[scale=0.32]{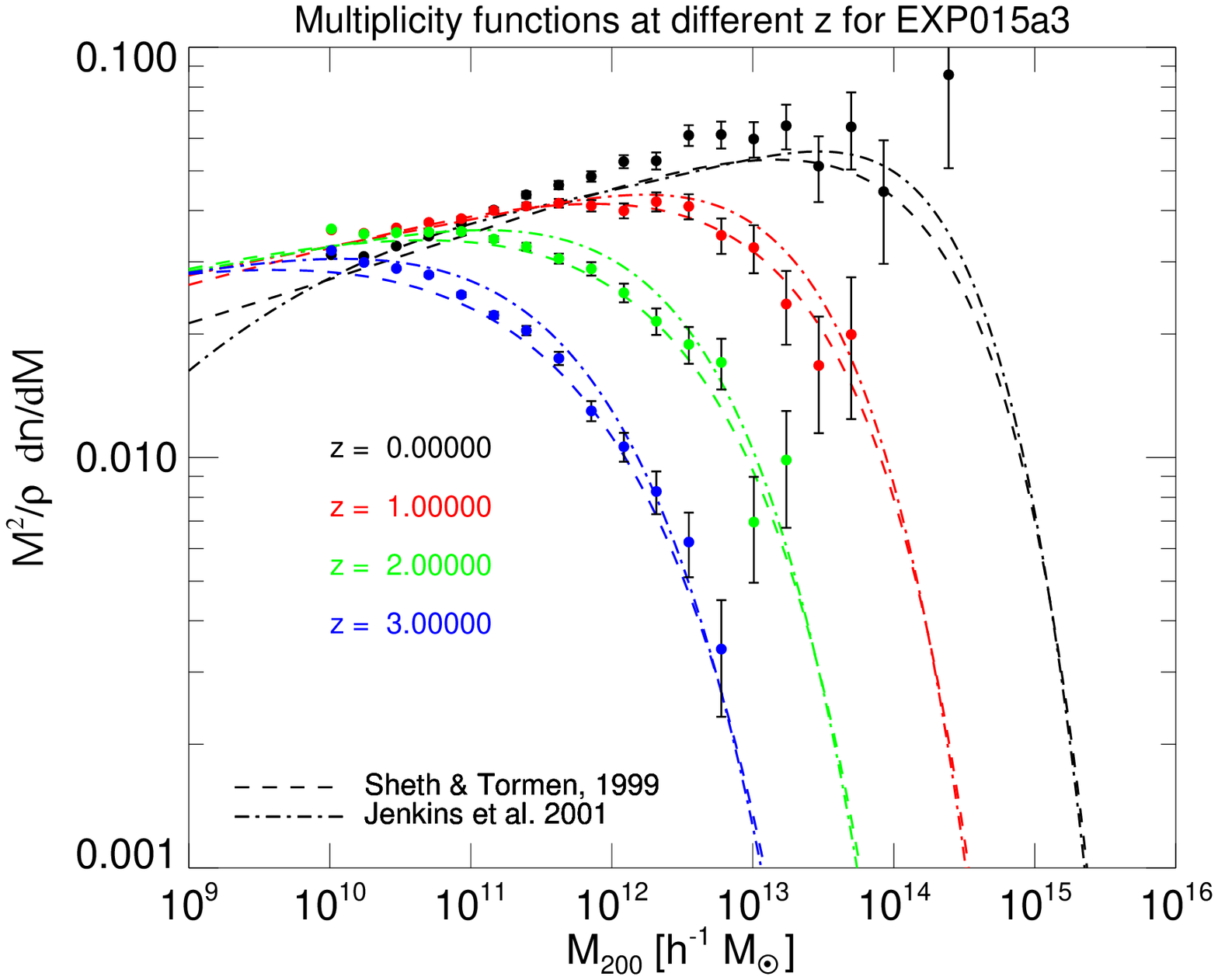}\\
\includegraphics[scale=0.32]{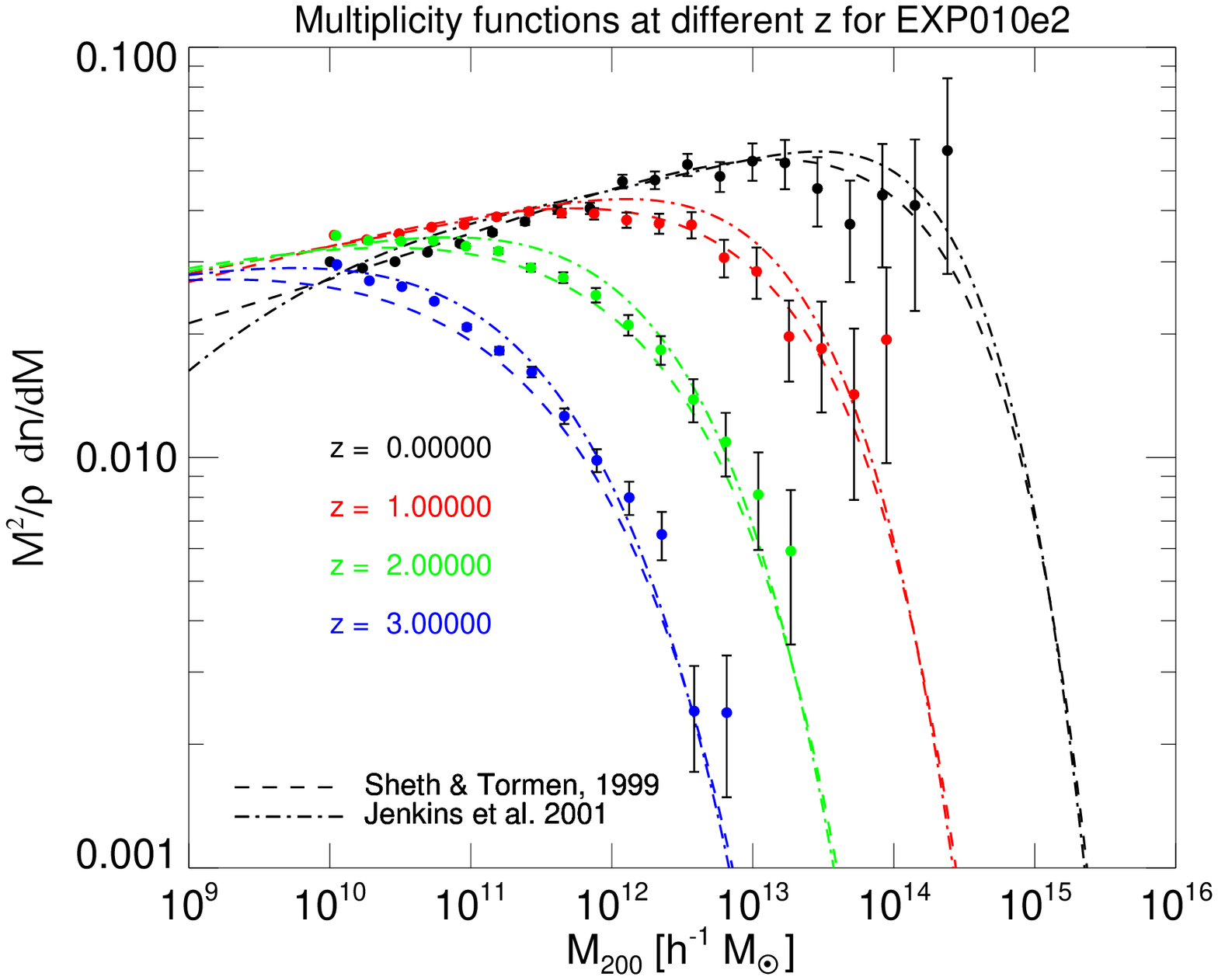}
\includegraphics[scale=0.32]{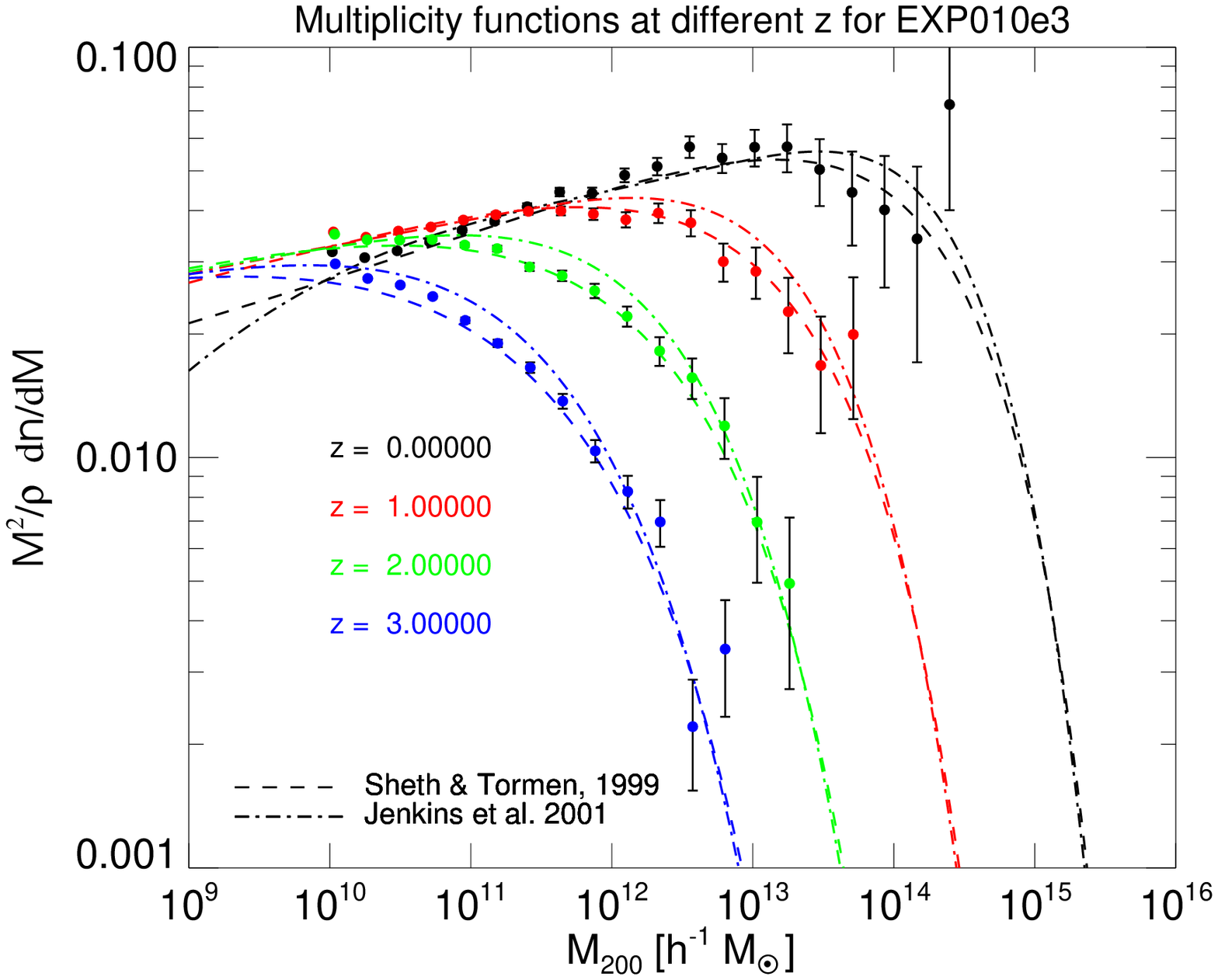}
\includegraphics[scale=0.32]{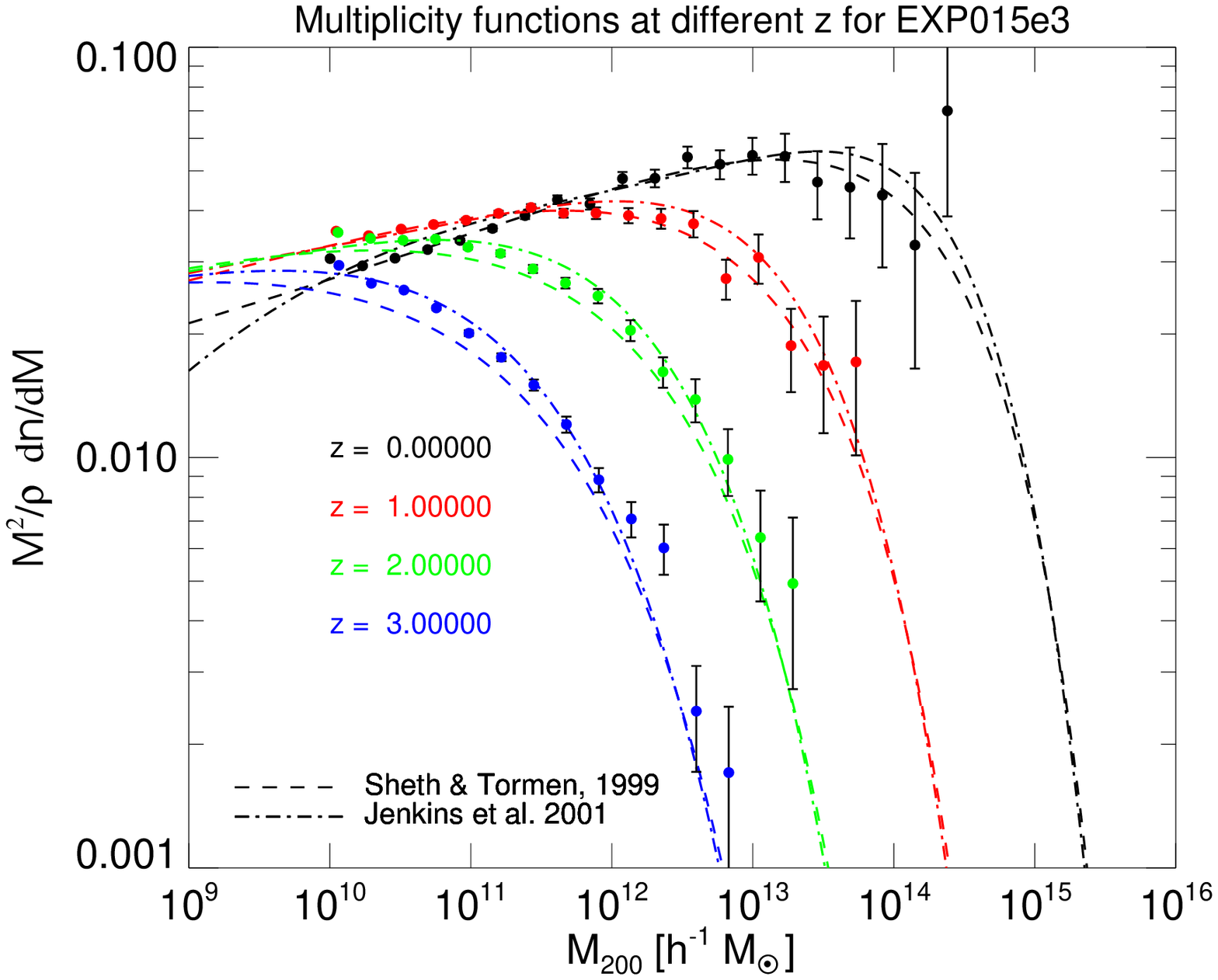}
\caption{The multiplicity functions for the six models investigated with N-body simulations. In each panel, the differently-coloured sets of data points, with the relative error bars, are the multiplicity functions evaluated in equally spaced logarithmic mass bins at four different redshifts. The dashed and dot-dashed lines represent the predictions for the multiplicity functions according to the \citet{Sheth_Tormen_1999} and \citet{Jenkins_etal_2000} fitting formulae, respectively. The plots show a fairly good agreement between the simulated multiplicity function and the theoretical predicitions, with the sole exception of the EXP015a3 model at $z=0$ ({\em right upper panel}), where both the fitting functions underestimate the simulated distribution.}
\label{fig:multiplicity}
\end{figure*}

We have also computed the multiplicity function (defined as $[M^{2}/\rho] \cdot $d$n(<M)/$d$M$) for all of our models, which is plotted in Fig.~\ref{fig:multiplicity} where we also plot for comparison the \citet{Sheth_Tormen_1999} and \citet{Jenkins_etal_2000} fitting formulae evaluated at different redshifts using the appropriate growth factor for each model, and with the standard value of the extrapolated linear density contrast at collapse $\delta _{c}=1.686$. We find that the usual mass function formalism reproduces fairly well the distribution of our simulated halos up tp $z=3$ in most of the investigated models, extending the validity of recent findings for early dark energy cosmologies \citep{Grossi_2008,Francis_etal_2008b,Pace_Waizmann_Bartelmann_2010} and for constant coupling interacting DE models (BA10), to the case of variable couplings. However, some sizeable discrepancy between the predicted multiplicity function and the outcomes of our simulations appears for the EXP015a3 model at $z=0$, where both the mass functions fitting formulae systematically underestimate the simulated distribution. This behavior might be related to the sudden and strong increase of halo masses at $z\sim 0.5-0$ in this model, which would explain why the offset appears only at $z=0$. However, a careful analysis of the spherical collapse formalism in the context of variable coupling models of interacting DE, extending to these scenarios the analysis recently carried out by \citet{Wintergerst_Pettorino_2010} for constant couplings, will be necessary in order to fully understand the reasons of this discrepancy. We defer such analysis to future work.

\normalsize

\begin{figure*}
\includegraphics[scale=0.45]{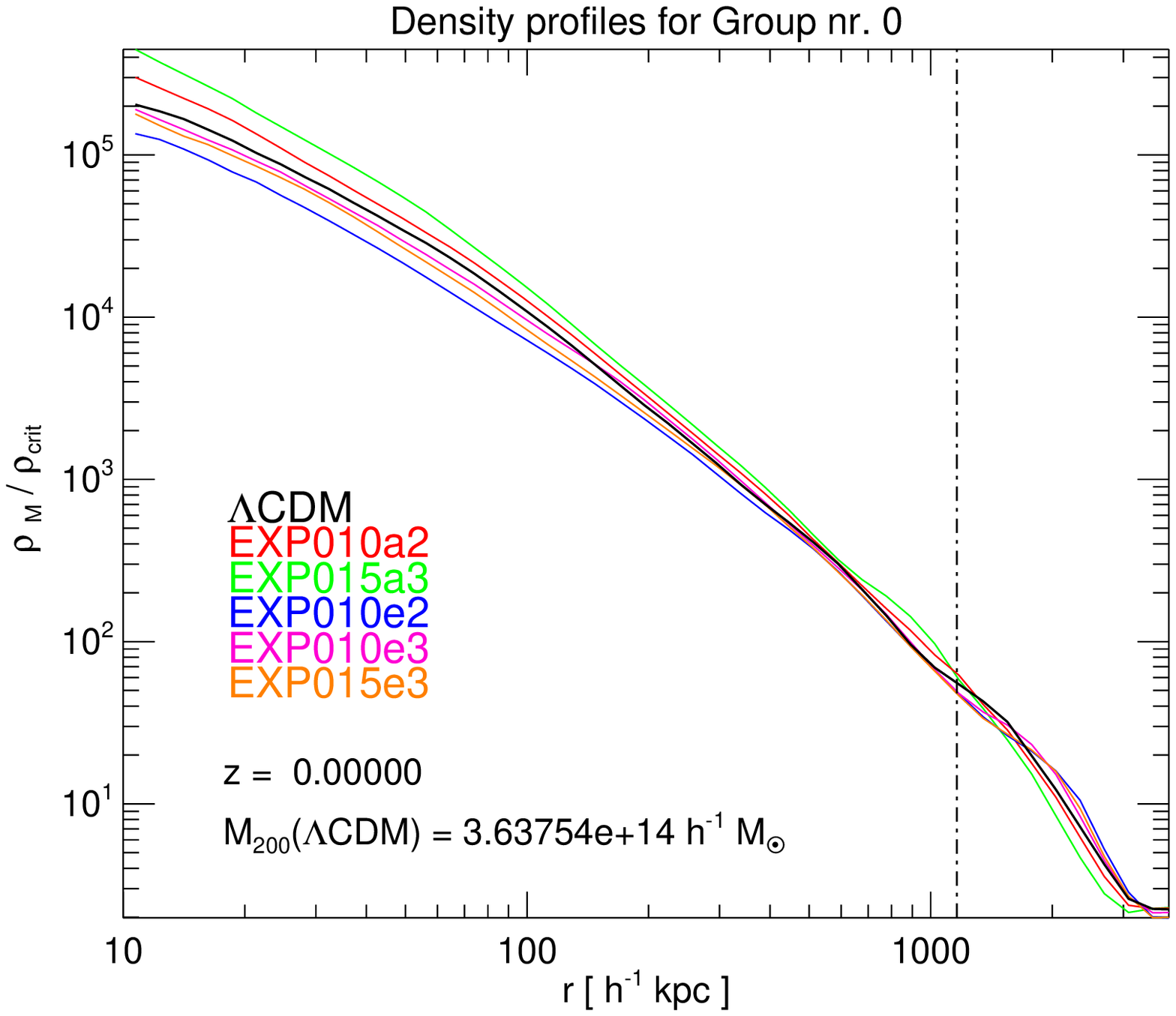}
\includegraphics[scale=0.45]{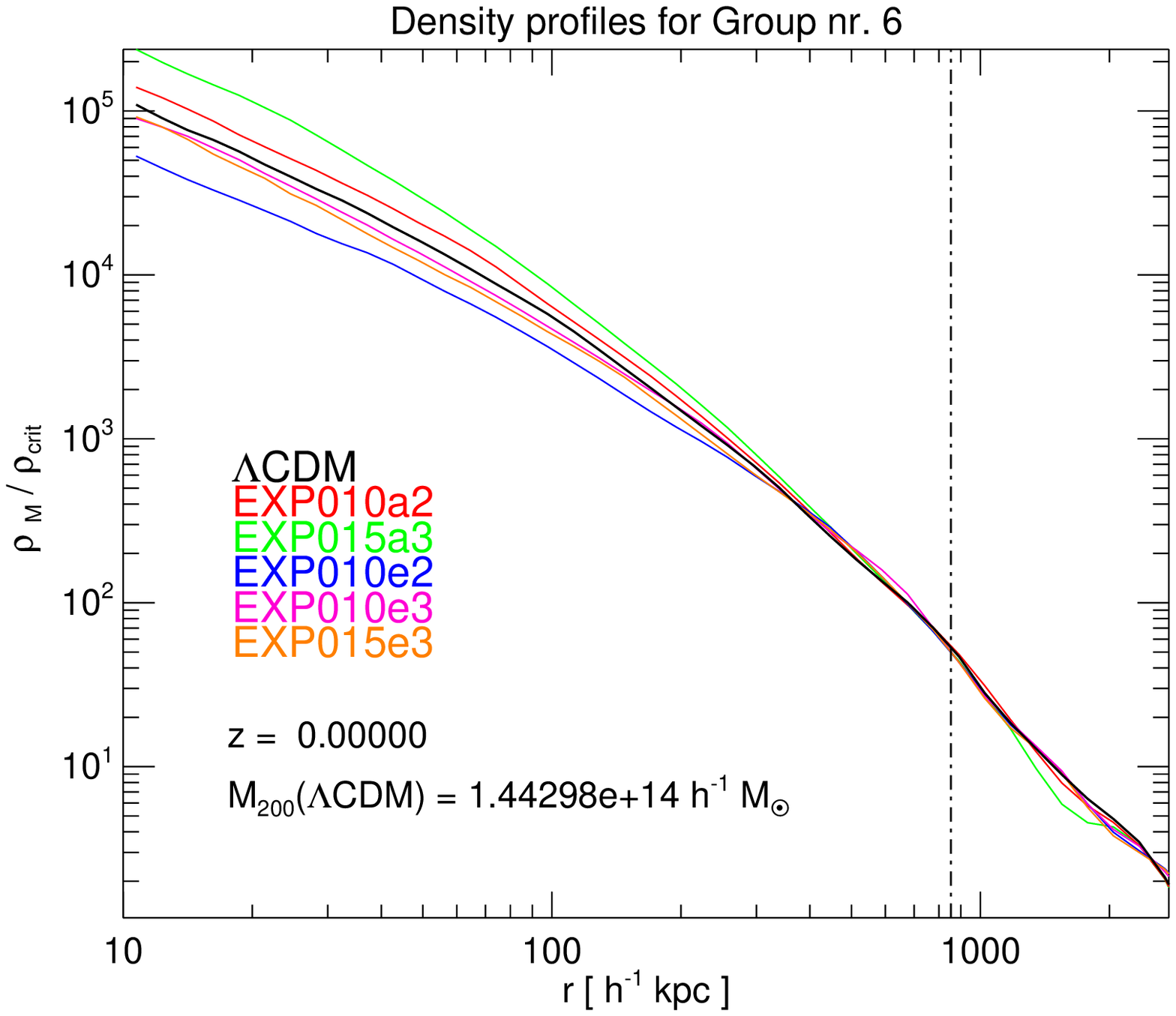}\\
\includegraphics[scale=0.45]{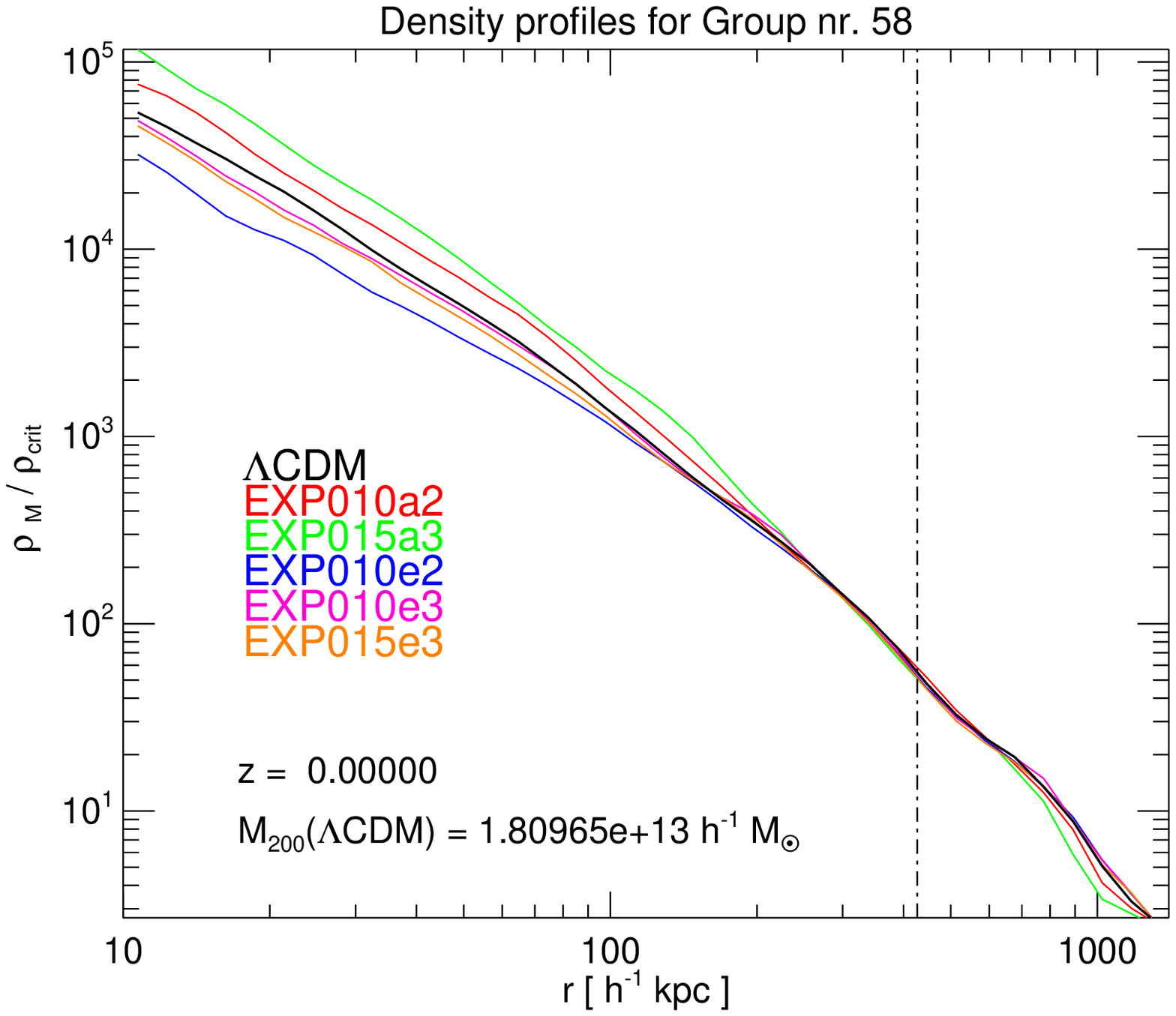}
\includegraphics[scale=0.45]{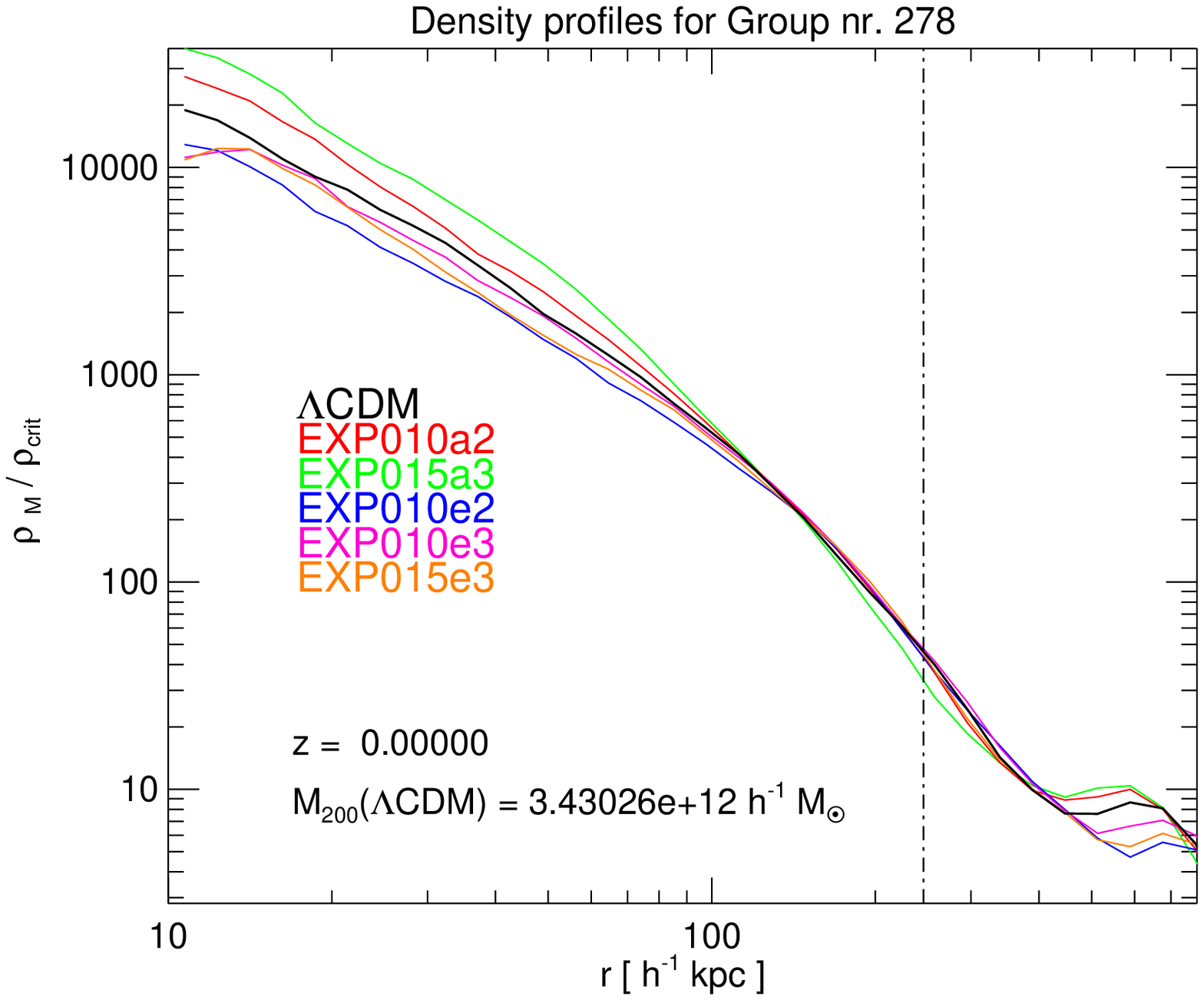}
\caption{Total matter (CDM + baryons) density profiles for four halos of different masses in the six models investigated with our set of N-body simulations. The left upper panel represents the most massive halo in our sample, while the right lower panel shows the lowest mass one. The vertical dot-dashed line indicates the location of $r_{200}$ for the $\Lambda $CDM halo. In all the four plots it is clear how the variable coupling models investigated in the present work, contrarily to what happens for constant couplings, do not always determine a decrease of the inner overdensity of halos. In particular, as discussed in full detail in the text, the models where a ``$\phi $MDE" or a ``Growing $\phi $MDE" is absent show a significant increase of the inner overdensity.}
\label{fig:profiles}
\end{figure*}
\normalsize

\subsection{Halo density profiles}
\label{profiles}

Cosmological simulations of structure formation have consistently shown that the density profiles of dynamically relaxed CDM halos have a universal shape that can be accurately fitted for any halo mass by the NFW fitting formula \citep{Navarro_Frenk_White_1995b,Navarro_Frenk_White_1995,NFW}:
\begin{equation}
\frac{\rho (r)}{\rho _{crit}} = \frac{\delta _{c}}{\left( \frac{r}{r_{s}}\right) \left( 1+ \frac{r}{r_{s}}\right) ^{2}} \,,
\end{equation}
where $\rho _{crit} = 3H_{0}^{2}M^{2}$ is the critical density of the Universe, $\delta _{c}$ is the characteristic halo density contrast, and $r_{s}$ is the halo scale radius.
However, several astrophysical observations of mass density profiles of dwarf {\em Low Surface Brightness} (LSB) galaxies \citep{Moore_1994,Flores_Primack_1994,Simon_etal_2003}, of Milky-Way type spiral galaxies \citep{Navarro_steinmetz_2000,Salucci_Burkert_2000,Salucci_2000,Binney_Evans_2001}, or even of large galaxy clusters \citep{Sand_etal_2002,Sand_etal_2004,Newman_etal_2009}, have shown at different levels that these objects have shallower density profiles than predicted by the theoretical universal NFW shape. This tension between simulations and observations is often referred to as the ``cusp-core" problem. Several attempts have been made in order to solve this discrepancy by invoking backreaction mechanisms of the baryonic component on the CDM density profiles \citep[see \eg][]{Duffy_etal_2010} or by different flavors of warm dark matter (WDM) \citep{Avila_Reese_etal_2001,Strigari_etal_2007, deNaray_etal_2009, deVega_etal_2010} and Self Interacting Dark Matter (SIMD) \citep{Spergel_Steinhardt_2000,Wandelt_etal_2000,Dave_etal_2001}. Here we want to investigate another independent possibility, namely the fact that a DE-CDM interaction could play a role in alleviating the ``cusp-core" tension.

One of the main results found in BA10 for the case of constant couplings consisted in the discovery that the nonlinear dynamics of interacting DE cosmologies determines a systematic reduction of the inner overdensity of CDM halos with respect to $\Lambda $CDM, with the effect growing for increasing coupling. This result, in stark contrast with previous works, was the first evidence of how interacting DE models could produce shallower density profiles and less concentrated halos, thereby providing a possible solution to the ``cusp-core" problem. Nevertheless, the present observational constraints on constant coupling models (BE08,LV09) put tight bounds on the maximum allowed amplitude of this effect, which turns out to be not strong enough to fully address the problem. In particular, it is worth reminding here that the largest coupling considered in BA10  ($\beta _{c}=0.25$), which was found to determine a reduction of the inner overdensity by $\sim 20\%$, is already observationally ruled out even by the weaker bounds derived by LV09 ($|\beta _{c}| < 0.17 $) for the case of  a large average neutrino mass. 

It is therefore very interesting to investigate whether variable coupling models -- where the coupling is relatively small for a large fraction of the expansion history of the Universe --  could produce stronger effects on the density profiles without running into conflict with the present observational bounds, as it is the case for all the models selected for N-body simulations in the present work, which are at least consistent (according to the selection criterion described in Sec.~\ref{obs}) with the constraints derived by LV09.

To this end, we have computed the spherically averaged total matter (baryons and CDM) density profile as a function of radius around the halo center (defined as the position of the particle with the minimum gravitational potential within the group) for all the halos in our group catalogs that can be safely identified as the same object arising in the different simulations.
In Fig.~\ref{fig:profiles} we plot the total matter density profiles for four halos of different mass in the six cosmological models under investigation. The left upper panel of Fig.~\ref{fig:profiles} shows the highest mass halo in our sample, while the right lower panel show the lowest mass one. 
Interestingly, we find that not all the models show a decrease of the inner halo overdensity as it was found for the case of constant couplings, where the $\Lambda $CDM model always showed a larger overdensity than any coupled DE cosmology. On the contrary, some of the models (the ones where the coupling depends on a power of the scale factor), are even found to have significantly larger values of the central overdensity with respect to $\Lambda $CDM, therefore showing an opposite trend to the one required in order to address the ``cusp-core" problem. On the other hand, the exponential coupling models show the expected systematic lowering of the inner density, although the effect is clearly not yet strong enough in these models to produce a cored profile. 
Once more, we have found a very different behavior of the two classes of variable couplings under investigation, and the interesting situation depicted in Fig.~\ref{fig:profiles} deserves to be extensively discussed.

The reason for these strikingly different evolutions of the nonlinear dynamics of CDM halos within the two different classes of models can be understood by having a look at Fig.~\ref{fig:friction_term}. In BA10 it was clearly shown how the reduction of the inner overdensity of halos in constant coupling models (and the consequent reduction of halo concentrations, that will be discussed in the next section) was primarily due to the effect of the friction term $2\beta _{c}x\vec{v}_{c}$ in Eqn.~\ref{accel_c}. In particular, it was shown that the energy gained by the collapsed systems due to the extra acceleration of CDM particles induced by the friction term could move the systems out of their virial equilibrium and determine a slight expansion of the halos. One of the consequences of such slight adiabatic expansion is the transfer of mass from the center of the halos towards the outskirts, thereby determining a reduction of the inner overdensity.
It is therefore clear, by looking at Fig.~\ref{fig:friction_term}, why this mechanism does not produce the same effects for the EXP010a2 and EXP015a3 models as it does for the constant coupling models studied in BA10: although having large values of the coupling at low redshift, these models feature a quite small friction term due to the absence of a ``$\phi $MDE". This determines a fast decay with redshift of the scalar field kinetic energy $x$ and a consequent strong suppression of the extra friction term during most of the cosmological evolution, as discussed in detail in Sec.~\ref{prt}. On the contrary, in the exponential coupling models EXP010e2, EXP010e3, and EXP015e3, the presence of what we called a ``Growing $\phi $MDE" is able to sustain a slower decay of the scalar field kinetic energy with redshift, and the friction term is therefore less suppressed and still capable of inducing the expansion of CDM halos and the consequent reduction of the inner overdensity.

We have therefore shown here that the presence of a ``$\phi $MDE" or of a ``Growing $\phi $MDE" is not only a desirable feature of an interacting DE model due to its capability of easing the ``fine tuning problem", but is also an essential ingredient in determining the type of impact that the interaction can have on the nonlinear  dynamics of coupled matter particles at small scales.\\
\begin{figure*}
\includegraphics[scale=0.37]{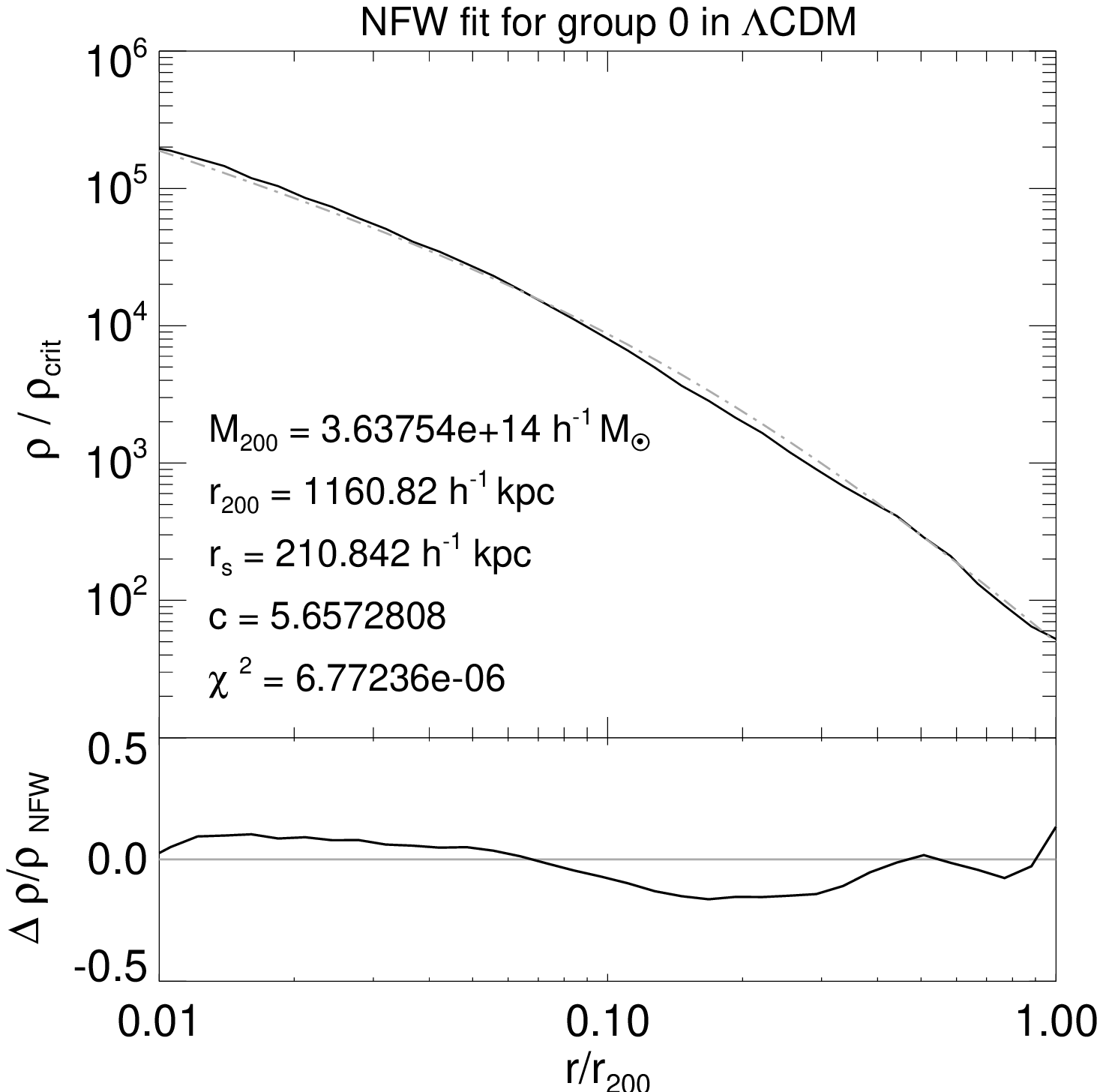}
\includegraphics[scale=0.37]{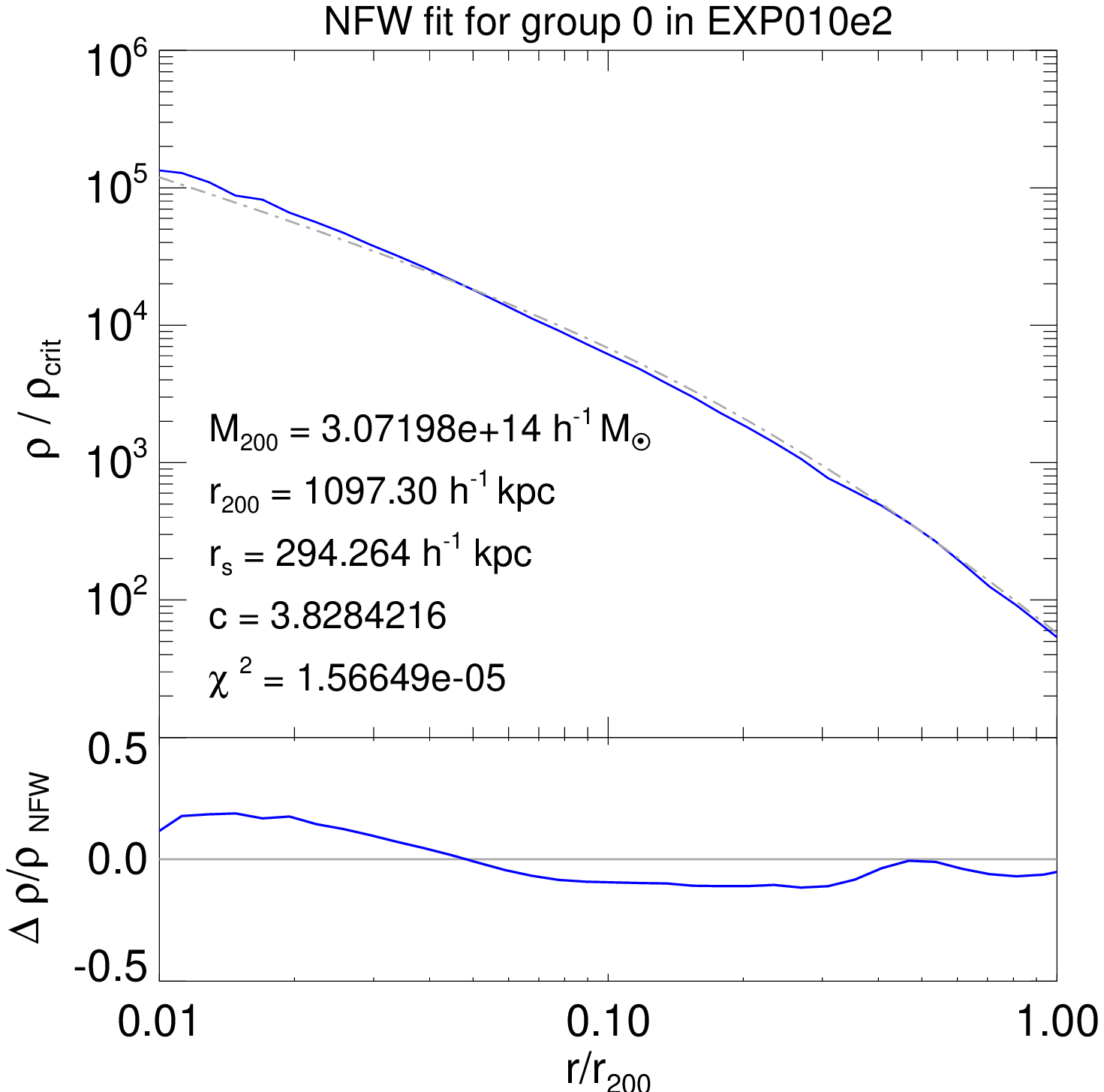}
\includegraphics[scale=0.37]{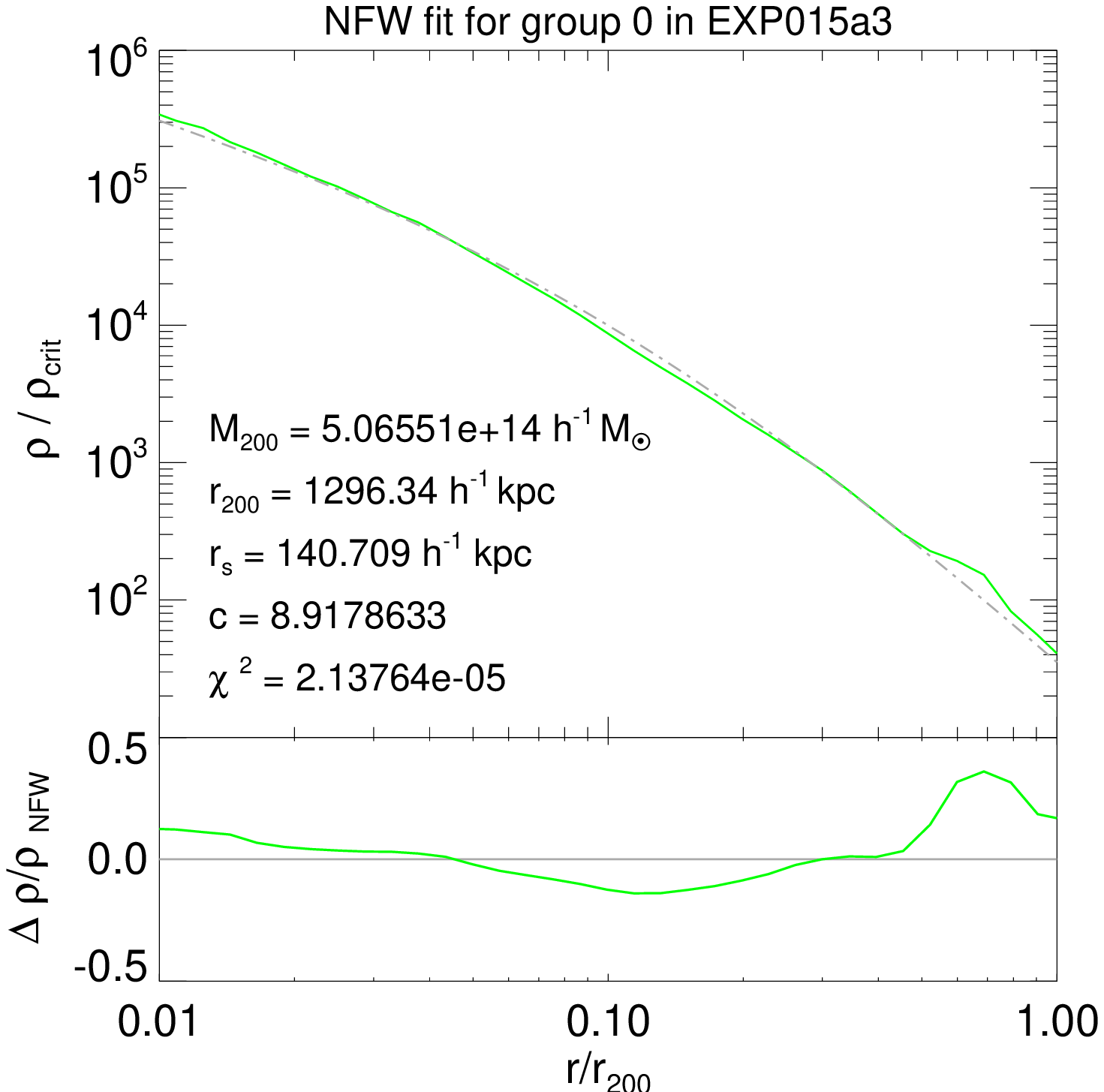}\\
\includegraphics[scale=0.37]{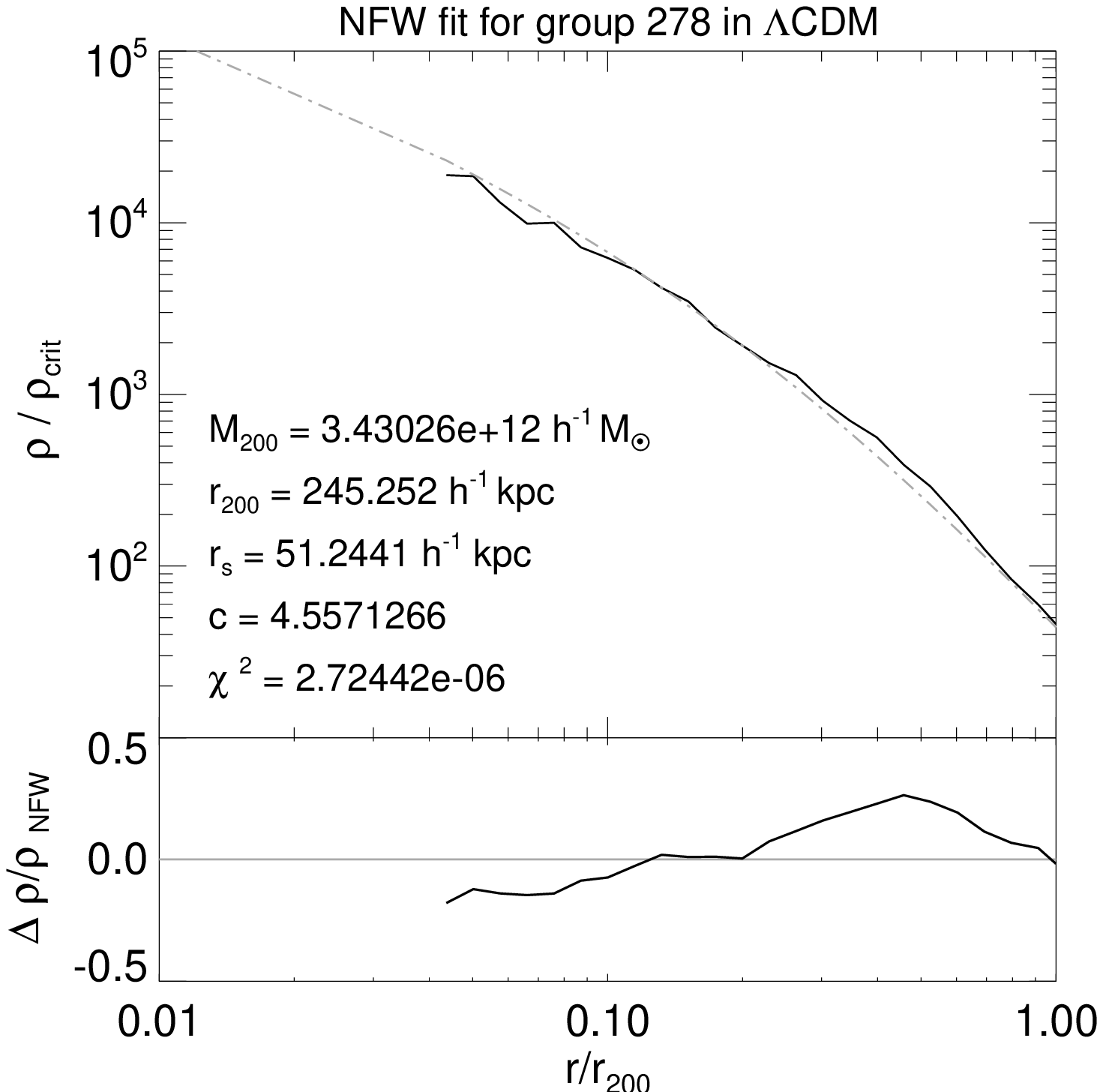}
\includegraphics[scale=0.37]{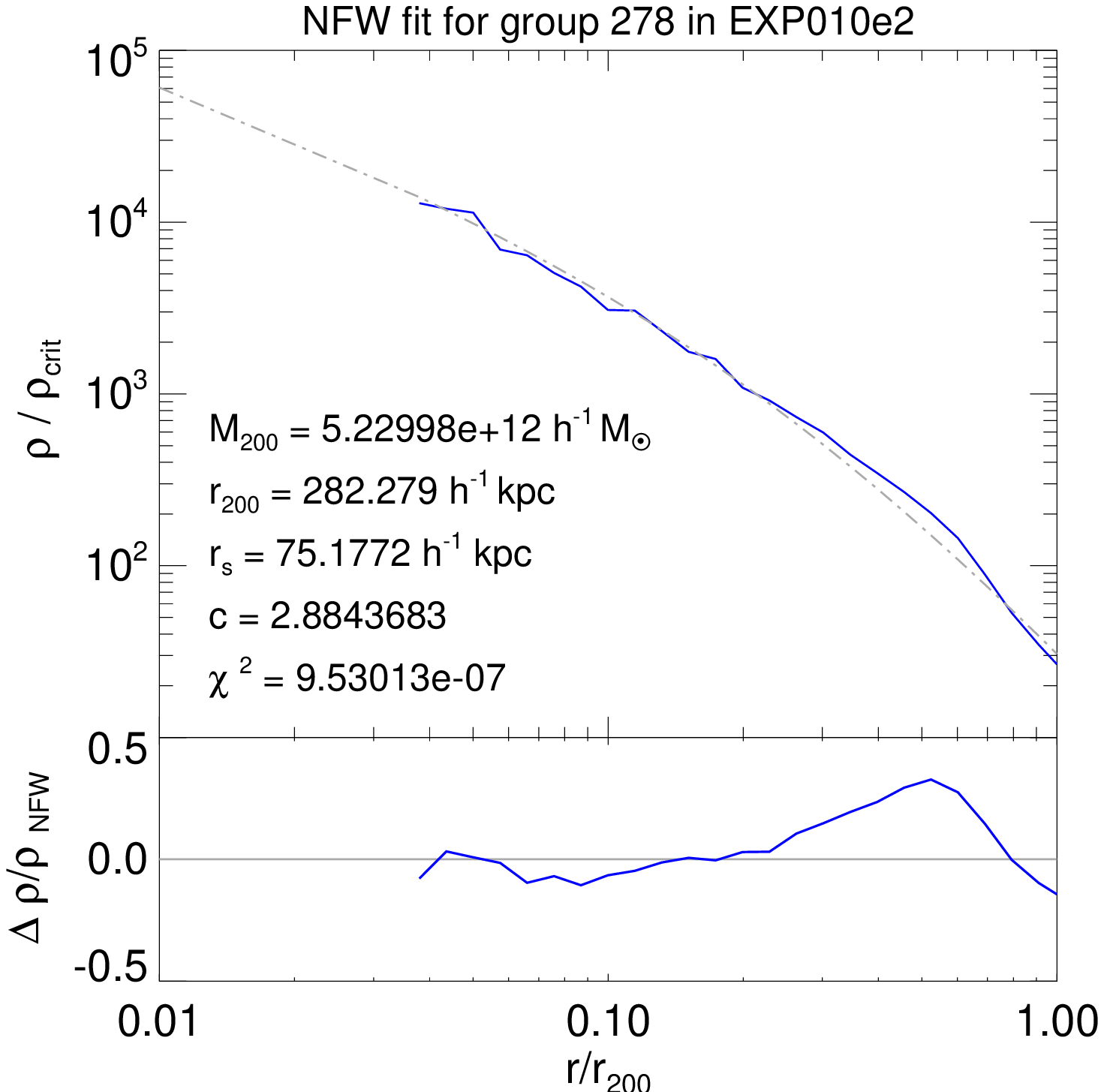}
\includegraphics[scale=0.37]{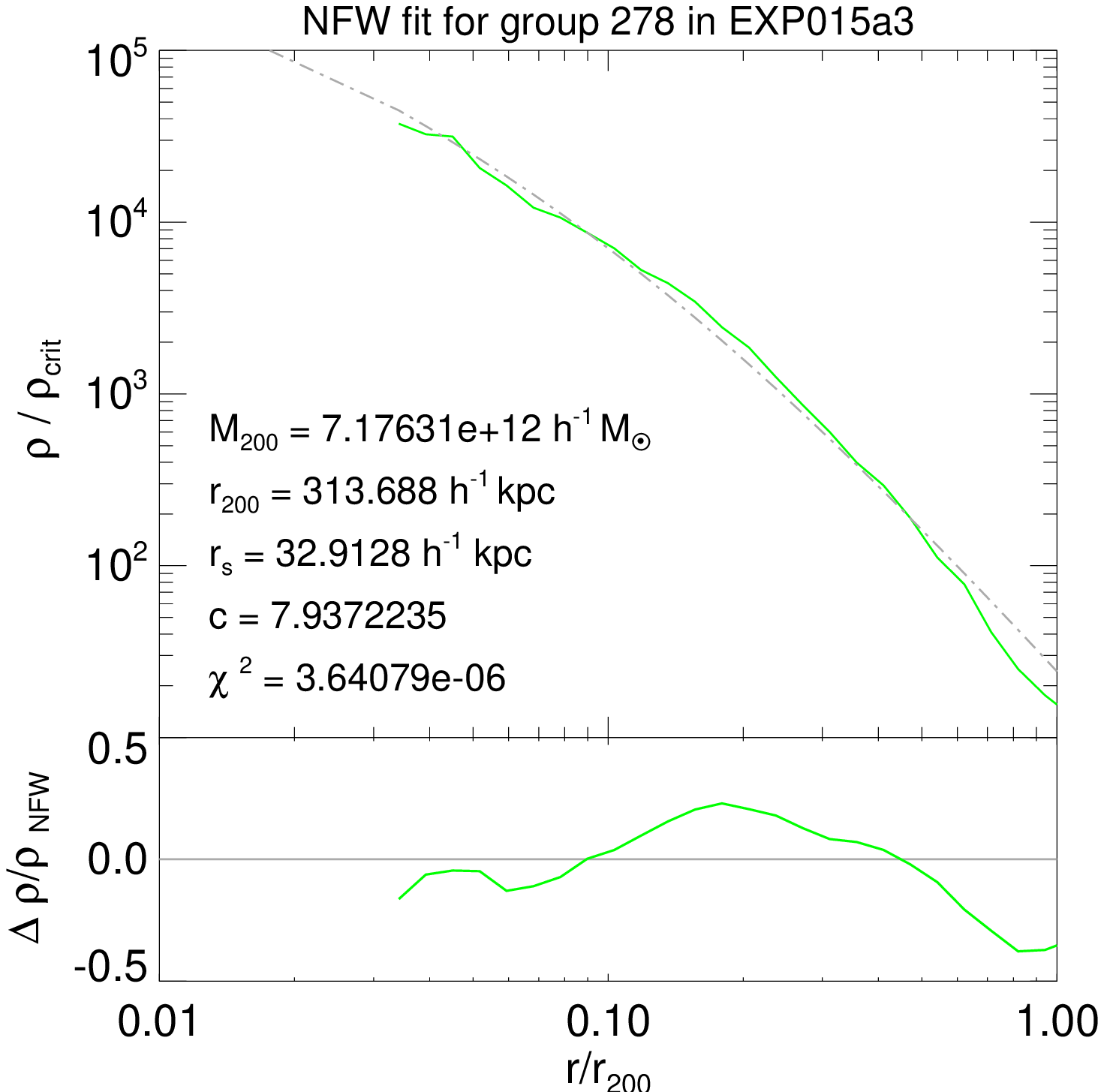}
\caption{NFW fit (grey, dot-dashed lines) to the density profiles (solid lines) of the most massive ({\em upper three panels}) and the least massive ({\em lower three panels}) halos in our sample in the $\Lambda $CDM ({\em left panels}), EXP010e2 ({\em middle panels}) and EXP015a3 ({\em right panels}) cosmological models. Each plot reports the halo mass $M_{200}$, the halo radius $r_{200}$, the scale radius $r_{s}$, the concentration $c$, and the $\chi ^{2}$ of the NFW fit. A clear trend of increase of the scale radius in the exponential coupling models, and of decrease in the scale factor models appears at both mass scales.}
\label{fig:NFW_profiles}
\end{figure*}
\normalsize
\begin{table*}
\begin{tabular}{ccccccccc}
Model & \begin{minipage}{45pt}
Group 0 \\ $r_{s}$ (h$^{-1}$ kpc)
\end{minipage} &  
\begin{minipage}{45pt}
Group 0 \\ $\frac{r_{s}}{r_{s}(\Lambda \mathrm{CDM})}$
\end{minipage} &
\begin{minipage}{45pt}
Group 6 \\ $r_{s}$ (h$^{-1}$ kpc)
\end{minipage} &
\begin{minipage}{45pt}
Group 6 \\ $\frac{r_{s}}{r_{s}(\Lambda \mathrm{CDM})}$
\end{minipage} &
\begin{minipage}{45pt}
Group 58 \\ $r_{s}$ (h$^{-1}$ kpc)
\end{minipage} &
\begin{minipage}{45pt}
Group 58 \\ $\frac{r_{s}}{r_{s}(\Lambda \mathrm{CDM})}$
\end{minipage} &
\begin{minipage}{45pt}
Group 278 \\ $r_{s}$ (h$^{-1}$ kpc)
\end{minipage} &
\begin{minipage}{45pt}
Group 278 \\ $\frac{r_{s}}{r_{s}(\Lambda \mathrm{CDM})}$
\end{minipage} \\
\hline
$\Lambda $CDM & 210.842 & 1.0 & 184.811 & 1.0 & 112.022 & 1.0 & 51.244 & 1.0\\
EXP010a2 & 181.134  & 0.859 & 151.660 & 0.821 & 85.959 & 0.767 & 44.900 & 0.876\\
EXP015a3 & 140.709  & 0.667 & 99.528 & 0.539 & 61.153 & 0.546 & 32.913 & 0.642\\
EXP010e2 & 294.264  & 1.396 & 295.849 & 1.601 & 174.661 & 1.559 & 75.177 & 1.467\\
EXP010e3 & 237.655  & 1.127 & 213.840 & 1.157 & 127.927 & 1.142 & 59.603 & 1.163\\
EXP015e3 & 246.118  & 1.167 & 223.194 & 1.211 & 130.655 & 1.166 & 62.153 & 1.213\\
\hline
\end{tabular}
\caption{Evolution of the scale radius $r_{s}$ for the four halos shown in
  Fig.~\ref{fig:profiles} with respect to the corresponding $\Lambda $CDM
  value. Contrarily to what happens for constant coupling models the scale radius can either increase or decrease with respect to $\Lambda $CDM in variable coupling scenarios according to the type of coupling evolution: models that do not present a ``Growing $\phi $MDE" phase feature a contraction of collapsed halos and a consequent decrease of their scale radius up to $\sim 45\%$, while the exponential coupling models show an increase of the halo scale radius up to $\sim 60\%$, a much more significant effect than for constant coupling models.}
\label{Table_scale_radii}
\end{table*}

Even though we have now explained why the absence of a ``Growing $\phi $MDE" suppresses the efficiency of the coupling in lowering the density profiles of collapsed structures, we still need to explain why some of our models even show a significant increase of the inner overdensity of halos over the whole mass range of our catalog with respect to the uncoupled $\Lambda $CDM case. The explanation of this new effect seems also clear if one considers the virial equilibrium of a collapsed halo. Along with the gain of energy due to the extra friction term, there are two other effects that can modify the virial state of a system within interacting DE cosmologies. On one side, the variation of the mass of CDM particles described by Eqn.~\ref{mass_variation} determines a progressive increase of the gravitational potential energy of the system. The effect of this mass loss would also be a slight expansion of the halos. However, as it was shown in BA10, this effect is expected to be rather small for models with an overall variation of the CDM particle mass comparable or even larger than for the specific models under investigation here. On the other side, the fast growth of the effective gravitational constant acting between CDM particles $\tilde{G}(\phi ) = \Gamma _{c}(\phi )G$ determines a corresponding fast decrease of the gravitational potential energy of the system that balances and exceeds the small increase due to the mass variation. 
This latter mechanism is not present at all in constant coupling models, but plays a crucial role here where the factor $\Gamma _{c}(\phi )$ can grow by $\sim $30 to 75 \% during cosmic evolution.
The total combined effect  is therefore a net decrease of the total energy, and a consequent contraction of the halos forming in the models where the friction term is suppressed and cannot provide to the systems the energy increase necessary to counteract this mechanism. Consistently with this picture, halos are found to be more overdense in the EXP010a2 and EXP015a3 models as compared to $\Lambda $CDM, and less overdense in the other exponential coupling models, where the friction term balances -- and in some cases exceeds -- the decrease of gravitational potential energy.

It is therefore clear that a significant growth in time of the effective gravitational constant in any cosmological modification of newtonian dynamics will result in an increase of the central density of CDM halos unless there are other mechanisms -- as \eg a large extra friction term in the case of the interaction of CDM with the DE -- able to balance the decrease of gravitational potential energy. 
Hence, it is natural to expect that also other cosmological models that introduce an effective  modification of the gravitational interaction due to a long or a short-range fifth-force \citep[as \eg the recently proposed {\em ReBEL} scenario:][]{Gubser2004,Farrar2004,Farrar2007,Hellwing_etal_2010} will produce larger values of the halo inner overdensity and more cuspy profiles with respect to $\Lambda $CDM if the strength of the additional fifth-force grows in time during the epoch of structure formation.
\begin{center}
\begin{figure}
\includegraphics[scale=0.45]{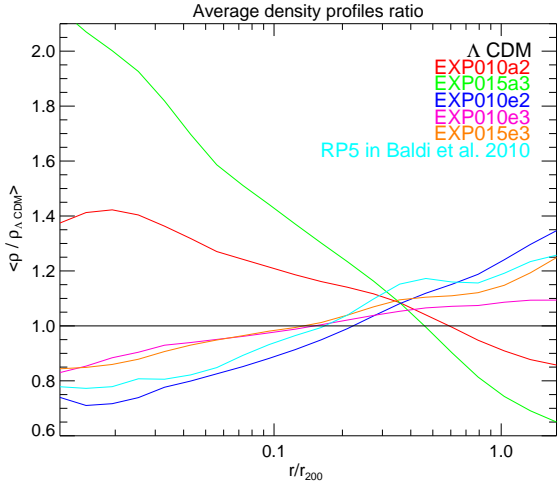}
\caption{The ratio of the spherically averaged matter density around the center of a halo to the density of the same halo in the $\Lambda $CDM cosmology, averaged over all the halos in our sample and plotted as a function of fractional radius with respect to $r_{200}$. The plot clearly shows how to a reduced inner density corresponds a higher density in the outer parts of the halos, due to the transfer of mass from the central regions to the outskirts. The plot also shows how in some cases (\eg for EXP010e2 model among the cosmologies considered here) a time dependent coupling can produce comparable or even stronger reductions of the inner overdensity of halos with respect to constant coupling models, without determining the same level of tension with present observational constraints on the background expansion of the Universe.}
\label{fig:profile_ratio}
\end{figure}
\normalsize
\end{center}

The last point that we still need to address in this section is whether or not the time variation of the coupling allows to have a more significant reduction of the inner overdensities of collapsed objects with respect to the constant coupling case, without running into conflict with observational constraints. We have already shown that none of the models investigated in this work actually presents cored density profiles. 
This leads us to the conclusion that a complete solution of the ``cusp-core" problem cannot be achieved by the simple classes of coupling evolution considered here.
Nevertheless, at least the EXP010e2 model is found to determine a significant lowering of the halo density profiles in the inner regions. In order to quantify this effect, and to compare it with the constant coupling case, we have computed for our sample the average radial density ratio with respect to $\Lambda $CDM, defined as: 
\begin{equation}
\left< \frac{\rho _{M}(r)}{\rho _{M,\Lambda CDM}(r)}\right> \,,
\end{equation}
where the average is taken over all the halos in the sample.
This quantity is plotted in Fig.~\ref{fig:profile_ratio} for all the models under investigation and for comparison also for the RP5 model studied in BA10 as a function of the fractional radius $r/r_{200}$, where $r_{200}$ is the radius enclosing a mean overdensity 200 times larger than the critical density $\rho _{crit}$.
By looking at Fig.~\ref{fig:profile_ratio} it is first of all interesting to notice how all the models that show an average lower density in the inner regions of CDM halos with respect to $\Lambda $CDM, show correspondingly larger values of the density in the outskirts. This behavior witnesses the fact that the lower inner overdensity is due to the transfer of mass form the core of the halos to the outer regions, as it was demonstrated in BA10 for the case of constant coupling. The opposite applies to the scale factor dependent models where the contraction of the halos brings mass from the outer regions to the center.

Finally, we notice how indeed the time dependence of the coupling can produce a comparable and in some cases a stronger reduction of the halo inner density with respect to constant coupling models, without determining the same impact on the overall background evolution of the Universe. In particular, the EXP010e2 model determines an average decrease of the inner overdensity with respect to $\Lambda $CDM roughly $1.5$ times larger than the RP5 constant coupling model. For the other exponential coupling models EXP010e3 and EXP015e3 the balance between the friction term and the decrease of gravitational potential energy is less favorable, and they are found to produce weaker effects with respect to RP5, although still having a much lower impact on the background expansion, which makes them still preferable over the constant coupling scenario.\\

Despite all these significant effects on the inner overdensity of CDM halos, we find that the density profiles can be still fitted extremely well with an NFW shape in all the models. This striking result confirms also for the case of time dependent couplings, as it was found for constant couplings, that the shape of the density profiles is not significantly modified by the additional physical processes related to the DE-CDM interaction, and that the change of inner overdensity is related to a change in the location of the scale radius $r_{s}$ in the different cosmologies. This can be clearly seen in  Fig.~\ref{fig:NFW_profiles} where we plot for the largest  and the smallest halos in our sample (upper and lower three panels, respectively) the density profile and the best-fit NFW function for $\Lambda $CDM and for the two most extreme models in each of the two different classes of coupling evolution, the EXP010e2 and EXP015a3 models. As Fig.~\ref{fig:NFW_profiles} shows, the NFW fit to the simulated halos is equally good (the $\chi ^{2}$ for the fit is indicated in each figure) at the two mass extremes of our sample in all the models. Consistently with our interpretation, there is a clear trend of the scale radius $r_{s}$ which always increases in the EXP010e2 and decreases in the EXP015a3 models. This effect is also clear from Table~\ref{Table_scale_radii}, where we list for the four halos considered in Fig.~\ref{fig:profiles} the scale radius in all our simulated cosmologies, and its ratio to the $\Lambda $CDM value. Although some exceptions to this general trend can be found in our halo sample, Table~\ref{Table_scale_radii} shows the tendency of the vast majority of the halos, with a decrease of $r_{s}$ up to $\sim 45\%$ in the scale factor dependent models, and an increase up to $\sim 60\%$ in the exponential coupling models, with respect to $\Lambda $CDM.
\begin{figure*}
\includegraphics[scale=0.45]{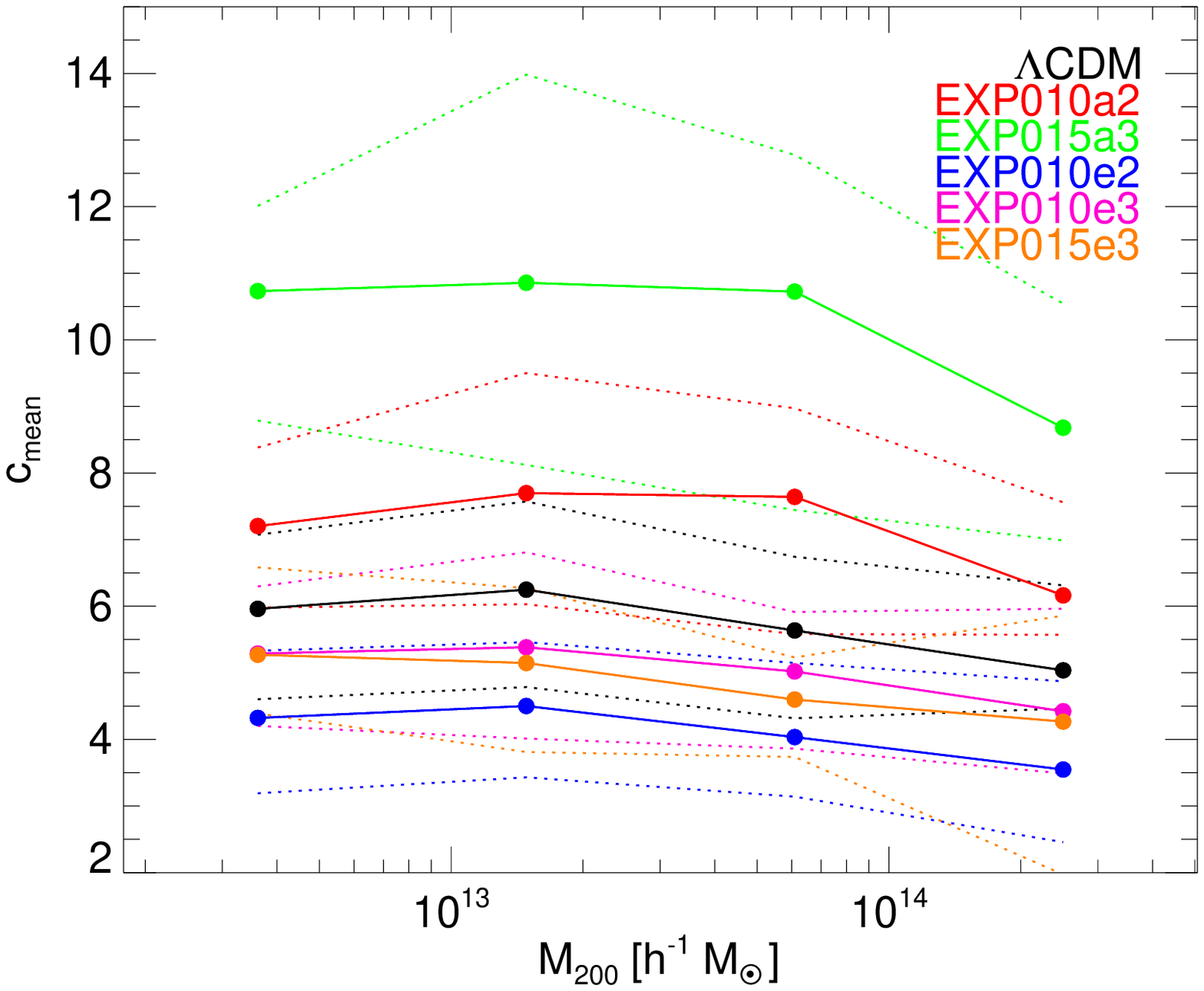}
\includegraphics[scale=0.45]{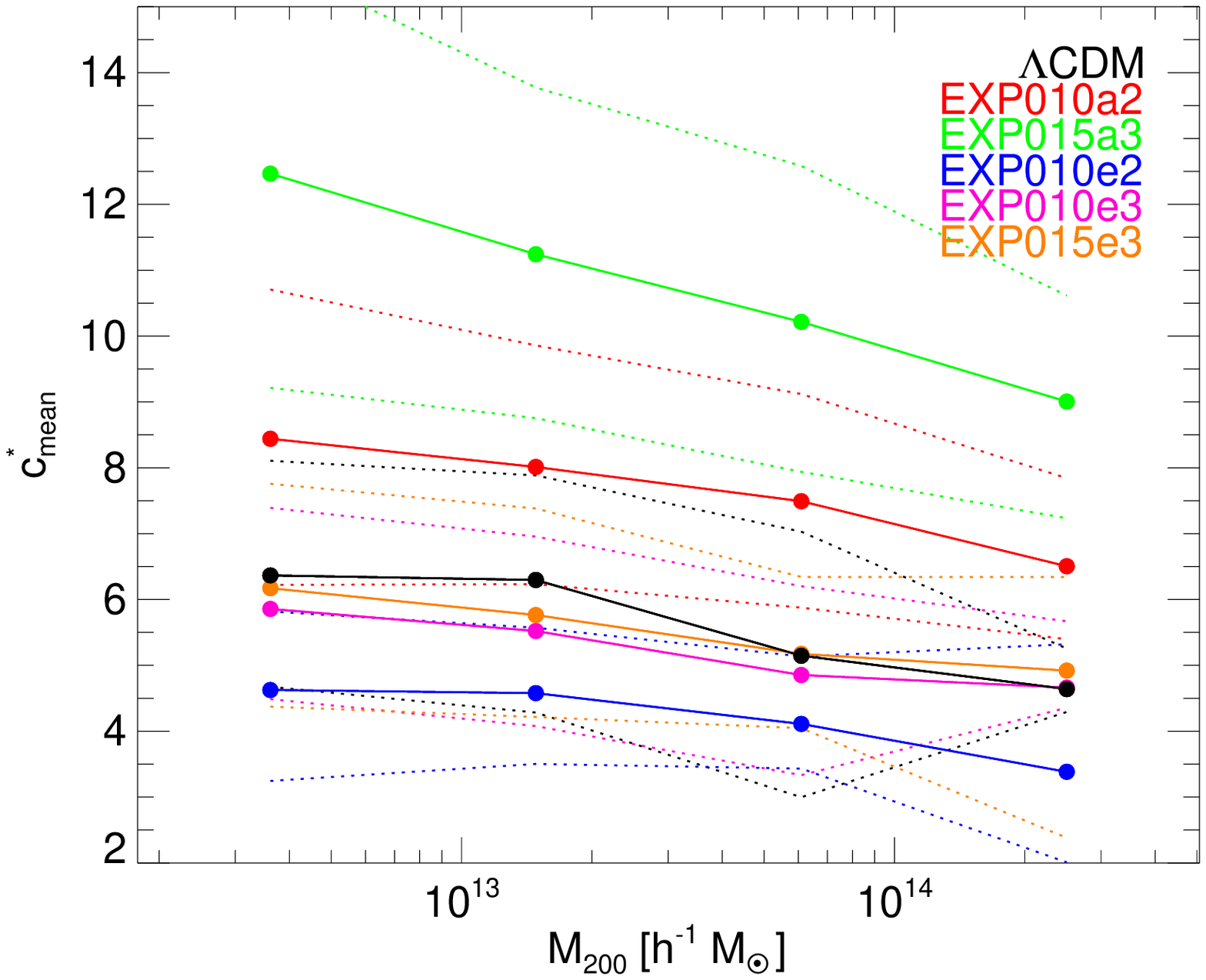}
\caption{Evolution of the mean halo concentrations as a function of mass for the 300 most massive halos in our group catalogs, in the different cosmological models under investigation. The concentrations have been computed by a direct fit of the density profiles with an NFW shape ({\em left panel}) or by using the independent method devised by \citet{Aquarius} and described in the text ({\em right panel}). The filled circles represent the mean halo concentrations in each of the four mass bins in which the mass range of our group catalogs has been divided. The correspondingly coloured dotted lines indicate the spread of 68\% of the halos in each mass bin. Both plots show the same trend of concentrations in the different cosmological models, with the exponential coupling models being less concentrated than $\Lambda $CDM while the remaining models show significantly larger values of concentrations over the whole mass range.} 
\label{fig:average_conc}
\end{figure*}
\normalsize

\subsection{Halo concentrations}

As anticipated in the previous section, we also compute for all the objects in our sample the halo concentrations with two independent methods. First, we compute the concentration of a halo as:
\begin{equation}
c=\frac{r_{200}}{r_{s}}
\end{equation}
based on a NFW fit of the density profiles. Then, we use the independent method devised by \citet{Aquarius} to compute halo concentrations according to the equation:
\begin{equation}
\frac{200}{3}\frac{c^{3}}{\ln (1+c) - c/(1+c)} = 7.213~\delta _{V}
\end{equation}
with $\delta _{V}$ defined as:
\begin{equation}
\delta _{V} = 2\left( \frac{V_{max}}{H_{0}r_{max}}\right) ^{2}
\end{equation}
where $V_{max}$ and $r_{max}$ are the maximum rotational velocity of the halo and the radius at which this velocity peak is located, respectively. 
Following the notation used in BA10 we will denote the concentrations computed with this method as $c^{*}$.

The results of these two independent methods to compute concentrations are shown in Fig.~\ref{fig:average_conc}, where concentrations are plotted as a function of the halo virial mass $M_{200}$ for all the high-resolution simulations carried out in the present work.
Despite the different methods used, the two panels of Fig.~\ref{fig:average_conc} show the same relative trend for the concentrations in the different cosmological models.
Consistently with what found for the halo density profiles in the previous section, the halos formed within one of the scale factor dependent coupling models are found to have significantly higher concentrations than $\Lambda $CDM halos at all masses. Once again, we see here that in case of time dependent couplings the $\Lambda $CDM model no longer represents an extreme for the range of models investigated, as it happens for constant couplings. On the contrary, deviations from the standard cosmological model are possible in both directions for different types of coupling evolution. The exponential coupling models, as expected, show lower halo concentrations with respect to $\Lambda $CDM, and the hierarchy of models follows the same order already shown for the inner overdensity of CDM halos in the previous section.

\subsection{Nonlinear bias and halo baryon fraction}
\label{nonlinear_bias}

Another source of tension between the predictions of the $\Lambda $CDM paradigm and a number of astrophysical observations concerns the baryonic budget of large galaxy clusters. The baryon fraction contained in massive clusters of galaxies is expected to be a fair sample of the background cosmological baryon fraction. However, several observational estimations of the baryonic content of X-ray clusters (as \eg \citet{Allen_etal_2004,Vikhlinin_etal_2006,LaRoque_etal_2006,Afshordi_etal_2007}, but see also \citet{Giodini_etal_2009} for a recent opposite claim) seem to indicate that these objects have a lower content of baryons as compared to the background baryon fraction estimated from cosmological observations as \eg CMB \citep{wmap7}. 

As we have already shown in Sec.~\ref{linear_bias}, one of the characteristic features of interacting DE cosmologies is the gravitational bias that develops between the amplitude of density fluctuations of baryons and CDM due to the different effective gravitational dynamic equations (see Eqs.~\ref{accel_c},\ref{accel_b}) that coupled and uncoupled particles obey. As a consequence of this different evolution, the baryonic fraction of any overdense region of the Universe is no longer expected to match the background cosmological value, even in the linear regime, as we showed in Sec~\ref{linear_bias}. 
\begin{figure}
\begin{center}
\includegraphics[scale=0.45]{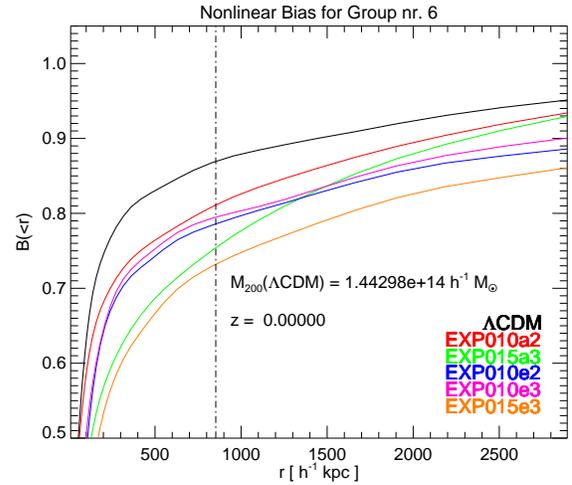}
\caption{Nonlinear bias between the overdensity of baryonic and CDM particles as a function of distance from the center of one of the halos shown in Fig.~\ref{fig:profiles}, in all the six different cosmologies considered in our set of N-body simulations. A similar trend is found for all the other halos in our sample. The enhancement of the bias in the innermost regions of the halo is evident for all the cosmologies, and becomes progressively stronger for larger values of the present coupling strength $\beta _{0}$. The vertical dot-dashed line represents the location of the virial radius $r_{200}$ of the halo in the $\Lambda $CDM cosmology.}
\label{fig:nonlinear_bias}
\end{center}
\end{figure}
\normalsize

We want to extend here the analysis of this effect to the nonlinear regime of structure formation, and give an estimate of how the halo baryon fraction at $z=0$ can be affected by time dependent couplings in the dark sector. To do so, we first compare the ratio of baryon to CDM overdensity as a function of radius around the center of a halo, defined as:
\begin{equation}
B(< r) \equiv \frac{\rho _{b}(< r) - \bar{\rho }_{b}}{\bar{\rho }_{b}} \cdot \frac{\bar{\rho }_{c}}{\rho _{c}(< r) -\bar{\rho }_{c}}\,,
\end{equation}
for all the halos in our catalog, as it was done in \citet{Maccio_etal_2004} and BA10 for the constant coupling case. 
In Fig.~\ref{fig:nonlinear_bias} we plot the ratio $B(< r)$ as a function of radius for one of the four halos already shown in Fig.~\ref{fig:profiles}. The strong enhancement of the bias when moving towards the center of the halo follows a very similar behavior to the case of constant coupling models studied in \citet{Maccio_etal_2004} and BA10, although the amplitude of the effect is generally larger, consistently with the higher values of the coupling at low redshift in the models considered here as compared to previous works. 

\begin{figure}
\begin{center}
\includegraphics[scale=0.45]{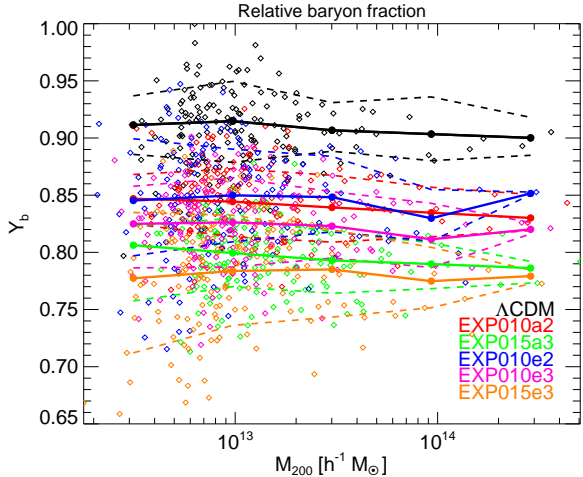}
\caption{Relative baryon fraction $Y_{b}$ within the virial radius $r_{200}$ for the 300 most massive halos in our group catalogs plotted as a function of virial mass $M_{200}$. The open diamonds represent the relative baryon fraction of individual halos for a uniform random sampling of our catalogs, while the filled circles show the mean  relative baryon fraction in each of the five mass bins in which the full available mass range has been subdivided. Dotted lines indicate the spread around the mean of 68\% of the halos for each cosmological model. A strong reduction of the relative baryon fraction is clearly visible for all the coupled DE models as compared to $\Lambda $CDM, with the total effect being approximately proportional to the different values of the coupling at the present time $\beta _{0}$.}
\label{fig:baryon_fraction}
\end{center}
\end{figure}
\normalsize

We then compute the relative baryon fraction $Y_{b}$ for all of our halos, defined as:
\begin{equation}
Y_{b} \equiv \frac{M_{b}(< r_{200})}{M_{tot}(< r_{200})} \cdot \frac{\Omega _{M}}{\Omega _{b}} \,,
\end{equation}
which is plotted in Fig.~\ref{fig:baryon_fraction} as a function of halo virial mass. While the $\Lambda $CDM baryon fraction is in full agreement with previous findings for the $\Lambda $CDM model \citep{Ettori_etal_2006, Gottloeber_Yepes_2007}, the coupled DE models show as expected a significant reduction of the relative baryon fraction over the full mass range covered by our catalogs of simulated halos, with average baryon fractions that decrease down to a value of $Y_{b} \sim 0.76-0.81$ for the most extreme cases represented by the models EXP015e3 and EXP015a3. These values are somewhat lower, as expected, than for the constant coupling scenarios investigated in previous works. It is also interesting to notice, as it can be seen from the spread of the 68\% of the halos (indicated in Fig.~\ref{fig:baryon_fraction} by the dotted curves above and below the mean) or from the location of single halos in the different models (represented by the open diamonds) that it is not particularly rare in these latter models to find halos with relative baryon fractions of the order of $\sim 0.7$, in particular at intermediate and small masses.

We stress again here that our simulations include hydrodynamical forces for the baryonic particles, but do not include other non-adiabatic processes like  \eg radiative cooling, star formation, and feedback. Therefore, we do not expect our predictions for the baryonic fraction of halos to be directly comparable with observations; nevertheless, it is clear that the strong reduction of the baryonic content in collapsed objects shown in Fig.~\ref{fig:baryon_fraction} would still be in place also in the presence of such non-adiabatic processes, which would therefore operate on a significantly smaller reservoir of baryonic mass. These results therefore indicate that a time dependent interaction between DE and CDM might be considered as one of the possible explanations for the observed low baryon fraction of galaxy clusters.

\section{Conclusions}
\label{concl}

In the context of interacting DE models we have studied a few general classes of time evolution of the interaction strength between DE and CDM, generalizing the widely studied case of constant couplings to the more natural scenario of a variable coupling. 

Following the idea -- already discussed in previous works -- that a large value of the coupling could leave distinctive features in the properties of observable structures and even possibly alleviate the tensions between the $\Lambda $CDM cosmological model and astrophysical observations at small scales, we have designed a few general forms of coupling functions $\beta _{c}(\phi )$ that grow in time, thereby having a significantly weaker impact on the overall background expansion of the Universe as compared to constant coupling models with the same interaction strength at $z=0$. 

In particular, we have investigated three classes of time evolution of the coupling, where the interaction strength is proportional either to a power of the cosmological scale factor $a(t)$, or to the fractional DE density $\Omega _{\phi }$, or to an exponential function of the DE scalar field $\phi $. 
The first two classes are purely phenomenological parametrizations of the time evolution of the coupling, while the latter one represents a more physical situation where the interaction depends on the dynamical evolution of the scalar field.\\

We have performed a complete numerical analysis of the background evolution for these three different types of coupling functions by solving the full system of coupled dynamic equations in the presence of a variable coupling, generalizing previous works. 
Even in the absence of analytic solutions for variable coupling models, our numerical integrations allow to identify the main background features of these cosmologies. 

More specifically, we have shown that the first two phenomenological parametrizations of the coupling evolution mentioned above, due to the very fast decrease of the coupling with increasing redshift, do not present the so called ``$\phi $MDE" scaling solution typical of constant coupling models, and for what concerns their background evolution are practically indistinguishable from $\Lambda $CDM, thereby suffering of the same level of fine tuning of the cosmological constant. On the contrary, the more physical form of a coupling that depends on the evolution of the scalar field shows a background evolution with an intermediate behavior between $\Lambda $CDM and a standard ``$\phi $MDE" phase which can be still well reproduced by the usual analytic solution for the fractional DE density $\Omega _{\phi }$ during the ``$\phi $MDE" phase, once generalized to the case of growing couplings. We have therefore called this intermediate type of background evolution a ``Growing $\phi $MDE" phase.\\

We have then studied the evolution of linear perturbations within variable coupling models, pointing out the main differences arising in the perturbations equations due to the time dependence of the coupling. In particular, we have shown how in general a growing coupling function could induce instabilities in the growth of scalar perturbations at large scales, due to the presence of a negative effective mass term in the linear scalar field equation, and we have discussed under which conditions these instabilities can be avoided. We have then numerically computed the growth factor of matter density perturbations at subhorizon scales for a few selected models. Among these, the computed growth factors again clearly shows two distinct classes: on one side the phenomenological couplings present no significant differences in the growth of density perturbations with respect to $\Lambda $CDM, except at very low redshifts, while on the other side the exponential coupling models have a faster growth of density fluctuations during most of the expansion history of the Universe.\\

The main focus of the present paper is on the effects of variable couplings on nonlinear structure formation. Exploiting the implementation of coupled DE cosmologies into the N-body code {\small GADGET-2} developed for previous works, we have run high-resolution hydrodynamical N-body simulations for some selected cosmological models belonging to different classes of coupling evolution. For all these cosmologies initial conditions have been generated discarding possible early effects of the coupling based on the consideration that these effects were shown by previous studies to have a minor impact on the final properties of nonlinear structures at the present level of numerical resolution. We have also chosen to normalize all the cosmologies to the same amplitude of the large-scale power at the present time. Although this is a common choice, other conventions are equally valid and could lead to different predictions for the same cosmological models.\\

We have shown that the power spectrum of matter density fluctuations evolves in a strikingly different way in the two different types of models with respect to $\Lambda $CDM. In particular, the phenomenological couplings have a very similar evolution to $\Lambda $CDM until $z\sim 0.5$, followed by a very fast growth of the power spectrum amplitude at scales below $k \sim 1.0~h~\text{Mpc}^{-1}$. On the contrary, the exponential models have a lower amplitude than $\Lambda $CDM at all scales during most of the cosmic evolution, and catch up with the $\Lambda $CDM power spectrum only at large scales at $z=0$, while at small scales they still show a significant lack of power also at the present time.

We have then confirmed that also for variable couplings the scalar fifth-force acting only between CDM particles induces a bias in the amplitude of density fluctuations in baryons and CDM at all scales, as it happens for constant coupling models. The bias shows a clear scale dependence that develops very quickly as the couplings approaches their large values at low redshift, and the enhancement of this effect when moving from the linear regime of very large scales to smaller and progressively more nonlinear scales is found to be stronger than for the constant coupling models studied in previous works.

This bias can be detected also in the very nonlinear regime characterizing the inner parts of collapsed objects, and has an impact on the total amount of baryons contained in massive halos that could therefore influence the determination of the baryon fraction from cluster measurements. We have therefore computed the evolution of the average baryon fraction within the virial radius $r_{200}$ for all the halos arising in the simulations of the different cosmological models, finding a generally stronger reduction of the baryon fraction as compared to constant couplings, with a decrease up to $\sim 14 - 16 \%$ with respect to $\Lambda $CDM. 

We have computed the mass functions at different redshifts for all of our selected models, showing how at $z=0$ all the cosmologies have similar shapes and amplitudes of the mass function, with relative differences of the order of $\sim 10\%$. However, at higher redshifts the mass functions of the exponential coupling models show a clear excess of small halos and a strong lack of large halos with respect to $\Lambda $CDM, consistently with the later onset of structure formation in these models. 
We have also computed the multiplicity function for all the models and compared it with the theoretical predictions according to the \citet{Sheth_Tormen_1999} and \citet{Jenkins_etal_2000} fitting formulae, evaluated with the appropriate growth factor for all the models, and with the standard value of the extrapolated linear overdenisty at collapse. We found good agreement between the multiplicity functions in our simulated cosmologies and the theoretical fitting functions, with the only exception of one of the scale factor dependent models at $z=0$. This discrepancy might suggest the need to reconsider the spherical collapse formalism in the presence of strongly variable couplings, will be investigated in future works. Nevertheless, the usual mass function fitting formulae were found to be fairly accurate also for most of our variable coupling models, generalizing previous results.

Finally, we have investigated the effects of variable couplings on the halo density profiles. As for the case of constant coupling models, we find that they are still remarkably well fit in all the different cosmologies by the NFW formula. However -- in contrast with what happens within constant coupling models -- we also find that variable coupling cosmologies do not always show a decrease of the inner overdensity of halos with respect to the standard $\Lambda $CDM case, but present opposite trends for the two different classes of coupling functions. The more realistic and physically motivated exponential coupling models show a significant decrease of the inner overdensity of halos with respect to $\Lambda $CDM, while the phenomenological models show on the contrary a clear increase of the density in the central regions. This strikingly different behavior can be explained by considering which are the main physical mechanisms that can account for a modification of the equilibrium state of a collapsed halo in the context of our variable coupling models. As it was shown before for the case of constant couplings, the  mass decrease and the extra friction term in the equation of motion of CDM particles can only induce an increase of the total energy of a virialized system, which therefore restores its virial equilibrium by slightly expanding. This effect is the source of the lower densities in the cores of CDM halos in the presence of constant couplings. However, if the coupling is changing in time, there is an additional mechanism coming into play: the total potential energy of the system decreases due to the increase of the effective gravitational constant as a consequence of the growing scalar fifth-force. In our models, the effective gravitational constant grows by $\sim $30 - 75 \% during the whole cosmic evolution, which determines a corresponding decrease of the gravitational potential energy of collapsed systems. This decrease of total energy then determines a contraction of the halos.

Interstingly, we have shown that the two opposite behaviors found for the inner overdensity of nonlinear structures are determined by the background evolution of the coupled DE-CDM system: in the models that do not present a ``Growing $\phi $MDE" phase the friction term, which is the main driver of the expansion of halos in constant coupling models, is strongly suppressed and cannot counteract the contraction induced by the strong increase of the effective gravitational constant. 
On the contrary, the exponential coupling models, due to the presence of a ``Growing $\phi $MDE", have a still efficient friction term that balances and sometimes overcomes the effect of the potential energy decrease, thereby determining an overall expansion of the halos. 

We have therefore shown how the nonlinear behavior of matter particles at small scales, in the context of coupled DE models with time dependent couplings, is directly influenced by the cosmological background evolution of the scalar field, and how the presence of a ``Growing $\phi $MDE" phase is essential to determine whether halos will contract or expand in these cosmologies.

These considerations apply also to halo concentrations, which are found to be higher with respect to $\Lambda $CDM in the phenomenological models due to the absence of a ``Growing $\phi $MDE", and lower in the exponential coupling models.\\

In conclusion, we have performed a  complete numerical study of interacting DE cosmologies for a few general types of time dependence of the DE-CDM coupling, concerning background evolution, linear perturbations evolution, and nonlinear structure formation. We have presented the first high-resolution hydrodynamical N-body simulations of structure formation in the context of variable coupling models to date. In our analysis, we have found that differently from the constant coupling case, halo density profiles and halo concentrations do not evolve in the same direction with respect to $\Lambda $CDM for all types of coupling evolution. In particular, depending on the type of background evolution determined by the coupling function, density profiles can be less overdense and correspondingly less concentrated than in $\Lambda $CDM, or vice versa.
Furthermore, the growth of structures at large scales is also affected in a significantly different way according to the different types of coupling evolution.
Finally, we find that the decrease of the halo baryon fraction already found for constant coupling models can be significantly enhanced in variable coupling cosmologies.
Some of these effects alleviate tensions between astrophysical observations and the $\Lambda $CDM cosmology at small scales, and arise in cosmological models that contrarily to the constant coupling scenarios are not in stark conflict with present observational constraints on the background evolution of the Universe even in the presence of a significant coupling strength at low redshifts. 
Therefore, cosmological models with time dependent couplings in the dark sector might represent -- for some specific forms of coupling evolution -- a viable alternative to the standard $\Lambda $CDM concordance model.

\section*{Acknowledgments}

I am deeply thankful to Jochen Weller for reading the manuscript and providing precious comments, and to Luca Amendola for very useful discussions on the variable coupling models.
This work has been supported by 
the DFG Cluster of Excellence ``Origin and Structure of the Universe''.
All the numerical simulations have been performed on the Power6 cluster at the RZG computing centre in Garching.

\bibliographystyle{mnras}
\bibliography{baldi_bibliography}

\label{lastpage}

\end{document}